\documentclass{aa}
\usepackage{txfonts}
\usepackage{amsmath}
\usepackage{siunitx}
\usepackage{graphicx}
\usepackage{booktabs}
\usepackage{multirow}
\usepackage{makecell}
\usepackage{multicol} 
\usepackage{natbib}
\bibpunct{(}{)}{;}{a}{}{,} 
\usepackage{xcolor}
\usepackage{xspace}
\newcommand*{\scocen}{\mbox{\textit{SC15}}\@\xspace}
\newcommand*{\elup}{\mbox{$e$\,Lup}\@\xspace}
\newcommand*{\philup}{\mbox{$\phi$\,Lup}\@\xspace}
\newcommand*{\sigdrei}{\mbox{$\sigma_\mathrm{3D}$}\@\xspace}

\makeatletter
\renewcommand*\aa@pageof{, page \thepage{} of \pageref*{LastPage}}
\makeatother
\usepackage{hyperref}
\hypersetup{colorlinks, allcolors=blue}

\begin{document} 

\title{The evolution of velocity dispersion in the Sco-Cen OB association}

\author{Josefa~E.~Gro\ss schedl\inst{1,2}
          \and
          Jo\~ao Alves\inst{3,4}
          \and
          Sebastian Ratzenböck\inst{5} 
          \and
          N\'uria Miret-Roig\inst{6,7} 
          \and
          Sebastian Hutschenreuter\inst{3}
          \and
          Laura Posch\inst{3} 
          \and
          Alvaro Hacar\inst{3}
          }

\institute{Astronomical Institute of the Czech Academy of Sciences, Boční II 1401, 141 31 Prague 4, Czech Republic, \\
    \email{grossschedl@asu.cas.cz}
    \and
    Universit\"at zu K\"oln, I.~Physikalisches Institut, Z\"ulpicher Str.~77, 50937 K\"oln, Germany
    \and
    University of Vienna, Department of Astrophysics, T\"urkenschanzstra{\ss}e 17, 1180 Vienna, Austria
    \and
    University of Vienna, Data Science at Uni Vienna Research Platform, Austria
    \and
    Center for Astrophysics | Harvard \& Smithsonian, 60 Garden St., Cambridge, MA 02138, USA 
    \and
    Departament~de Física Quàntica i Astrofísica (FQA), Univ.~de Barcelona (UB), Martí i Franquès, 1, 08028 Barcelona, Spain
    \and
    Institut de Ciències del Cosmos (ICCUB), Univ.~de Barcelona (UB), Martí i Franquès, 1, 08028 Barcelona, Spain
    }
    
\date{Received May 14, 2025; Accepted March 4, 2026}

\abstract
{
We study how the stellar velocity dispersion within the Scorpius-Centaurus OB association (Sco-Cen) has evolved over approximately 20 million years, from its formation to the present day, by investigating 32 stellar clusters in Sco-Cen. Using data from the \textit{Gaia} mission along with supplementary stellar radial velocities, we identified a surprising sequence of abrupt jumps and intervening plateaus in the evolution of velocity dispersion correlating with times of star formation bursts. 
We find that the association is almost isotropically expanding and that star formation propagated from inside-out with a speed of about 5--6\,km\,s$^{-1}$.
We measure a present-day expansion rate of about 10--12\,pc\,Myr$^{-1}$ and observe that younger star clusters within the association exhibit higher velocities compared to older ones. This result, along with the stepwise increase in velocity dispersion over time, suggests a structured and sequential star formation process rather than a random one. This phased evolution suggests that stellar feedback is the primary driver of Sco-Cen's star formation history, expansion, and eventual dispersal. Our findings emphasise the value of precisely characterising stellar populations within OB associations, particularly through the creation of detailed, high-resolution age maps.
}
\keywords{Methods: data analysis -- Stars: kinematics and dynamics -- Galaxy: open clusters and associations: individual: Sco-Cen}
\maketitle
\defcitealias{Ratzenboeck2023a}{R23a}
\defcitealias{Ratzenboeck2023b}{R23b}
\defcitealias{MiretRoig2025}{MR25}
\defcitealias{Posch2023}{P23}
\defcitealias{Posch2025}{P25}
\defcitealias{ElmegreenEfremov1998}{El\&Ef98}
\defcitealias{EfremovElmegreen1998}{Ef\&El98}

\section{Introduction}\label{sec:intro}

Understanding the temporal evolution of OB associations is crucial for understanding the role of massive stars in shaping their environments \citep[e.g.,][]{Brown1997}. OB associations represent an essential, yet transient phase in the life cycle of massive stars and star-forming regions \citep[e.g.,][]{Blaauw1964a, Wright2023}. The traditional view of OB associations has been significantly refined with \textit{Gaia} data \citep{Brown2016}. Recent \textit{Gaia}-based studies reveal that OB associations comprise dozens of largely unbound stellar populations with ages up to roughly 20 Myr \citep[e.g.,][]{Kos2019, Chen2020, Kerr2021, Ratzenboeck2023b, Hunt2023}, rather than just a few subgroups. Moreover, they are likely linked to even larger and older cluster families, which could trace their origins to spiral arm activity \citep{Swiggum2024, Swiggum2025}.     

A comprehensive characterisation of stellar groups within OB associations is crucial for understanding their formation mechanisms and boundness state, as they ultimately disperse and merge with the field star population \citep[e.g.,][]{LyngaPalous1987, Kamaya2004}. OB associations also provide insights into star formation mechanisms, stellar feedback, and early dynamical evolution. 
Their velocity dispersion is a key factor, which provides insights into the internal motions, dynamical states, and evolutionary histories of stellar populations, eventually shaping the structure of the Galactic field population \citep[e.g.,][]{Lada1984, LadaLada2003, Kroupa1995, Kroupa2008, deLaFuenteMarcos2008, Kuhn2019}. 

Although velocity dispersion is widely recognised as a fundamental population property, its temporal evolution during the formation of a single stellar association remains largely unexplored. Previous observational studies provide only snapshots in time, which limit our understanding of how stellar populations evolve dynamically. The lack of observational data on the temporal evolution of velocity dispersion represents a major gap in our understanding of stellar population formation. 

To address this, we use high-precision \textit{Gaia} DR3 data \citep{GC-Vallenari2023}, supplemented with ancillary radial velocity (RV) measurements, to investigate the temporal evolution of velocity dispersion in the Scorpius-Centaurus OB association \citep[Sco-Cen; e.g.,][]{Blaauw1964a, Blaauw1964b, Preibisch2008}. This study builds on the recent identification of more than 30 coeval and comoving stellar clusters\footnote{We use the term ``cluster'' in a statistical sense, as in \citetalias{Ratzenboeck2023a}.} within Sco-Cen with ages of approximately 3 to 21 Myr \citep[see][hereafter, \citetalias{Ratzenboeck2023a}, \citetalias{Ratzenboeck2023b}, \citetalias{MiretRoig2025}]{Ratzenboeck2023a, Ratzenboeck2023b, MiretRoig2025}. These studies identified spatio-temporal patterns, using high-resolution age maps, indicative of sequential star formation. This enabled the identification of cluster chains with well-defined age, mass, position, and velocity gradients extending outward at the periphery of the association (see \citealt{Posch2023, Posch2025}, hereafter, \citetalias{Posch2023}, \citetalias{Posch2025}). In this paper, we aim to reconstruct, for the first time, the temporal evolution of velocity dispersion and internal motions of an OB association by analysing 32 well-defined clusters in the 6D phase space over a time span of about 20\,Myr.

\section{Data} \label{sec:data}

We use the Sco-Cen cluster sample from \citetalias{Ratzenboeck2023a}, which was selected using the unsupervised machine-learning tool \texttt{SigMA} (Significance Mode Analysis), containing 34 clusters related to Sco-Cen. We add two clusters from the TW Hydrae association (TWA), which are connected to Sco-Cen as a cluster chain \citepalias[see][]{{MiretRoig2025}}. The combined sample contains a total of 13,011 stellar members.
\citetalias{Ratzenboeck2023b} determined cluster ages by fitting PARSEC model isochrones \citep{Bressan2012} with a Bayesian inference approach. We use the isochronal ages fitted to the \textit{Gaia} colour--absolute-magnitude diagram $G_\mathrm{abs}$ vs $G_\mathrm{BP}-G_\mathrm{RP}$ (BPRP-PARSEC ages). \citetalias{MiretRoig2025} estimate the cluster ages of TWA-a,b consistently with the same method as \citetalias{Ratzenboeck2023b}. The Sco-Cen clusters cover ages from about 3 to 21\,Myr, and are used as time information in our analysis. A detailed data description is given in Appendices~\ref{apx:cluster-sample} \& \ref{apx:parameters}.

To study the clusters in 6D phase space, we combine the astrometric 5D data with RV data. \textit{Gaia} DR3 provides RV measurements \citep{Katz2023} for about 38\% of our sources, however, only about 11\% have relatively low uncertainties (\mbox{$e_{\mathrm{RV},\mathit{Gaia}}<3.1$\,km\,s$^{-1}$})\footnote{
As in our quality criteria (Appendix~\ref{apx:quality}); chosen to ensure that also sparse clusters contain several stars with measured RVs.} 
with a median error of  about 1.5\,km\,s$^{-1}$. Moreover, some of the relatively sparse clusters contain no or very few stars with \textit{Gaia} RVs. Hence, we add supplementary RV data from 22 spectroscopic surveys (see Appendix~\ref{apx:rv_comparison} \& Table~\ref{tab:ref}). After the cross-match, about 50\% of the sources have RV measurements from at least one survey. 
To determine robust 3D space motions, we apply several cleaning steps, including a cut using the cluster stability value from \citetalias{Ratzenboeck2023a}, an RV error cut at $e_{\mathrm{RV}}<3.1$\,km\,s$^{-1}$, removal of binary candidates, and a global outlier cut and 3$\sigma$-clipping, as outlined in detail in Appendix~\ref{apx:quality}. Finally, our RV sample contains about 25\% of the original stellar sample (3240/13,011).
The median RV error of this sample is about 0.4\,km\,s$^{-1}$ (min/max = 0.010/3.098~km\,s$^{-1}$). A detailed overview of the data statistics per cluster is given in Table~\ref{tab:overview}. 
 
Eventually, we use 32 out of the 36 Sco-Cen clusters. We find that the sparse cluster $\mu$\,Sco has very poor RV statistics, and we remove it from further analysis in this work. Moreover, we remove three clusters from the Chamaeleon-Musca-Coalsack region (Chamaeleon I \& II, Centaurus-Far). They are slightly detached from Sco-Cen and probably belong to a background structure, called the ``C'' \citep{Edenhofer2024b}. The remaining 32 analysed clusters contain a total of 12,612 stellar candidate members (containing 3123 sources with valid RVs), while we also report statistics for the four removed clusters in Table~\ref{tab:overview}.

\section{Methods} \label{sec:methods}

We calculate the velocity dispersion from the stellar members of the clusters in 3D, using the Galactic Cartesian velocities (\textit{UVW}, see Appendix~\ref{apx:parameters}), after applying the quality criteria from Appendix~\ref{apx:quality}:
\begin{small}
\begin{equation}
    \sigma_\mathrm{3D} \, (\mathrm{km}\,\mathrm{s}^{-1}) = 
        \sqrt{
            \sigma_{U}^2 + 
            \sigma_{V}^2 + 
            \sigma_{W}^2 } 
    \label{eq:sigma3D}
\end{equation}
\end{small}
The one-dimensional velocity dispersions ($\sigma_{U}$, $\sigma_{V}$, $\sigma_{W}$) were calculated via the standard deviation in the three velocity spaces.
We use the cluster ages as time information to investigate the evolution of velocity dispersion in 3D. This is possible for the first time, as we have a coherent sample of clusters that formed in a single OB association with a wide range of ages and with 3D information. 

We calculate \sigdrei cumulatively by progressively incorporating member stars of the next younger cluster at each time step for its calculation. In other words, we start the calculation of the cumulative \sigdrei with only the stellar members of the oldest cluster (\elup), then add the stars of the next youngest cluster at the next step, and finally, we end with all member stars from all studied 32 clusters. We calculate the cumulative \sigdrei by iteratively picking equal subsamples from each cluster at each step, to account for the different cluster sizes (number of stellar members), as detailed in Appendix~\ref{apx:methods:calc-veldisp}.

Next, we estimate the present-day spatial arrangement of the clusters in Sco-Cen in chronological order by ordering the clusters by age, named cumulative size ($S$, pc). To this end, we determine the maximum cluster extent as observed today by measuring the minimal distance across the connections using a k-nearest neighbour graph of the member stars in the $XYZ$ space. We construct this graph from individual stellar members as nodes, and we use Dijkstra's algorithm to compute the shortest paths between any two sources along the edges of this graph. We do this again cumulatively by starting with member stars from the oldest cluster and ending with all member stars.
Thus, at each step, we compute the maximally possible path distance that you can take between any two pairs of Sco-Cen member stars older than the given age step.
More details are given in Appendix~\ref{apx:methods:calc-size}.

The here calculated present-day cumulative $S$ does not reflect the evolution of size over time, since we do not include orbital tracebacks of the clusters at this stage, but only order the clusters by their decreasing age.
The cumulative sizes calculated in this manner can be interpreted as an upper limit of the region’s size at a given age (“time”), while it does not show the true physical extent before the present day, which was likely smaller. 
We will look into a more detailed traceback analysis in future work, where various second-order effects and required assumptions will be taken into account (e.g., cluster expansion, unknown acceleration from feedback, non-sphericity, Galactic potential, possible internal gravitational effects, orbit integration with different Solar parameters; see Sect.~\ref{sec:discuss-larson}).

We further investigate the clusters' 3D bulk positions and motions, using the cluster medians in Galactic Cartesian coordinates ($XYZ$) and velocities ($UVW$). The corresponding uncertainties are determined by bootstrap sampling from the cluster members (Appendix~\ref{apx:methods-average-3D}). 
We calculate the position and velocity vectors for each cluster relative to a chosen centre (\textit{XYZ}$_\mathrm{ctr}$) and  rest velocity (\textit{UVW}$_\mathrm{rest}$): 
\begin{small}
\begin{equation}
\begin{aligned}
    & \textit{\textbf{r}} = ((X-{X}_\mathrm{ctr}), (Y-{Y}_\mathrm{ctr}), (Z-{Z}_\mathrm{ctr})) \\
    & \textit{\textbf{v}} = ((U-{U}_\mathrm{rest}), (V-{V}_\mathrm{rest}), (W-{W}_\mathrm{rest})) 
    \label{eq:vel-pos-vector}
\end{aligned}
\end{equation} 
\end{small} 
Next, we calculate the relative distances of the clusters ($r$, vector norm, in pc) and the relative velocities  ($v$, or speed, in $\mathrm{km}\,\mathrm{s}^{-1}$), via the Euclidean distance in position and velocity space to the given reference point.
{\small
\begin{equation}
  \begin{aligned} 
    & r = \|\textit{\textbf{r}}\| = \sqrt{(X-{X}_\mathrm{ctr})^2 + (Y-{Y}_\mathrm{ctr})^2 + (Z-{Z}_\mathrm{ctr})^2 }  \\
    & v = \|\textit{\textbf{v}}\| = \sqrt{(U-{U}_\mathrm{rest})^2 + (V-{V}_\mathrm{rest})^2 + (W-{W}_\mathrm{rest})^2 } 
    \label{eq:v-d-rel}
  \end{aligned}
\end{equation}
}
We determine the radial component of the velocities relative to the chosen reference point as follows: 
\begin{equation}
    \label{eq:radial_component}
    v_r = \frac{\textit{\textbf{r}} \cdot \textit{\textbf{v}}}{r}
\end{equation}
The tangential component is then computed with:
\begin{equation}
    \label{eq:tangential_component}
    v_\mathit{tan} = \sqrt{v^2 - v_r^2}
\end{equation}

To define the reference point (centre and rest velocity), we use three approaches to investigate possible systematics. The centre of Sco-Cen is not a well-defined point and depends also on the question at hand (e.g., centre of feedback, centre of mass, or using older clusters to find the point of first star formation). 
First, we use the median position and motion of the oldest cluster in the region, \elup (age $\sim$21\,Myr), which is located at a central position. We assume that its velocity is reminiscent of the original velocity of the primordial star-forming molecular cloud when the first stars formed. 
Additionally, we use the cluster \philup, another central cluster in Sco-Cen with an age of $\sim$17\,Myr, which is connected to two chains of clusters \citepalias[see][]{Posch2023, Posch2025}.
Finally, we use the median motion of the stellar members of the oldest clusters in Sco-Cen with ages\,$>15$\,Myr, denoted as \scocen, to get another estimate of the bulk motion of the early star-forming region (see similar approaches in \citetalias{Posch2025} and \citetalias{MiretRoig2025}).
More details on the determination of \scocen and a comparison of the three chosen centres (\elup, \philup, \scocen) are given in Appendix~\ref{apx:methods-bulk}. 

We further investigate the biasing effects from choosing different reference points in our subsequent analysis by using each of the 32 Sco-Cen clusters once as a reference point. This creates cases that are not ordered by cluster age or that are starting from a non-central location, as detailed in Appendix~\ref{apx:test-random-speed}.

\begin{figure}[!t]
    \centering
        \includegraphics[width=1\linewidth]{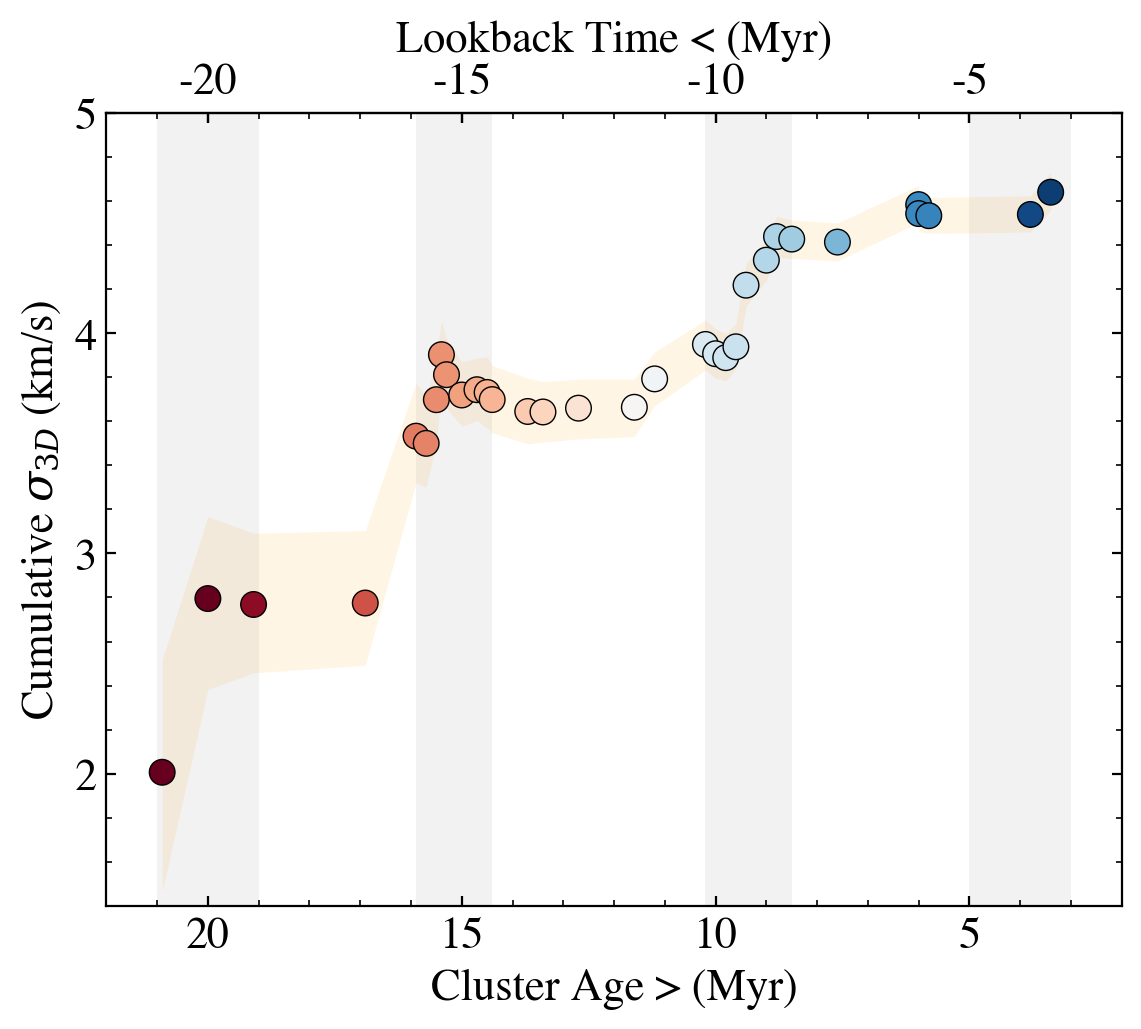}
    \caption{
    Cumulative 3D velocity dispersion of Sco-Cen, ordered by decreasing cluster age.
    The orange shaded area shows the 95\% interquartile ranges (2$\sigma$ bound), highlighting the uncertainties of the trend (see Appendix~\ref{apx:methods:calc-veldisp}). The symbols are coloured by the formation time of the youngest cluster included in the cumulative calculation (colour correlates with x-axis). We indicate the Lookback Time at the top of the x-axis, since the cumulative \sigdrei could also be interpreted as the evolution of velocity dispersion over time.
    The four vertical grey bars indicate the four main star formation events in Sco-Cen, marking periods of increased star formation rate, as discussed in \citetalias{Ratzenboeck2023b} (see their Fig.~3).
    }
    \label{fig:cumul-vel-disp}
\end{figure}

\section{Results} \label{sec:results}

\subsection{Cumulative velocity dispersion} \label{sec:results:cumulative-veldisp}

Figure~\ref{fig:cumul-vel-disp} shows the cumulative 3D velocity dispersion (\sigdrei) of Sco-Cen as a function of cluster age (or lookback time). We observe a general increase in velocity dispersion over time, with a pattern of jumps and plateaus, reaching a final value of 4.64$\pm$0.04\,km\,s$^{-1}$. This cumulative value is determined by iterating over equally sized sub-samples within each cluster (see details in Appendix~\ref{apx:methods:calc-veldisp}).
For comparison, the total \sigdrei, calculated using all stellar members of Sco-Cen that meet our quality criteria (without sub-sampling), yields a value of 3.97$\pm$0.03\,km\,s$^{-1}$. The different values of the total 3D velocity dispersion are caused by the sub-sampling approach that we use for the cumulative \sigdrei, to give similar weight to each cluster, while in the case when using all available stars, the more massive clusters might dominate the total \sigdrei. Moreover, the iterative sub-sampling approach likely creates some under-sampling of the velocity space, which gives more weight to individual measurement errors, artificially inflating the velocity dispersion, while the shape of the cumulative trend appears unaffected, as demonstrated in the following paragraph and in more detail in Appendix~\ref{apx:methods:calc-veldisp}.

We test the shape and robustness of the cumulative trend by calculating the cumulative \sigdrei, first, by applying stricter error cuts ($e_\mathrm{RV} < 1$\,km\,s$^{-1}$), and second, by using the $UVW$ medians of the 32 clusters instead of individual stars. In the first case, we get a total velocity dispersion of 4.46$\pm$0.03\,km\,s$^{-1}$, and in the second case 4.13$\pm$0.06\,km\,s$^{-1}$. 
Additionally, we test the effect of binaries that might be contained in our sample, using the {\it Gaia} RUWE parameter to roughly exclude binary candidates, which results in effectively the same trend as in Fig.~\ref{fig:cumul-vel-disp}. Generally, all tests produce similar increasing trends of the cumulative \sigdrei over time, while only the magnitude of \sigdrei gets shifted (see  Appendix~\ref{apx:methods:calc-veldisp} \& Fig.~\ref{fig:cumul-vel-clusters})

We conclude that the whole Sco-Cen association has a 3D velocity dispersion of about 4--4.7\,km\,s$^{-1}$ (see Table~\ref{tab:sig3D}), whereas individual clusters within the association have significantly smaller dispersions, on the order of 1--2\,km\,s$^{-1}$ per cluster\footnote{We discuss the individual clusters' velocity dispersions in future work, since this requires a more careful selection of RVs and ideally a higher quantity of high quality RVs. See also the caveats outlined in Appendix~\ref{apx:rv_comparison}).}. 
We find that the shape of the cumulative \sigdrei is largely unaffected by measurement errors or binaries, since only the magnitude of the \sigdrei values shifts consistently to somewhat lower values when applying more conservative cuts (see Fig.~\ref{fig:cumul-vel-clusters}).

\begin{figure*}[!ht]
    \centering
        \includegraphics[width=0.84\linewidth]{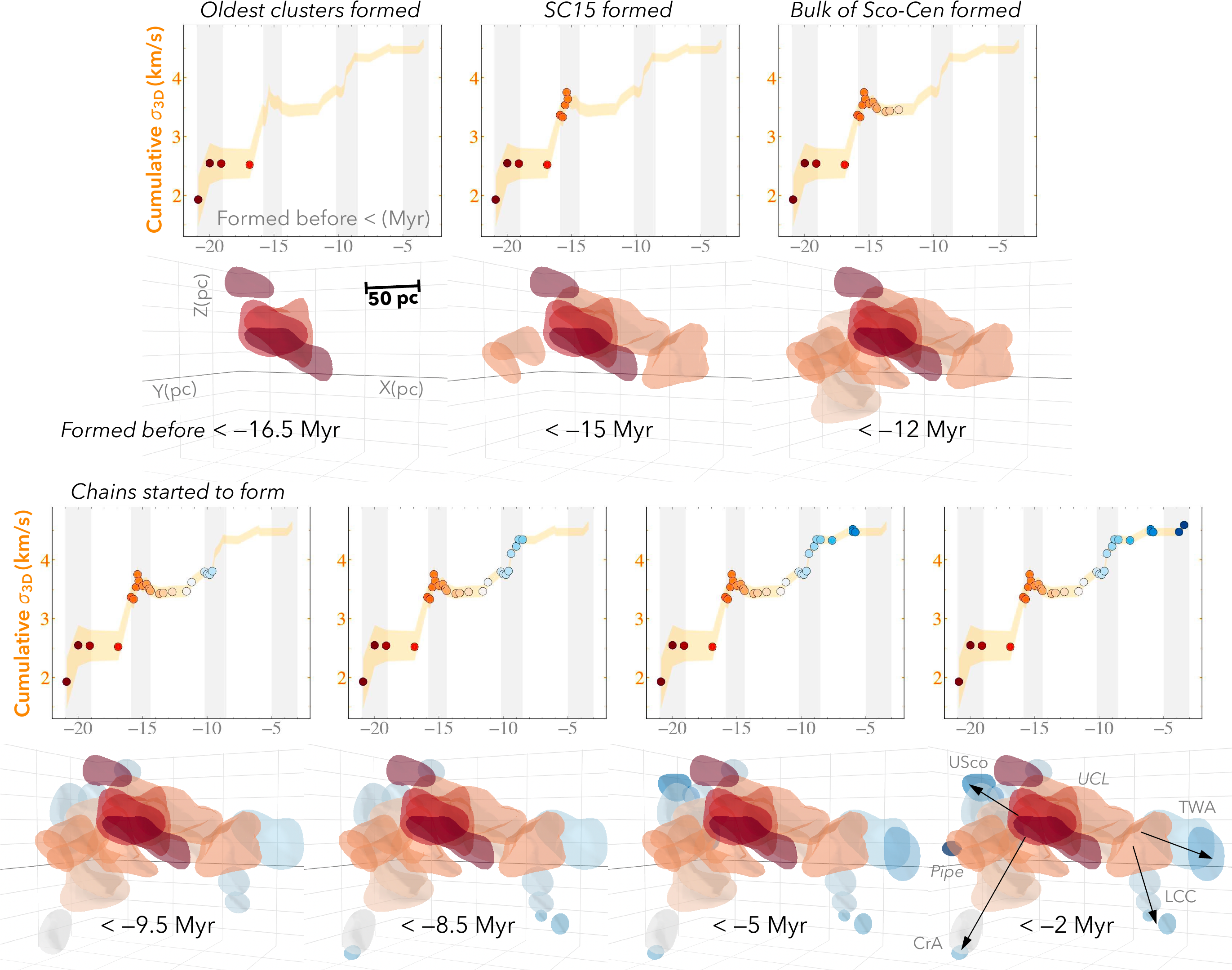}
    \caption{
    3D spatial distribution of clusters in Sco-Cen is shown together with the cumulative \sigdrei. The seven panels show seven age ranges, which dissect the chronological build-up of Sco-Cen.
    The upper panels display the cumulative 3D velocity dispersion (as in Fig.~\ref{fig:cumul-vel-disp}, {\it left panel}).
    Below each graph, seven 3D age maps depict the present-day spatial distribution of clusters (in $XYZ$).
    The clusters are represented by their enveloping surfaces and colour-coded by age.
    Each of the seven panels only shows the clusters that formed before the indicated cluster formation times. 
    The shown cluster sizes can be interpreted as upper limits of the region’s size at a given age and not as the true physical extent before the present-day, which was likely smaller (see also Sect.~\ref{sec:discuss-larson}).
    The figure illustrates which clusters contribute to which jumps or steps in the cumulative \sigdrei.
    Sub-regions and cluster chains are labelled in the final panel. The 3D visualisations originate from the studies \citetalias{Ratzenboeck2023b} and \citetalias{MiretRoig2025}; interactive versions are available in the respective publications.  
    }
    \label{fig:build-up}
\end{figure*}
\begin{figure}[!h]
    \centering
        \includegraphics[width=0.92\linewidth]{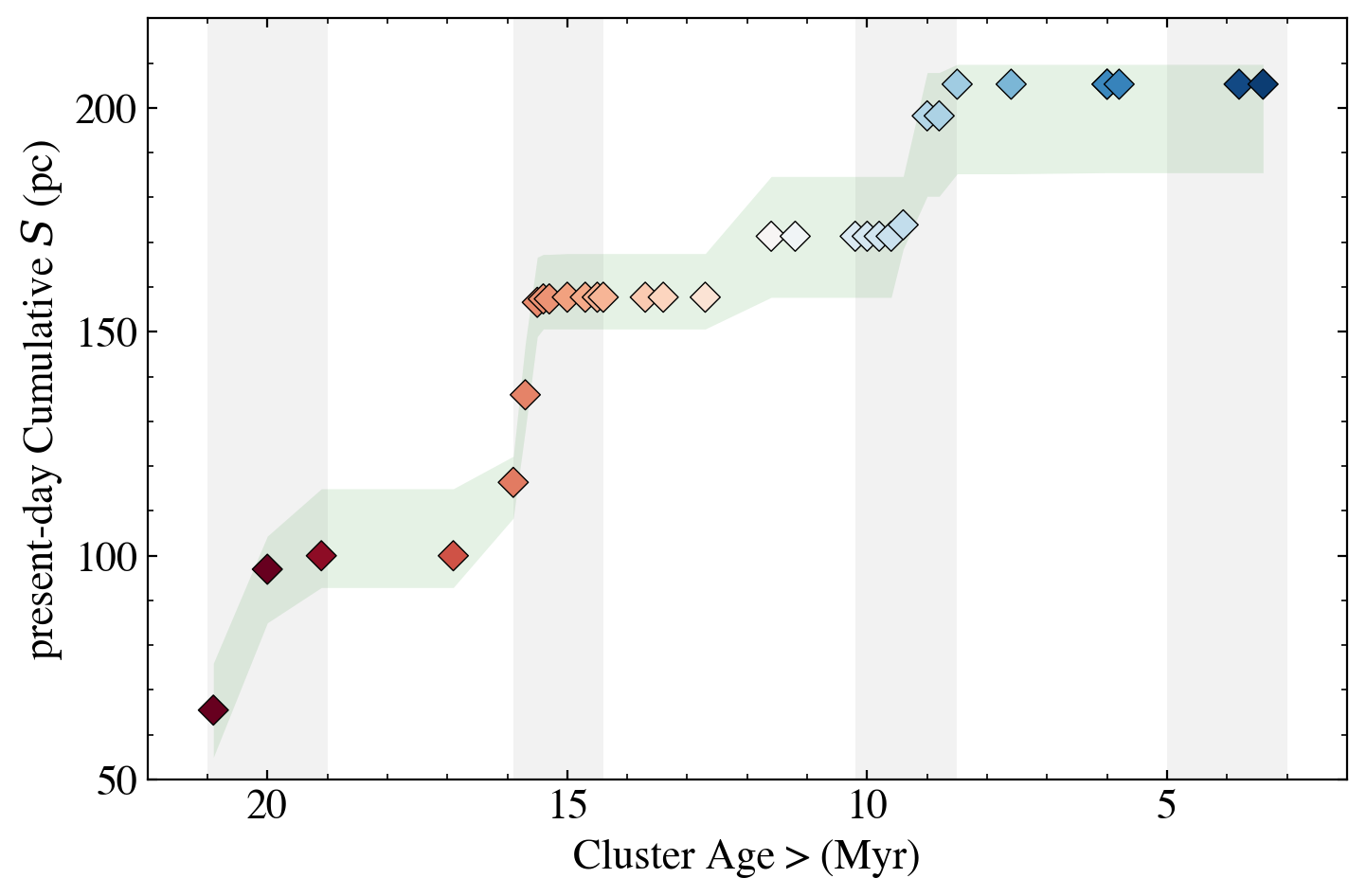}
    \caption{
    Present-day cumulative size of Sco-Cen, when ordering and adding the stellar cluster members by their decreasing age (without considering orbital trace-backs). The sizes can be seen as upper limits of the region’s size at a given age, but not as the true physical extent before the present-day, which was likely smaller (see Sect.~\ref{sec:discuss-larson}).
    The green shaded area shows the 95\% interquartile range (2$\sigma$) (see Appendix~\ref{apx:methods:calc-size}). The colours and grey bars are as in Fig.~\ref{fig:cumul-vel-disp}.
    }
    \label{fig:cumul-size}
\end{figure}

Figure~\ref{fig:build-up} highlights the present-day 3D distribution of the Sco-Cen clusters, and it shows which clusters contribute to the patterns in the cumulative \sigdrei. This figure visualises the arrangement of Sco-Cen clusters at seven age ranges, systematically including younger clusters. The cluster positions in 3D represent the present-day locations, since we do not consider orbital tracebacks, which will be considered in future work to better understand the true spatial evolution of Sco-Cen. Hence, the shown cluster sizes can be interpreted as upper limits of the region’s size at a given age and not as the true physical extent before the present-day, which was likely smaller (see also Sect.~\ref{sec:discuss-larson}).
Nevertheless, already in the visualisation shown in Fig.~\ref{fig:build-up} it becomes clear that the increasing velocity dispersion within Sco-Cen is connected to an evolutionary pattern, from the inside out. Changes in \sigdrei likely appear when star formation proceeds to an adjacent gas reservoir; notably, the onset of the formation of the cluster chains correlates with an increase and jump in \sigdrei.
To further investigate the origin of the increase in the velocity dispersion during the evolution of Sco-Cen, we look in more detail into the relative 3D space motions and positions of the Sco-Cen clusters in the following Sect.~\ref{sec:results:speed}.

To quantify the age-ordered, inside-out patterns visible in the 3D distribution of clusters in Fig.~\ref{fig:build-up}, we calculate Sco-Cen's size chronologically in a cumulative manner.
Figure~\ref{fig:cumul-size} shows the present-day cumulative size ($S$, pc); hence, it highlights the cumulative chronological arrangement of Sco-Cen clusters, when ordering and adding the member stars by the age of their parent clusters, calculated cumulatively with decreasing cluster age. We like to emphasise that the presented trend does not show the evolution of size, since we do not consider orbital tracebacks at this stage. Thus, as mentioned above, the presented sizes at given ages can be rather seen as upper limits, since the region was likely more compact in the past, based on a preliminary traceback analysis (see Sect.~\ref{sec:discuss-larson}).
The total present-day extent of Sco-Cen is 205$^{+4}_{-20}$\,pc, when calculated with the method described in Appendix~\ref{apx:methods:calc-size} and when including all clusters. 
The size increases when adding stars from clusters ordered by decreasing age, which is consistent with the visible spatio-temporal patterns identified in \citetalias{Ratzenboeck2023b} (Fig.~\ref{fig:build-up}) and the cluster chains discussed in \citetalias{Posch2023}, \citetalias{Posch2025}, and \citetalias{MiretRoig2025}. These studies propose an inside-out formation history, which was also suggested earlier for Sco-Cen \citep[e.g.,][]{Preibisch1999, Gaczkowski2017, Krause2018}. 
Interestingly, we see that the present-day cumulative size increases similarly to the cumulative \sigdrei, which we discuss further in Sect.~\ref{sec:discuss-larson}.

\begin{table*}[!ht]
\begin{small}
\centering
\caption{
Linear fitting results (slope $a$, intercept $b$) of the speed ($v$) versus time ($t$) and radial-motion ($v_\mathrm{r}$) versus distance ($r$) relations when using three different reference points (see Figs.\,\ref{fig:age-speed-distance}, \ref{fig:hubble-flow}, and \ref{fig:speed+hubble-other}). 
}
\renewcommand{\arraystretch}{1.3}
\begin{tabular}{lrrrrrr}

\hline \hline
\multicolumn{1}{c}{} &
\multicolumn{3}{c}{speed--time relation} &
\multicolumn{3}{c}{radial-motion--distance relation} \\

\multicolumn{1}{c}{} &
\multicolumn{3}{c}{$v = b + a \cdot t$} &
\multicolumn{3}{c}{$v_\mathrm{r} = b + a \cdot r$} \\
\cmidrule(lr){2-4}
\cmidrule(lr){5-7}

\multicolumn{1}{l}{} &
\multicolumn{1}{c}{$a = \frac{dv}{dt}$} &
\multicolumn{1}{c}{$b$} &
\multicolumn{1}{c}{$b$} &
\multicolumn{1}{c}{$a = \frac{dv_\mathrm{r}}{dr}$} &
\multicolumn{1}{c}{$b$} &
\multicolumn{1}{c}{$a^{-1}$} \\

\multicolumn{1}{l}{Reference Point} &
\multicolumn{1}{c}{km\,s$^{-1}$\,Myr$^{-1}$} &
\multicolumn{1}{c}{km\,s$^{-1}$} &
\multicolumn{1}{c}{pc\,Myr$^{-1}$} &
\multicolumn{1}{c}{km\,s$^{-1}$\,pc$^{-1}$} &
\multicolumn{1}{c}{km\,s$^{-1}$} & 
\multicolumn{1}{c}{Myr} \\

\cmidrule(lr){1-7}

$e$\,Lup (20.9\,Myr) & 
0.45$^{+0.08}_{-0.07}$ & 10.55$^{+0.87}_{-0.84}$ & 10.8$\pm$0.9 
& 0.092$^{+0.009}_{-0.009}$ & -2.02$^{+0.56}_{-0.60}$ & 
10.6$^{+1.2}_{-1.0}$ \\

$\phi$\,Lup (16.9\,Myr) & 
0.42$^{+0.07}_{-0.06}$ & 9.54$^{+0.84}_{-0.82}$ & 9.8$\pm$0.9 
& 0.082$^{+0.010}_{-0.010}$ & -1.31$^{+0.58}_{-0.54}$ & 
11.9$^{+1.6}_{-1.3}$ \\

\scocen (>15\,Myr) &
0.34$^{+0.06}_{-0.06}$ &  8.44$^{+0.71}_{-0.71}$ & 8.6$\pm$0.7  
& 0.071$^{+0.011}_{-0.010}$ & -0.91$^{+0.58}_{-0.66}$ & 
13.7$^{+2.4}_{-1.8}$ \\

\hline
\end{tabular}
\label{tab:acceleration}
\tablefoot{
Col.\,1 lists the three chosen reference points, with the cluster age in brackets or the age-limit in the case of \scocen.
The slope of the speed--time relation gives the expansion over time (Col.\,2). The intercept of the speed--time relation gives the expansion speed of Sco-Cen at the present day, given in km\,s$^{-1}$ in Col.\,3, and converted to pc\,Myr$^{-1}$ in Col.\,4.   
In Col.\,7, the inverse of the slope of the radial-motion--distance relation gives the time of the possible onset of the expansion.
}
\renewcommand{\arraystretch}{1}
\end{small}
\end{table*}

\begin{figure}[!t]
    \centering
    \includegraphics[width=0.95\linewidth]{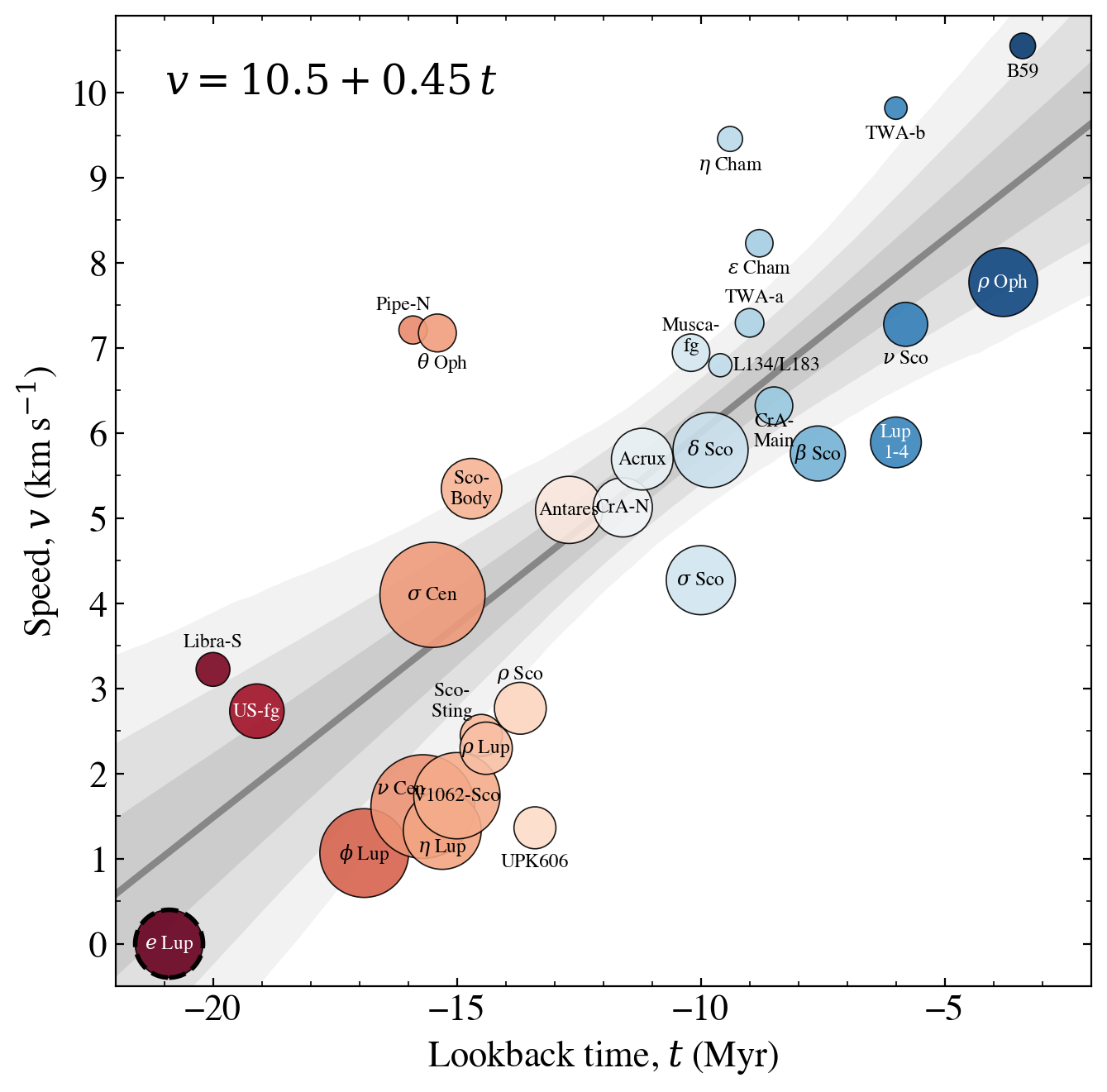}
    \caption{Speed--time relation. Relative cluster speed ($v$) versus lookback time ($t$), with the oldest cluster (\elup) as reference point that is excluded from the linear fit (black, dashed circle). The symbols are colour-coded by formation time (or cluster age, see x-axis) and scaled by number of sources per cluster. The linear fit (grey, solid line) is obtained via bootstrapping, with the median fitting parameters given in the upper left corner. The fit uncertainties are plotted as grey shaded areas (1-2-3-$\sigma$).
    } 
    \label{fig:age-speed-distance}
\end{figure}
\begin{figure}[!t]
    \centering
    \includegraphics[width=0.95\linewidth]{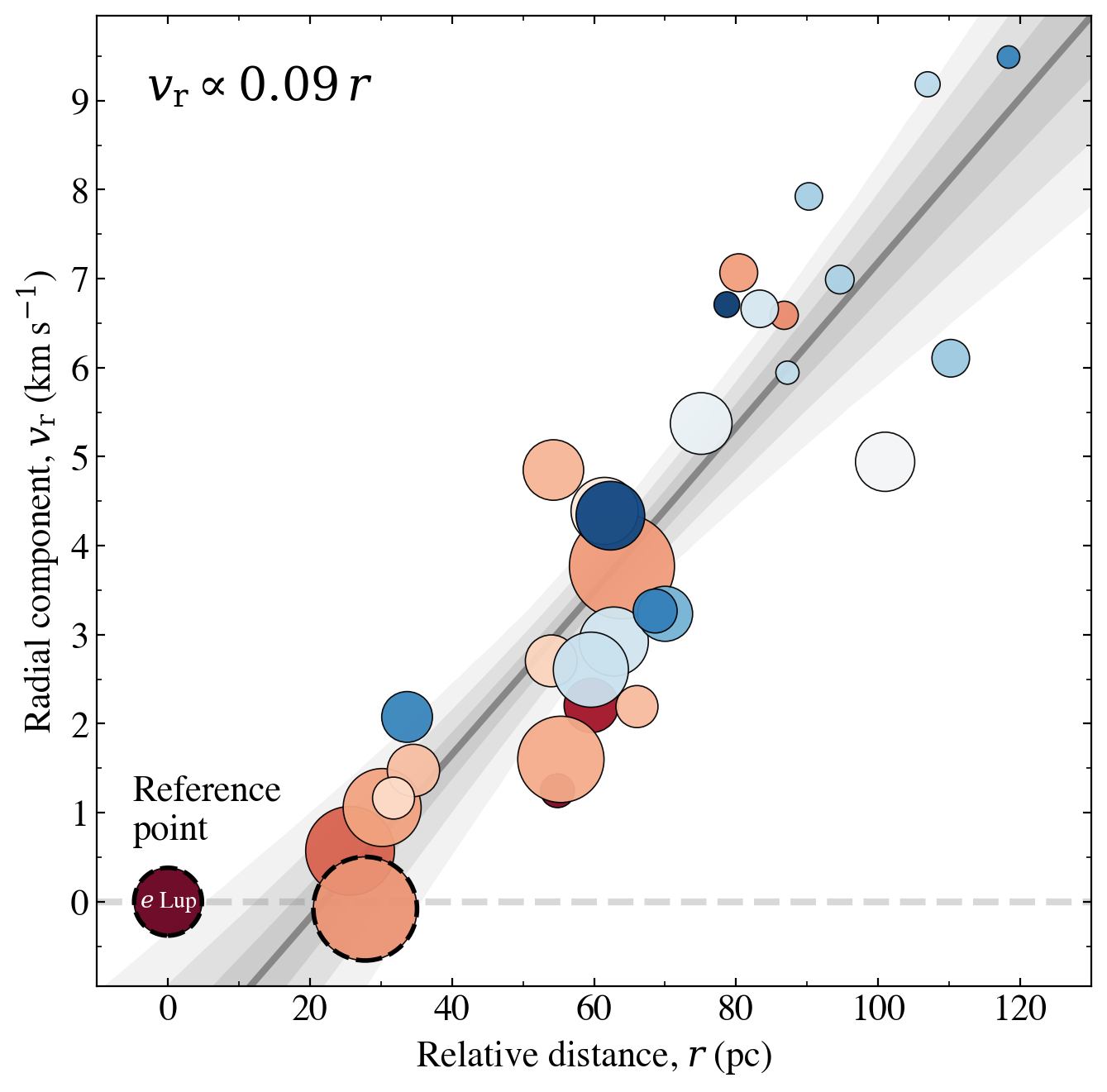}
    \caption{Radial-motion--distance relation. Radial component of the relative cluster motions ($v_r$) versus relative cluster distance ($r$), with \elup as reference point. The symbols are colour-coded by formation time, as in Fig.~\ref{fig:age-speed-distance}. A linear fit to the data is shown as a grey solid line, with the best fitting slope given in the upper left corner, and the 1-2-3-$\sigma$ fit uncertainties are plotted as grey shaded areas. The two clusters marked by black, dashed circles (\elup, $\nu$\,Cen) are excluded from the linear fit.
    } 
    \label{fig:hubble-flow}
\end{figure}

\begin{figure*}[!t]
    \centering
        \includegraphics[width=0.95\linewidth]{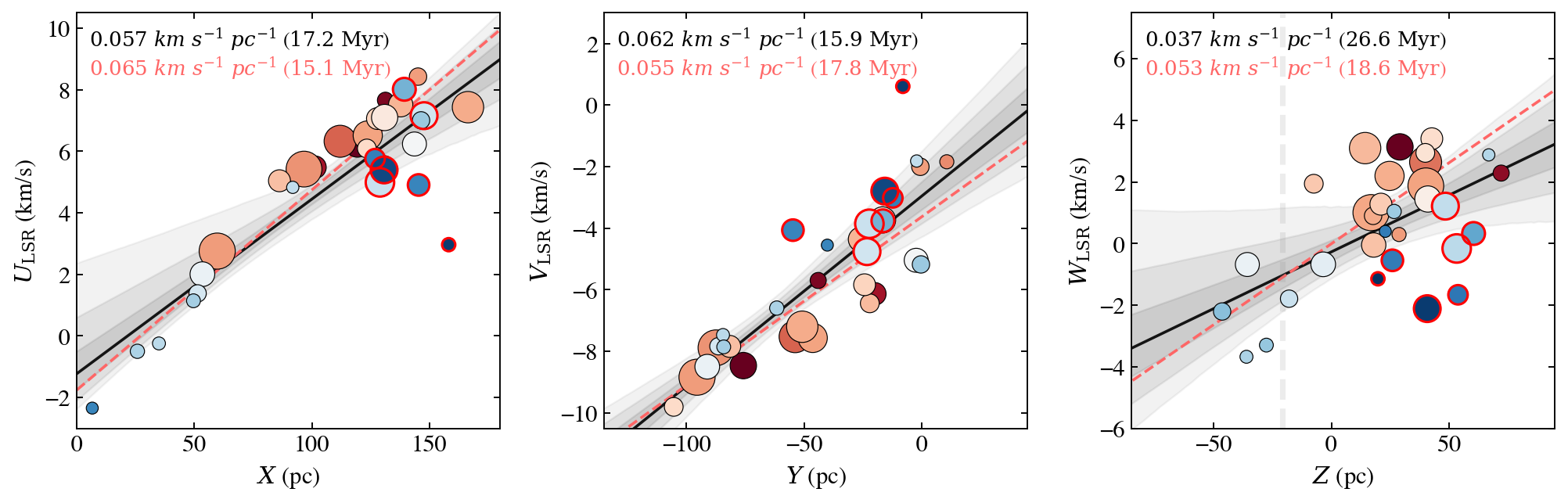}
    \caption{Position velocity (PV) diagrams for the 32 Sco-Cen clusters using the $XU$, $YV$, and $ZW$ spaces. Symbols and colours as in Fig.~\ref{fig:age-speed-distance}. The median slopes of the fitted linear regressions (black, solid lines) are given in the upper left corners. The grey shaded areas are the 1-2-3-$\sigma$ uncertainty ranges. The inverse of the slope gives a time in Myr, in brackets, marking the possible onset of expansion. The red, dashed slopes are the linear fitting results when using only 25 clusters, after removing seven clusters with peculiar tangential motions, marked with red circles (see Sect.~\ref{sec:tangential}). In the $ZW$ diagram ({\it right panel}), the vertical grey, dashed line indicates the approximate location of the Galactic mid-plane, assuming that the Sun is located about 21\,pc above the plane \citep[e.g.,][]{BennettBovy2019}.
    }
    \label{fig:PV}
\end{figure*}

\subsection{Correlation of time, cluster motion, and position} \label{sec:results:speed}

Figure~\ref{fig:age-speed-distance} shows the cluster speed ($v$) versus the time of cluster formation ($t$), denoted as the look-back time, with the oldest cluster, \elup, as a reference point. 
We find a linear relation between speed and time, with the resulting fitting parameters reported in Table~\ref{tab:acceleration}. 
The slope of the relation is about 0.45\,km\,s$^{-1}$\,Myr$^{-1}$, where younger clusters have a systematically higher relative speed. The intercept gives about 11\,pc\,Myr$^{-1}$ (units converted from km\,s$^{-1}$), which could be interpreted as expansion velocity at the present day. However, the speed calculated in this manner (Eq.~(\ref{eq:v-d-rel})) does not give information about the direction of motion; therefore, we also dissect the relative motions into a radial and tangential component (Eqs.~(\ref{eq:radial_component}--\ref{eq:tangential_component})). 

Figure~\ref{fig:hubble-flow} shows the radial component of the relative cluster motions ($v_r$) versus relative distance ($r$), again with \elup as reference point. 
First, we find that the majority of clusters have positive values in $v_r$, which shows that most of them are moving away from the central cluster \elup, while younger clusters tend to have higher radial outward motions. Only one of the older and the most massive clusters at the centre of Sco-Cen shows a slightly negative component, which is $\nu$\,Cen with $v_r \sim -0.1$\,km\,s$^{-1}$. 
Second, we find a clear linear trend, where $v_r$ increases with $r$. We fit a linear regression to the data points, only using clusters with $v_r > 0$ (excluding \elup and $\nu$\,Cen). This delivers a slope of about 0.09\,km\,s$^{-1}$\,pc$^{-1}$, named radial-motion--distance relation (see Table~\ref{tab:acceleration}). Hence, clusters at larger distances from the centre tend to have higher velocities radially away from it, suggesting an outward expansion, reminiscent of a Hubble flow. 
The inverse of the slope gives a time of about 11\,Myr. This can be interpreted, to first order, as the approximate time of the onset of significant radial expansion, assuming no external forces.

We repeat the same analysis with two alternative reference frames---using  \philup and \scocen, instead of \elup (see Sect.~\ref{sec:methods} and Appendix~\ref{apx:methods-bulk})---to test the robustness of the identified relations. For these two cases, we find similar trends with slightly shallower slopes, showing that the expansion patterns are valid from different reference points that are located at central locations (see Fig.~\ref{fig:speed+hubble-other} \& Table~\ref{tab:acceleration}).

We further test the linear relations by setting each of the 32 clusters once as a reference point (see Appendix~\ref{apx:test-random-speed}). With this, we check if the older clusters at central locations (such as \elup or \philup) are likely a centre of expansion, or if the expansion pattern is independent of the central area or the age of the reference cluster (hence, independent of time). In most cases, the resulting correlations are less clear, showing more scatter and flatter slopes when compared to \elup as a reference point.  
We also test if the radial motions of the clusters are positive for the majority of the test cases, when using each cluster once as a reference point, since negative values indicate infalling motions. In some cases (mostly when using younger clusters as reference points), we find that several clusters now have negative $v_r$, which means that those reference points appear to create relatively in-falling motions instead of predominantly expanding motions, which suggests that these clusters are less likely to be the point of expansion. 
These tests show us that independent of the reference frame, the association appears to be expanding radially in most cases, in particular when using clusters with ages~>~12\,Myr as reference, while the cleanest trend is created when using \elup (see more details in Appendix~\ref{apx:test-random-speed}). 

To further test the statistical significance of the correlation of speed with time (Fig.~\ref{fig:age-speed-distance}), we randomise the order of the time-axis, as detailed in the Appendix~\ref{apx:test-random-speed}. We find that the probability of obtaining a similar linear relation is below $4\sigma$. 
In other words, random (not time-ordered) cases tend to produce shallower or reverse slopes and/or larger scatter. This suggests that the Sco-Cen clusters did not form randomly and independently from each other, but rather sequentially via propagated star formation. 

Additionally, we explore the expansion patterns using three position-velocity diagrams (PVD) in the three Cartesian directions, shown in Fig.~\ref{fig:PV}. We use the clusters' median positions, $XYZ$, and space motions relative to the local standard of rest (LSR), $UVW_\text{LSR}$, named $XU$, $YV$, and $ZW$ diagrams. The $XU$ and $YV$ diagrams show positive correlations of position with velocity, which again suggests a large-scale expansion of the whole region. The $ZW$ diagram shows more scatter and a less clear trend, while position and velocity are generally positively correlated. We fit linear regressions to the data in the three spaces and report the medians and uncertainties in Table~\ref{tab:pv-digram-fits} (determined via bootstrapping). 

\begin{table}[!t]
\begin{small}
\centering
\caption{Linear regression results for the three PV diagrams (PVD), as shown in Fig.~\ref{fig:PV}.}
\renewcommand{\arraystretch}{1.3}
\resizebox{1\columnwidth}{!}{
\begin{tabular}{ l l r r r }

\hline \hline

\multicolumn{1}{c}{} &
\multicolumn{1}{c}{PVD} &
\multicolumn{1}{c}{Slope ($a$)} &
\multicolumn{1}{c}{Intercept ($b$)} &
\multicolumn{1}{c}{Time ($a^{-1}$)}  \\

\multicolumn{1}{c}{} &
\multicolumn{1}{c}{} &
\multicolumn{1}{c}{km~s$^{-1}$~pc$^{-1}$} &
\multicolumn{1}{c}{km~s$^{-1}$} &
\multicolumn{1}{c}{Myr} \\

\hline

\multirow{3}{*}{\makecell{using 32 \\ clusters}} 
& $XU$ & 0.057$^{+0.006}_{-0.007}$ & $-1.24^{+0.73}_{-0.51}$ & $17.2^{+2.4}_{-1.9}$ \\
& $YV$ & 0.062$^{+0.006}_{-0.006}$ & $-2.97^{+0.46}_{-0.45}$ & $15.9^{+1.9}_{-1.5}$ \\
& $ZW$ & 0.037$^{+0.009}_{-0.010}$ & $-0.26^{+0.35}_{-0.32}$ & $26.6^{+9.3}_{-5.5}$ \\
\hline
\multirow{3}{*}{\makecell{using 25 \\ clusters}} 
& $XU$ & 0.065$^{+0.004}_{-0.005}$ & $-1.76^{+0.49}_{-0.35}$ & $15.1^{+1.1}_{-1.0}$ \\
& $YV$ & 0.055$^{+0.006}_{-0.007}$ & $-3.63^{+0.46}_{-0.50}$ & $17.8^{+2.4}_{-1.9}$ \\
& $ZW$ & 0.053$^{+0.009}_{-0.008}$ & $-0.00^{+0.32}_{-0.30}$ & $18.6^{+3.7}_{-2.7}$ \\

\hline
\end{tabular}
} 
\renewcommand{\arraystretch}{1}
\label{tab:pv-digram-fits}
\tablefoot{
Top three rows: Results obtained by using all 32 clusters. Bottom three rows: Results obtained by using only 25 clusters, after removing seven clusters with peculiar motions.  
}
\end{small}
\end{table}

Looking at the $ZW$ diagram in more detail ({\it right panel} in Fig.~\ref{fig:PV}), we find that some clusters appear to be outliers, located mainly in the Upper Scorpius (USco) region. The discrepancy in the motion of these clusters correlates with an additional tangential component of motion relative to the reference point, observed for seven clusters, which is discussed in more detail in Sect.~\ref{sec:tangential}. When we remove these seven clusters, we find a clearer correlation and expansion pattern in $ZW$, while the trends in $XU$ and $YV$ remain similar (see Fig.~\ref{fig:PV} \& Table~\ref{tab:pv-digram-fits}). Nevertheless, in Fig.~\ref{fig:hubble-flow} we find that the radial component of motion versus relative distance of these seven odd clusters follows the same trend as the rest of the Sco-Cen clusters.

\section{Discussion} \label{sec:discussions}

In this paper, we present the first measurement of the temporal evolution of stellar velocity dispersion within a young stellar population. By using high-precision \textit{Gaia} data supplemented with additional RV data, we identified an unexpected series of jumps and plateaus in the evolution of the velocity dispersion over the formation period of the Sco-Cen OB association (approximately 20 Myr). These findings suggest that star formation within Sco-Cen did not occur randomly but is structured and sequential. The simplest interpretation of our results suggests a scenario in which stellar feedback plays a significant role in influencing the observed dynamic evolution of stellar populations within the association.  

The surprising sequence of abrupt jumps and intervening plateaus in the cumulative \sigdrei of the Sco-Cen association, the central finding of this paper, correlates with four distinct star formation bursts. The four periods of heightened star formation rate, as identified in \citetalias{Ratzenboeck2023b}, are marked by four vertical grey bars in Figs.~\ref{fig:cumul-vel-disp}, \ref{fig:build-up}, and \ref{fig:cumul-size}. These periods are separated by about 5\,Myr.  
These findings suggest a structured and sequential formation process for Sco-Cen, likely driven by stellar feedback. 
As illustrated in Fig.~\ref{fig:build-up}, the jumps occur when star formation transitions spatially to an adjacent region, as suggested by the present-day distribution of clusters. The observed jumps can be interpreted as the addition of a new ensemble of younger clusters; these additional young clusters have been formed from a nearby gas reservoir (close to the existing populations of stars) with a marginally different velocity. 

The concept of bursts in star formation and the complex age structures within Sco-Cen have been a subject of extensive research in previous works. For instance, \citet{Slesnick2008} investigated the USco subgroup, concluding that its low-mass population formed in a single burst approximately 5\,Myr ago, with an age spread of less than 3\,Myr, after accounting for observational uncertainties. While this suggested a relatively coeval formation at the time, the apparent age spread observed in their Hertzsprung-Russell diagrams highlighted the complexities of age determination. More recently, \citet{Pecaut2016} conducted a comprehensive study across all three traditional Sco-Cen subgroups, demonstrating that none of them is consistent with simple, coeval populations formed in single bursts. Instead, they found strong evidence for age gradients and a ``multitude of smaller star formation episodes'' throughout the association, suggesting a more complex and prolonged star formation history, as recently confirmed, for instance, by \citetalias{Ratzenboeck2023b} (see also the discussion in Sect.~5 in \citetalias{Ratzenboeck2023b}). \citet{Pecaut2016} emphasised the presence of substructure and found larger intrinsic age spreads (e.g., $\pm 7$\,Myr for USco) when using revised age scales and accounting for various observational effects. Our current findings provide novel kinematic evidence for these multi-episodic or burst-like star formation events, demonstrating how these bursts manifest as distinct `stepwise' increases in the collective velocity dispersion of the association over time. This directly reinforces the understanding that Sco-Cen's evolution is not monolithic but rather a phased assembly of spatially and kinematically distinct stellar populations.

This sequential progression in space and time, characterised by radially ordered outward motions rather than random (Figs.~\ref{fig:age-speed-distance} \& \ref{fig:hubble-flow}), suggests the presence of a coordinating agent. 
A plausible agent to create the observed order in cluster positions, ages, and motions, particularly in a region of star formation known for producing massive stars, is stellar feedback. This mechanism can elucidate the inside-out arrangement of events, 
which is not easily reproduced. For instance, we find that the correlations break down if we arrange the clusters randomly and not by cluster age (formation time), or if we use other reference points instead of the older clusters that are largely located at central locations (see randomised trials in Appendix~\ref{apx:test-random-speed}). 
Furthermore, by effectively accelerating nearby gas reservoirs, the feedback scenario accounts for the observation that the youngest clusters exhibit the highest velocities relative to the older, more massive clusters (Fig.~\ref{fig:age-speed-distance}). 
This scenario is already outlined for the four cluster chains, as discussed in \citetalias{Posch2023, Posch2025}; or \citetalias{MiretRoig2025}, which find acceleration from older to younger clusters, with the youngest moving away the fastest from the centre of Sco-Cen. 
The simplest scenario that aligns with these observations posits that feedback from prior episodes of star formation compresses and accelerates adjacent gas reservoirs of the primordial Sco-Cen gas cloud, ultimately leading to the formation of new stars (e.g.,  \citealt{deAvillezBreitschwerdt2005, Grossschedl2021}; \citetalias{Posch2023, Posch2025}; \citealt{Alves2025}). 
This mechanism essentially embodies the \citet{Elmegreen1977} scenario in action, which naturally ends when the capacity to compress molecular gas into collapsing dense cores is exhausted. 

The plateaus observed in the cumulative velocity dispersion (Fig.~\ref{fig:cumul-vel-disp}) warrant further investigation. These plateaus are not perfectly flat and exhibit, on average, subtle increases in velocity dispersion between the transitions. In this context, plateaus suggest the presence of a relatively coherent reservoir of gas forming a group of clusters, analogous to ``peas in a pod'' (see also Fig.~\ref{fig:build-up}). In this analogy, the jumps in \sigdrei over time can be seen as star formation transitioning into a new ``gas pod'', at slightly different average velocity, either primordial or, likely, affected by feedback. It remains to be studied how much of the next ``pod'' was created by fragmentation of the primordial gas cloud and how much of it was assembled by feedback from the previous star formation episode. The CrA, LCC, and USco chains of clusters, representing the later stages of the formation of the association, clearly require a more dominant role for feedback, to be able to explain the observed accelerations \citepalias{Posch2023, Posch2025, MiretRoig2025}. It seems reasonable to posit that the impact of feedback on the formation is not constant over the formation of an OB association, but is coupled to the simultaneous presence of massive stars and their feedback output. 

The three to four periods of enhanced star-formation rate found in the star formation history of Sco-Cen, as reported in \citetalias{Ratzenboeck2023b}\footnote{The youngest period of enhanced star formation is less pronounced in our work, since we excluded the two, young Chamaeleon clusters, which are part of \citetalias{Ratzenboeck2023b}.}, could be interpreted as observational evidence in support of the so-called ``Type-I triggering'', discussed in \citet{Dale2015a}. This type of star formation triggering, although posited to be theoretically unlikely, describes a temporal increase in the star formation rate, which is attributed to the presence of massive stars able to shape their environments, not unlike the scenario proposed above.

\subsection{Expansion of Sco-Cen}

Using \textit{Gaia} DR1, \citet{Wright2018} found no evidence of expansion in Sco-Cen. With \textit{Gaia} DR3, we revisit this question by examining the relative motions of the 32 Sco-Cen clusters. This updated view of Sco-Cen covers a larger area than previous studies \citep[such as][]{Wright2018}, which allows us, together with the higher precision of {\it Gaia} DR3, to reinvestigate the internal motions of the region.
We find clear evidence for expansion, with a present-day speed of about 10\,pc\,Myr$^{-1}$, which likely started around 11--14\,Myr ago (Figs.~\ref{fig:age-speed-distance}--\ref{fig:hubble-flow}, Table~\ref{tab:acceleration}). Similar expansion patterns have been found for individual cluster chains \citepalias{Posch2023, Posch2025, MiretRoig2025}, while we confirm expansion for the whole Sco-Cen region. 

Additionally, we find that not only the speed but in particular the radial component of the motions shows a clear expansion of the whole region, similar to a Hubble flow (Fig.~\ref{fig:hubble-flow}). These spatio-kinematic patterns in Sco-Cen, as presented in Figs.~\ref{fig:build-up}--\ref{fig:hubble-flow}, are consistent with the hypothesis that the assembly of the Sco-Cen association happened sequentially and propagated from the inside out \citepalias{Ratzenboeck2023b}. The onset of expansion fits the age of the older clusters in Sco-Cen. Moreover, these older clusters are more massive and are the main source of stellar feedback in Sco-Cen; hence, they probably influenced the accumulation or even the momenta of the remaining clouds before they started to form stars \citep[e.g.,][]{Fuchs2006, Breitschwerdt2016, Krause2018, Grossschedl2021}

Other mechanisms, like the relaxation due to gas dispersal and the general influence of Galactic dynamics, should also be taken into account (e.g., differential rotation or shear).  
We investigated the PV-diagrams in the three Galactic directions, $XYZ$ (see Fig.~\ref{fig:PV} \& Table~\ref{tab:pv-digram-fits}). We find that the expansion patterns are largely comparable with each other within the uncertainties in the three dimensions. Especially when removing seven clusters with somewhat peculiar motions (see details in Sect.~\ref{sec:tangential}), we see almost isotropic expansion. If anything, there appears to be a slightly faster expansion in the $X$-direction when removing the seven odd clusters. Considering the effect of Galactic dynamics, one would rather expect a larger value in the $Y$-direction. This suggests that the Sco-Cen association is not (yet) strongly affected by Galactic dynamics and that we still observe the dynamics imprinted from the association's formation history. A similar conclusion is presented in \citetalias{MiretRoig2025}, where no strong effects of Galactic dynamics are observed for the TWA cluster chain. Detailed modelling is warranted to disentangle the different influences.

\subsection{Peculiar tangential motions of seven clusters} \label{sec:tangential}

\begin{figure}[!t]
    \centering
    \includegraphics[width=1\linewidth]{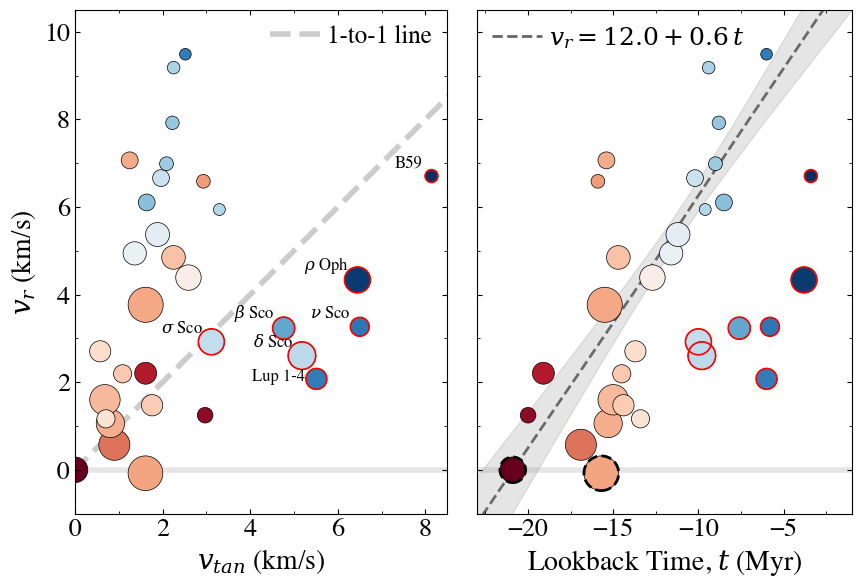}
    \caption{Radial component of the relative cluster motions ($v_r$) versus the tangential component ($v_\mathit{tan}$) ({\it left panel}), and versus formation time ({\it right panel}), with \elup as reference point. Symbols and colours are as in Fig.~\ref{fig:age-speed-distance}. The grey, dashed line in the left panel is a 1-to-1 line. The dark-grey, dashed line in the right panel shows a linear fit to data of 23 clusters, after removing the seven red-circled and the two black-dashed-circled clusters. The solid, light-grey line at $v_r = 0$ in both panels marks the transition from radial outward motion ($>0$) to inward motion ($<0$).   
    } 
    \label{fig:tangential}
\end{figure}

Figure~\ref{fig:tangential} compares the radial component and the tangential component of the motions relative to \elup ($v_r$ versus $v_\mathit{tan}$, {\it left panel}). We see that for most clusters the radial component dominates over the tangential component, while older clusters tend to have generally lower relative velocities in both radial and tangential directions. The older clusters are largely located at central locations, similar to \elup, and their formation was likely less influenced by feedback, since they are the source of most massive stars. Hence, the expansion pattern of the older clusters is more likely dominated by gas dispersal and dynamical relaxation, probably influenced by their internal feedback. 

There are seven younger clusters (ages\,<\,12\,Myr), in and around USco, which also have a significant tangential component, concerning $\rho$\,Oph, $\nu$\,Sco, $\delta$\,Sco, $\beta$\,Sco, Lup\,1-4, and also B59. This is not completely unexpected, since massive stars exist in USco, and there is some evidence of additional forces in this region \citep[e.g.,][]{Neuhaeuser2020, Squicciarini2021, MiretRoig2022b, Alves2025}. 

The right panel of Fig.~\ref{fig:tangential} is similar to the speed--time relation in Fig.~\ref{fig:age-speed-distance}, while here only the radial component is plotted versus time. We find a similar increase in relative velocity with time, while there appear to be some outliers. These outliers are the same clusters that also have a significant tangential component, as marked by the red circles. We fit a linear regression to the data of 23 clusters, after removing the seven odd clusters, and the two clusters \elup (reference point) and $\nu$\,Cen. The latter is one of the most massive clusters in Sco-Cen, and it shows here a minor negative value in $v_r$. 
The linear fit delivers a slope of $0.58\pm0.10$\,km\,s$^{-1}$\,Myr$^{-1}$. This expansion speed is somewhat higher than the one from the speed--time relation (cf., $0.45\pm0.08$\,km\,s$^{-1}$\,Myr$^{-1}$), while consistent within the uncertainties. 
An expansion between 5--7\,km\,s$^{-1}$\,Myr$^{-1}$ was also found for the individual cluster chains in \citetalias{Posch2023}, \citetalias{Posch2025}, or \citetalias{MiretRoig2025}. This highlights that the expansion is consistent in different directions of the association.
The intercept in Fig.~\ref{fig:tangential} delivers a velocity of about $12.0\pm1.1$\,km\,s$^{-1}$ ($12.3\pm1.2$\,pc\,Myr$^{-1}$), which could be interpreted as the present-day outward expansion speed of the association, when ignoring peculiar motions. This value is higher but consistent within the uncertainties when compared to the value delivered from the speed-time relation (cf., $10.8\pm0.9$\,pc\,Myr$^{-1}$, see Fig.~\ref{fig:age-speed-distance} \&  Table~\ref{tab:acceleration}). 

As mentioned above (Sect.~\ref{sec:results:speed}), we find that the seven clusters with peculiar tangential motions also appear to be outliers in the $ZW$ PV-diagram in Fig.~\ref{fig:PV}. If we remove these seven clusters, we see a cleaner linear trend in $ZW$ for the remaining 25 clusters. Fitting linear regressions to the data in the three PV-diagrams using all 32 clusters delivers expansion velocities between about 0.04--0.07\,km\,s$^{-1}$\,pc$^{-1}$. When using only the 25 clusters, we get more consensus, with expansion velocities around 0.06\,km\,s$^{-1}$\,pc$^{-1}$ (Table~\ref{tab:pv-digram-fits}), suggesting isotropic expansion. 

The inverse of the slopes from the three PV-diagrams gives, in most cases, a time of about 15--19\,Myr (especially when only using 25 clusters), which could be tentatively interpreted as the onset of expansion. This is slightly earlier than the time delivered by the radial expansion in Figs.~\ref{fig:hubble-flow} or \ref{fig:speed+hubble-other} (with about 11--14\,Myr), due to the somewhat higher expansion velocity. One reason could be that the radial-motion--distance relation depicts all outward motions relative to a rest frame, while the PV-diagrams are showing only the three selected directions of motion in $XYZ$. 
Regardless of the method, we see in all cases that the various estimated expansion velocities are generally in agreement with the age of Sco-Cen, where the oldest cluster has an age of about 21\,Myr. Hence, the expansion has probably begun a few million years after the first clusters formed.  

Taking a closer look at the $ZW$ diagram in Fig.~\ref{fig:PV}, we find that the seven odd clusters toward USco appear to be moving relatively ``downward''; hence, they appear to be moving relatively faster toward the Galactic mid-plane compared to the older bulk clusters of Sco-Cen, while at the same time being at relatively high $Z$ positions (see vertical grey, dashed line in Fig.~\ref{fig:PV}). This could be an indication of Galactic dynamics, where the pull of the gravitational potential was able to reverse the motions of clusters at higher Galactic $Z$, which would imply that these clusters might have been at higher $Z$ positions in the past. These relative motions are not observed for older clusters at similar $Z$ positions. On the other hand, it could be a sign of a more complex formation history towards USco, Lupus, and Pipe (B59), which was also suggested in recent studies (e.g., \citealt{MiretRoig2022b}; \citetalias{Posch2025}; \citealt{Alves2025}). Moreover, when considering the other two clusters toward the Pipe nebula (Pipe-North and $\theta$\,Oph), we find that they also appear to be standing out by having somewhat higher relative motions when compared to the rest of the older clusters with ages~>~15\,Myr. A more detailed analysis of the peculiar motions in these regions is part of future work (see also \citealt{Hutschenreuter2026}).

\subsection{Radial propagation speed of star formation} \label{sec:radial-propagation-speed}

\begin{figure}[!t]
    \centering
        \includegraphics[width=0.98\linewidth]{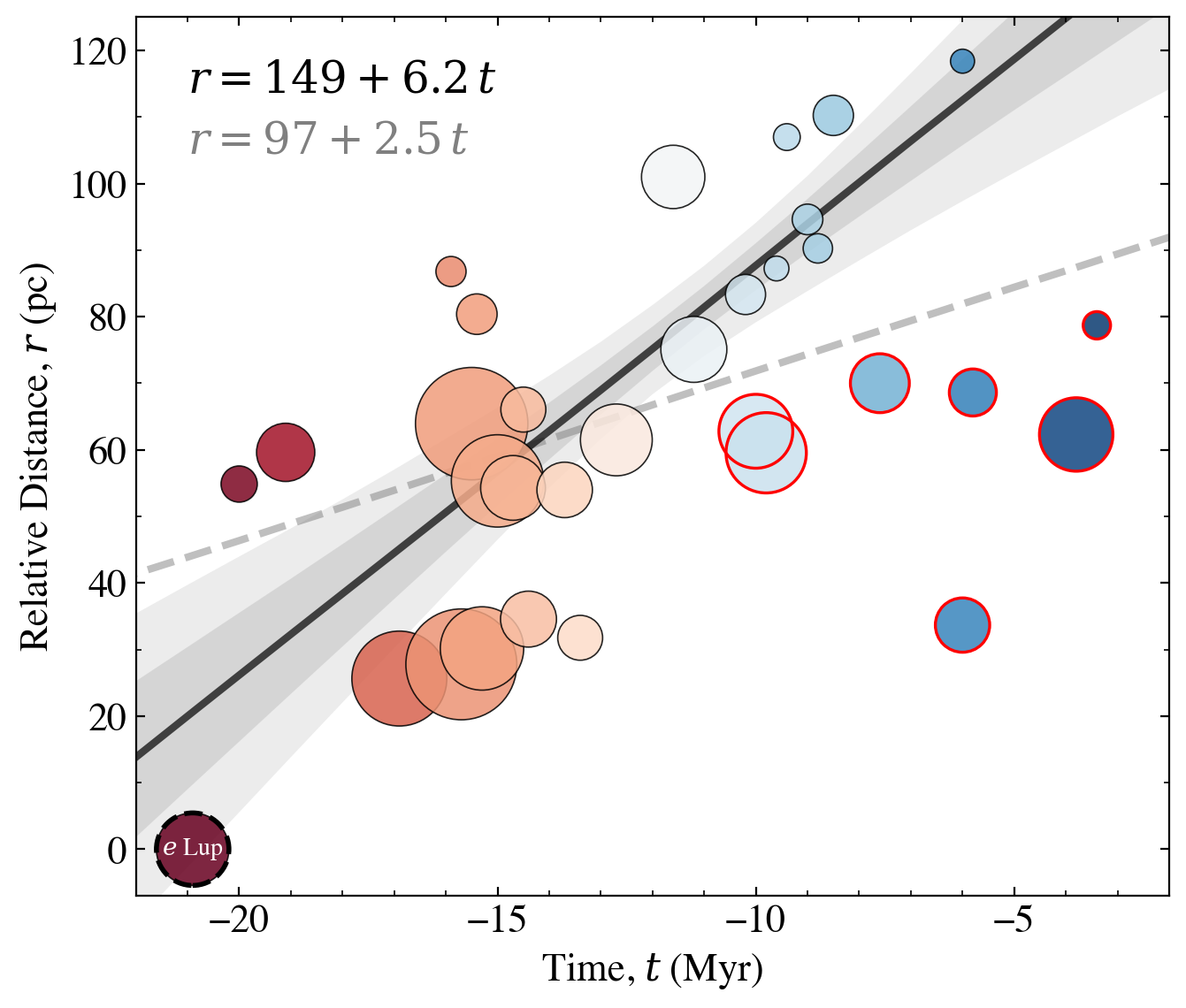}
    \caption{
        Relative cluster distances versus formation time, with \elup as reference point. Symbols and colours as in Fig.~\ref{fig:age-speed-distance}. Two linear relations are fitted, once for all clusters (dashed, light-grey line), and once for 25 clusters (solid, black line) after removing the seven, red-circled clusters; see also Fig.~\ref{fig:tangential}. The grey shaded areas show the 1-2-$\sigma$ uncertainty ranges around the median slope when using 25 clusters. 
    }
    \label{fig:radial-propagation}
\end{figure}

Looking at the relations in Figs.~\ref{fig:age-speed-distance} and \ref{fig:hubble-flow}, we find that we can combine the resulting linear relations from the speed--time ($dv/dt$) and radial-motion--distance ($dv_{r}/dr$) relations, to determine the radial propagation speed of star formation ($dr/dt$). In this case, we are ignoring that we first have the speed ($v$) and second only the radial component of the motions ($v_r$); then we can write:
\begin{small}
\begin{equation} \label{eq:drdt_1}
\frac{dr}{dt} \approx \frac{dv/dt}{dv_r/dr} \approx \frac{0.45\,\mathrm{(km/s)/Myr}}{0.09\,\mathrm{(km/s)/pc}} \approx 4.9\,\frac{\mathrm{pc}}{\mathrm{Myr}} \approx 4.8\,\frac{\mathrm{km}}{\mathrm{s}}
\end{equation}
\end{small}
When using instead the $dv_r/dt$ relation from Fig.~\ref{fig:tangential} ({\it right panel}) we can compare the same quantities; we get: 
\begin{small}
\begin{equation} \label{eq:drdt_2}
\frac{dr}{dt} = \frac{dv_r/dt}{dv_r/dr} = \frac{0.58\,\mathrm{(km/s)/Myr}}{0.09\,\mathrm{(km/s)/pc}} \approx 6.3\,\frac{\mathrm{pc}}{\mathrm{Myr}} \approx 6.1\,\frac{\mathrm{km}}{\mathrm{s}}
\end{equation}
\end{small}
However, in the latter case, the seven odd clusters are not included in the value of $dv_r/dt$.
Additionally, we directly plot the relative cluster distances versus time to get $dr/dt$, shown in Fig.~\ref{fig:radial-propagation}. Again, we find that the seven odd clusters appear to be outliers, while the rest of the clusters follow roughly a linear relation. When ignoring the odd clusters, we get a relation of about 6.2\,pc\,Myr$^{-1}$ (6.0\,km\,s$^{-1}$), similar to Eq.~\eqref{eq:drdt_2}. 

We find that the radial propagation speed of star formation (assuming isotropic expansion) roughly matches or is higher than the total, present-day velocity dispersion of the entire system with \sigdrei$\sim$\,(4 to 4.7)\,km/s.
This implies that the system is dominated by expansion, rather than random motions, and that it was initially dynamically cold.

\subsection{Discussing and comparing the cumulative trends} \label{sec:discuss-larson}

The visible steps seen in the cumulative size in Fig.~\ref{fig:cumul-size} suggest structure in the chronological, spatial cluster arrangement, when considering their present-day $XYZ$ positions.  
However, this measurement does not include orbital tracebacks, as also outlined above (Sects.~\ref{sec:methods} \& \ref{sec:results:cumulative-veldisp}). In this work, we have refrained from including orbital tracebacks due to several limitations and assumptions that have to be made and because it would go beyond the scope of this paper,  while more tracebacks will be addressed in future work. 
For instance, assumptions have to be made concerning the various literature-reported values for the Galactic potential or the Solar parameters, and one also needs to consider possible internal gravitational effects. 
Additionally, when measuring the size with the here presented method (Appendix~\ref{apx:methods:calc-size}), we would need to trace back the individual stars; however, the majority of the member stars have no RVs or too large velocity uncertainties, which makes it unfeasible to use the same method. 
If, instead, using the cluster average positions only, it would also require additional assumptions (e.g., changing cluster volumes due to cluster expansion, non-sphericity).
Finally, orbital tracebacks are sensitive to the initial conditions. While the cluster median positions and velocities are generally more robust than those of single stars, any uncertainties would still inflate the traceback errors for longer integration times.

We preliminarily test the influence of the tracebacks on the cumulative size estimate, to get an understanding of the ``true'' evolution of the association's size. For this, we use the clusters' median $XYZ$ \& $UVW$ as initial conditions to calculate orbital tracebacks (using \texttt{galpy}, \texttt{MWPotential2014}, an axisymmetric potential from \citealt{Bovy2015}). 
This allows us to use the trace-backed positions at each time step for the cumulative size instead of the present-day positions, while the individual cluster extents are ignored.
We find that the region was more compact in the past, consistent with the general expansion we find in this paper. 
This preliminary backward integration demonstrates that the absolute size at a given lookback time decreases substantially, while this measurement is sensitive to the adopted dynamical model and uncertainties of the initial conditions (e.g., assumed acceleration history, Galactic potential, internal gravity, internal expansion, or individual cluster extent). 
Considering that the clusters have relative velocities of several km\,s$^{-1}$, backward integration over 10--20\,Myr leads to position shifts of order 50--150 pc.
However, while the absolute scale of the trace-backed cumulative size is model-dependent and sensitive to uncertainties, its qualitative behaviour is largely governed by the relative spatial offsets between successively formed cluster groups. 
Backward integration compresses the overall structure but preserves the age-ordered, inside-out arrangement of clusters, because these offsets are coupled to the measured expansion pattern (Figs.~\ref{fig:age-speed-distance}--\ref{fig:PV}).
This test also shows that the given present-day cumulative size in Fig.~\ref{fig:cumul-size} represents an upper limit
to the true physical extent of Sco-Cen at earlier times.

On the other hand, the stepwise increases in cumulative size largely disappear when using the trace-back positions.
Nevertheless, looking at both trends in Figs.~\ref{fig:cumul-vel-disp} and \ref{fig:cumul-size}, we see that the cumulative \sigdrei and present-day cumulative size are increasing similarly when ordered by decreasing cluster age, with similar jumps and plateaus. 
The cumulative \sigdrei could be interpreted as the evolution of velocity dispersion over time during the star formation history of Sco-Cen, since the velocity dispersion of the stars should not have changed significantly since their formation, considering the relevant timescales; the velocity dispersion is invariant under uniform translation and insensitive to expansion offsets between cluster centroids since it measures velocity scatter.
However, as pointed out, the size is sensitive to spatial offsets; thus, the cumulative size in Fig.~\ref{fig:cumul-size} only represents the present-day arrangement of clusters. Therefore, the two measurements are not strictly equivalent and can not be compared at face value.
Still, the similarities between the two trends are intriguing, and their parallel behaviour might be physically suggestive. It appears that Sco-Cen's dynamical history might also be encoded in the present-day cluster arrangement within the region. 
Modelling is needed to test how, in regions with massive stars, feedback (e.g., acceleration of cloud parts before stars form) might influence the eventual cluster configuration of an OB association, to better understand if we can learn more about an association's history via such diagrams.

\subsection{The importance of high resolution age maps}

Our work underscores the importance of identifying the substructure and age distribution within a stellar association. When populations within an association are mixed, which is often unavoidable when studying more distant regions, the observed velocity dispersion may be misconstrued. Even Sco-Cen, the closest OB association to Earth, was traditionally divided into only three subgroups with three different ages (see Appendix~\ref{apx:cluster-sample}). Thanks to \textit{Gaia}, we now know that Sco-Cen comprises more than 30 individual stellar populations, each with distinct motions and ages, and with relatively low internal velocity dispersions per cluster. 

For example, \citet{Comeron1998} reported line-of-sight velocity dispersions of up to 60\,km\,s$^{-1}$ for the Cygnus superbubble and up to 15\,km\,s$^{-1}$ for the Canis Major OB1 association. They interpret these values as evidence for energetic expansion during the formation of these OB associations. Although this scenario is qualitatively consistent with our findings, we measure a significantly lower velocity dispersion for the entire Sco-Cen region, around 4--5\,km\,s$^{-1}$. 
These relatively high velocity dispersions, reported by \citet{Comeron1998}, could be real or could be due to various factors, such as the selection of radial velocities, instrumental limitations, or treatment of outliers and contamination. A likely contributor to the discrepancy is the presence of binary systems. In more distant clusters, the observed sample is typically dominated by massive stars, which are more likely to be multiples, even with careful selection, and thus introduce additional dispersion in the measured velocities. 

Regardless of the differences of the final, total values of \sigdrei, we highlight in our work the importance of dissecting an association into its individual stellar clusters, with low velocity dispersions individually (of the order 1--2\,km\,s$^{-1}$) and
with different ages. This allows us to produce high-resolution age maps and study the evolution of velocity dispersion in detail (Fig.~\ref{fig:build-up}). We conclude that the relatively high velocity dispersion of an OB association is the product of mixing stellar clusters within one region. Still, the discrepancy in the measured velocity dispersions needs more attention in future studies. 

In conclusion, the presence of subpopulations and detailed age gradients found for the closest OB association to Earth should be considered when studying more distant regions or when modelling the evolution of OB associations. Our results, along with those of others \citep[e.g.,][]{Kerr2021, Chen2020, Hunt2023}, indicate that OB associations should not be treated as a single stellar population \citep[e.g.,][]{Brown1997}, nor simply as a collection of a few subgroups \citep[e.g.,][]{deZeeuw1999}, but rather as complex structures \citep[e.g.,][]{Pecaut2016, Wright2018} composed of sequentially forming subpopulations \citepalias{Ratzenboeck2023b, Posch2025, MiretRoig2025}.

\section{Summary and conclusions} \label{sec:summary}

This study presents the first reconstruction of the time evolution of stellar velocity dispersion in a young OB association, using Sco–Cen as a case study. By combining high-precision \textit{Gaia} DR3 astrometry with supplementary radial velocity measurements, we analysed the kinematics of 32 stellar clusters spanning ages from about 3 to 21 Myr.
We find that the stellar members of the whole Sco-Cen association yield a total, present-day 3D velocity dispersion of about 4--5\,km\,s$^{-1}$.

Moreover, we find a stepwise increase in the cumulative 3D velocity dispersion over time, together with a systematic inside-out age sequence that is present in the 3D distribution of clusters in the Sco-Cen association. These patterns support a structured, sequential formation scenario, in which feedback from massive stars originating from older clusters likely shaped the formation and early kinematics of younger populations. The observed jumps in velocity dispersion align with known bursts of star formation (approximately separated by 5\,Myr), indicating that Sco-Cen assembled in phases from spatially and kinematically distinct gas reservoirs.

The motions of the Sco-Cen clusters reveal a well-defined expansion pattern, with a present-day rate of approximately 10--12~pc~Myr$^{-1}$, and a propagation speed of star formation of about 5--6\,km\,s$^{-1}$. The expansion likely began 11--14\,Myr ago, as indicated by the outward motions, which are correlated with distance from the centre. The nearly isotropic velocity distribution suggests that internal dynamics and stellar feedback dominate over Galactic shear or differential rotation on the relevant timescales ($\sim$20\,Myr) and spatial scales ($\sim$200\,pc).

Our findings highlight the importance of high-resolution age maps and detailed kinematic substructure analysis in the study of OB associations. Simplistically treating these associations as single, homogeneous populations risks obscuring their complex formation pathways and underestimating their internal kinematic diversity. The sequential formation and expanding structure of Sco-Cen demonstrates the value of resolved, multi-epoch analyses for tracing the dynamical evolution and feedback processes in star-forming regions.


\begin{acknowledgements}
    We thank the two referees for their constructive comments and feedback that improved the final version of this manuscript. We further thank Bruce Elmegreen and Jan Palou{\v s} for valuable comments on our analysis.
    JG acknowledges funding from the European Union, the Central Bohemian Region, and the Czech Academy of Sciences, as part of the MERIT fellowship (MSCA-COFUND Horizon Europe, Grant agreement 101081195); the Collaborative Research Center 1601 (SFB 1601) funded by the Deutsche Forschungsgemeinschaft (DFG, 500700252); and the Austrian Research Promotion Agency (FFG, \url{https://www.ffg.at/}), project number 873708.
    SR acknowledges funding by the Federal Ministry of the Republic of Austria for Climate Action, Environment, Energy, Mobility, Innovation, and Technology (BMK, \url{https://www.bmk.gv.at/}) and FFG under project number FO999892674; SR performed this work as an SAO postdoctoral fellow and acknowledges the Smithsonian Institution for their support.
    Co-funded by the European Union (ERC, ISM-FLOW, 101055318). Views and opinions expressed are, however, those of the author(s) only and do not necessarily reflect those of the European Union or the European Research Council. Neither the European Union nor the granting authority can be held responsible for them.
    This work has made use of data from the European Space Agency (ESA) mission {\it Gaia} (\url{https://www.cosmos.esa.int/gaia}), processed by the {\it Gaia} Data Processing and Analysis Consortium (DPAC, \url{https://www.cosmos.esa.int/web/gaia/dpac/consortium}). Funding for the DPAC has been provided by national institutions, in particular, the institutions participating in the {\it Gaia} Multilateral Agreement. 
    This work has made use of Python (\url{https://www.python.org}); 
    Astropy \citep{Astropy2013, Astropy2022};   
    NumPy \citep{Walt2011}; 
    Matplotlib \citep{Hunter2007}; 
    SciPy \citep{2020SciPy};
    Galpy \citep{Bovy2015}; 
    TOPCAT \citep{Taylor2005}; 
    and the 
    VizieR catalog access tool \citep{Ochsenbein2000} 
    and Aladin sky atlas \citep{Bonnarel2000, Boch2014} operated at CDS, Strasbourg Observatory, France.
    
\end{acknowledgements}


\bibliographystyle{aa} 
\bibliography{biblio.bib} 

@ARTICLE{BennettBovy2019,
       author = {{Bennett}, Morgan and {Bovy}, Jo},
        title = "{Vertical waves in the solar neighbourhood in Gaia DR2}",
      journal = {\mnras},
         year = 2019,
        month = jan,
       volume = {482},
       number = {1},
        pages = {1417-1425},
          doi = {10.1093/mnras/sty2813},
archivePrefix = {arXiv},
       eprint = {1809.03507},
 primaryClass = {astro-ph.GA},
       adsurl = {https://ui.adsabs.harvard.edu/abs/2019MNRAS.482.1417B}
}

@ARTICLE{Hutschenreuter2026,
       author = {{Hutschenreuter}, S. and {Alves}, J. and {Posch}, L. and {Gro{\ss}schedl}, J. and {Piecka}, M. and {Miret-Roig}, N. and {Ratzenb{\"o}ck}, S. and {Swiggum}, C.},
        title = "{The velocity field of the Scorpius-Centaurus OB association: I. Method and general properties}",
      journal = {\aap},
         year = 2026,
        month = jan,
       volume = {705},
          eid = {A108},
        pages = {A108},
          doi = {10.1051/0004-6361/202557273},
       adsurl = {https://ui.adsabs.harvard.edu/abs/2026A&A...705A.108H}
}

@INPROCEEDINGS{Kollmeier2019,
       author = {{Kollmeier}, Juna and {Anderson}, S.~F. and {Blanc}, G.~A. and {Blanton}, M.~R. and {Covey}, K.~R. and {Crane}, J. and {Drory}, N. and {Frinchaboy}, P.~M. and {Froning}, C.~S. and {Johnson}, J.~A. and {Kneib}, J. -P. and {Kreckel}, K. and {Merloni}, A. and {Pellegrini}, E.~W. and {Pogge}, R.~W. and {Ramirez}, S.~V. and {Rix}, H.~W. and {Sayres}, C. and {S{\'a}nchez-Gallego}, Jos{\'e} and {Shen}, Yue and {Tkachenko}, A. and {Trump}, J.~R. and {Tuttle}, S.~E. and {Weijmans}, A. and {Zasowski}, G. and {Barbuy}, B. and {Beaton}, R.~L. and {Bergemann}, M. and {Bochanski}, J.~J. and {Brandt}, W.~N. and {Casey}, A.~R. and {Cherinka}, B.~A. and {Eracleous}, M. and {Fan}, X. and {Garc{\'\i}a}, R.~A. and {Green}, P.~J. and {Hekker}, S. and {Lane}, R.~R. and {Longa-Pe{\~n}a}, P. and {Mathur}, S. and {Meza}, A. and {Minchev}, I. and {Myers}, A.~D. and {Nidever}, D.~L. and {Nitschelm}, C. and {O'Connell}, J.~E. and {Price-Whelan}, A.~M. and {Raddick}, M.~J. and {Rossi}, G. and {Sankrit}, R. and {Simon}, J.~D. and {Stutz}, A.~M. and {Ting}, Y. -S. and {Trakhtenbrot}, B. and {Weaver}, B.~A. and {Willmer}, C.~N.~A. and {Weinberg}, D.~H.},
        title = "{SDSS-V Pioneering Panoptic Spectroscopy}",
    booktitle = {Bulletin of the American Astronomical Society},
         year = 2019,
       volume = {51},
        month = sep,
          eid = {274},
        pages = {274},
       adsurl = {https://ui.adsabs.harvard.edu/abs/2019BAAS...51g.274K}
}

@ARTICLE{Edenhofer2024a,
       author = {{Edenhofer}, Gordian and {Zucker}, Catherine and {Frank}, Philipp and {Saydjari}, Andrew K. and {Speagle}, Joshua S. and {Finkbeiner}, Douglas and {En{\ss}lin}, Torsten A.},
        title = "{A parsec-scale Galactic 3D dust map out to 1.25 kpc from the Sun}",
      journal = {\aap},
         year = {2024a},
        month = may,
       volume = {685},
          eid = {A82},
        pages = {A82},
          doi = {10.1051/0004-6361/202347628},
archivePrefix = {arXiv},
       eprint = {2308.01295},
 primaryClass = {astro-ph.GA},
       adsurl = {https://ui.adsabs.harvard.edu/abs/2024A&A...685A..82E}
}

@ARTICLE{Edenhofer2024b,
       author = {{Edenhofer}, Gordian and {Alves}, Jo{\~a}o and {Zucker}, Catherine and {Posch}, Laura and {En{\ss}lin}, Torsten A.},
        title = "{The ``C'': The large Chameleon-Musca-Coalsack cloud}",
      journal = {\aap},
         year = {2024b},
        month = jul,
       volume = {687},
          eid = {L9},
        pages = {L9},
          doi = {10.1051/0004-6361/202450374},
archivePrefix = {arXiv},
       eprint = {2404.09592},
 primaryClass = {astro-ph.GA},
       adsurl = {https://ui.adsabs.harvard.edu/abs/2024A&A...687L...9E}
}

@ARTICLE{Dijkstra1959,
       author = {{Dijkstra}, E.~W.},
        title = "{A note on two problems in connexion with graphs}",
      journal = {Numerische Mathematik},
         year = {1959},
        month = dec,
       volume = {1},
        pages = {269-721},
          doi = {10.1007/BF01386390},
       adsurl = {https://doi.org/10.1007/BF01386390}
}

@ARTICLE{Swiggum2024,
       author = {{Swiggum}, Cameren and {Alves}, Jo{\~a}o and {Benjamin}, Robert and {Ratzenb{\"o}ck}, Sebastian and {Miret-Roig}, N{\'u}ria and {Gro{\ss}schedl}, Josefa and {Meingast}, Stefan and {Goodman}, Alyssa and {Konietzka}, Ralf and {Zucker}, Catherine and {Hunt}, Emily L. and {Reffert}, Sabine},
        title = "{Most nearby young star clusters formed in three massive complexes}",
      journal = {\nat},
         year = 2024,
        month = jul,
       volume = {631},
       number = {8019},
        pages = {49-53},
          doi = {10.1038/s41586-024-07496-9},
archivePrefix = {arXiv},
       eprint = {2406.06510},
 primaryClass = {astro-ph.GA},
       adsurl = {https://ui.adsabs.harvard.edu/abs/2024Natur.631...49S}
}

@ARTICLE{2020SciPy,
  author  = {Virtanen, Pauli and Gommers, Ralf and Oliphant, Travis E. and
            Haberland, Matt and Reddy, Tyler and Cournapeau, David and
            Burovski, Evgeni and Peterson, Pearu and Weckesser, Warren and
            Bright, Jonathan and {van der Walt}, St{\'e}fan J. and
            Brett, Matthew and Wilson, Joshua and Millman, K. Jarrod and
            Mayorov, Nikolay and Nelson, Andrew R. J. and Jones, Eric and
            Kern, Robert and Larson, Eric and Carey, C J and
            Polat, {\.I}lhan and Feng, Yu and Moore, Eric W. and
            {VanderPlas}, Jake and Laxalde, Denis and Perktold, Josef and
            Cimrman, Robert and Henriksen, Ian and Quintero, E. A. and
            Harris, Charles R. and Archibald, Anne M. and
            Ribeiro, Ant{\^o}nio H. and Pedregosa, Fabian and
            {van Mulbregt}, Paul and {SciPy 1.0 Contributors}},
  title   = {{{SciPy} 1.0: Fundamental Algorithms for Scientific
            Computing in Python}},
  journal = {Nature Methods},
  year    = {2020},
  volume  = {17},
  pages   = {261--272},
  adsurl  = {https://rdcu.be/b08Wh},
  doi     = {10.1038/s41592-019-0686-2},
}

@ARTICLE{Luhman2023,
       author = {{Luhman}, K.~L.},
        title = "{A Census of the TW Hya Association with Gaia}",
      journal = {\aj},
         year = 2023,
        month = jun,
       volume = {165},
       number = {6},
          eid = {269},
        pages = {269},
          doi = {10.3847/1538-3881/accf19},
archivePrefix = {arXiv},
       eprint = {2305.03557},
 primaryClass = {astro-ph.GA},
       adsurl = {https://ui.adsabs.harvard.edu/abs/2023AJ....165..269L}
}

@ARTICLE{Fuchs2006,
       author = {{Fuchs}, B. and {Breitschwerdt}, D. and {de Avillez}, M.~A. and {Dettbarn}, C. and {Flynn}, C.},
        title = "{The search for the origin of the Local Bubble redivivus}",
      journal = {\mnras},
         year = 2006,
        month = dec,
       volume = {373},
       number = {3},
        pages = {993-1003},
          doi = {10.1111/j.1365-2966.2006.11044.x},
archivePrefix = {arXiv},
       eprint = {astro-ph/0609227},
 primaryClass = {astro-ph},
       adsurl = {https://ui.adsabs.harvard.edu/abs/2006MNRAS.373..993F}
}

@ARTICLE{Poleski2013,
       author = {{Poleski}, Rados{\l}aw},
        title = "{Transformation of the equatorial proper motion to the Galactic system}",
      journal = {arXiv e-prints},
         year = 2013,
        month = jun,
          eid = {arXiv:1306.2945},
        pages = {arXiv:1306.2945},
          doi = {10.48550/arXiv.1306.2945},
archivePrefix = {arXiv},
       eprint = {1306.2945},
 primaryClass = {astro-ph.IM},
       adsurl = {https://ui.adsabs.harvard.edu/abs/2013arXiv1306.2945P}
}

@ARTICLE{Comeron1998,
       author = {{Comeron}, F. and {Torra}, J. and {Gomez}, A.~E.},
        title = "{Kinematic signatures of violent formation of galactic OB associations from HIPPARCOS measurements}",
      journal = {\aap},
         year = 1998,
        month = feb,
       volume = {330},
        pages = {975-989},
       adsurl = {https://ui.adsabs.harvard.edu/abs/1998A&A...330..975C}
}

@ARTICLE{LyngaPalous1987,
       author = {{Lynga}, G. and {Palous}, J.},
        title = "{The local kinematics of open star clusters.}",
      journal = {\aap},
         year = 1987,
        month = dec,
       volume = {188},
        pages = {35-38},
       adsurl = {https://ui.adsabs.harvard.edu/abs/1987A&A...188...35L}
}

@ARTICLE{Kamaya2004,
       author = {{Kamaya}, Hideyuki},
        title = "{Velocity Dispersion of Dissolving OB Associations Affected by External Pressure of the Formation Environment}",
      journal = {\aj},
         year = 2004,
        month = aug,
       volume = {128},
       number = {2},
        pages = {761-764},
          doi = {10.1086/422487},
archivePrefix = {arXiv},
       eprint = {astro-ph/0406574},
 primaryClass = {astro-ph},
       adsurl = {https://ui.adsabs.harvard.edu/abs/2004AJ....128..761K}
}

@ARTICLE{Kroupa1995,
       author = {{Kroupa}, Pavel},
        title = "{Star cluster evolution, dynamical age estimation and the kinematical signature of star formation}",
      journal = {\mnras},
         year = 1995,
        month = dec,
       volume = {277},
        pages = {1522},
          doi = {10.1093/mnras/277.4.1522},
archivePrefix = {arXiv},
       eprint = {astro-ph/9507017},
 primaryClass = {astro-ph},
       adsurl = {https://ui.adsabs.harvard.edu/abs/1995MNRAS.277.1522K}
}

@INCOLLECTION{Kroupa2008,
       author = {{Kroupa}, Pavel},
        title = "{Initial Conditions for Star Clusters}",
    booktitle = {The Cambridge N-Body Lectures},
    publisher = {Springer Dordrecht},
         year = 2008,
       editor = {{Aarseth}, Sverre J. and {Tout}, Christopher A. and {Mardling}, Rosemary A.},
       volume = {760},
        pages = {181},
          doi = {10.1007/978-1-4020-8431-7_8},
       adsurl = {https://ui.adsabs.harvard.edu/abs/2008LNP...760..181K}
}

@ARTICLE{deLaFuenteMarcos2008,
       author = {{de la Fuente Marcos}, R. and {de la Fuente Marcos}, C.},
        title = "{From Star Complexes to the Field: Open Cluster Families}",
      journal = {\apj},
         year = 2008,
        month = jan,
       volume = {672},
       number = {1},
        pages = {342-351},
          doi = {10.1086/524028},
       adsurl = {https://ui.adsabs.harvard.edu/abs/2008ApJ...672..342D}
}

@ARTICLE{Dale2015a,
       author = {{Dale}, J.~E. and {Haworth}, T.~J. and {Bressert}, E.},
        title = "{The dangers of being trigger-happy}",
      journal = {\mnras},
         year = {2015a},
        month = jun,
       volume = {450},
       number = {2},
        pages = {1199-1211},
          doi = {10.1093/mnras/stv396},
archivePrefix = {arXiv},
       eprint = {1502.05865},
 primaryClass = {astro-ph.GA},
       adsurl = {https://ui.adsabs.harvard.edu/abs/2015MNRAS.450.1199D}
}

@ARTICLE{LadaLada2003,
       author = {{Lada}, Charles J. and {Lada}, Elizabeth A.},
        title = "{Embedded Clusters in Molecular Clouds}",
      journal = {\araa},
         year = 2003,
        month = jan,
       volume = {41},
        pages = {57-115},
          doi = {10.1146/annurev.astro.41.011802.094844},
archivePrefix = {arXiv},
       eprint = {astro-ph/0301540},
 primaryClass = {astro-ph},
       adsurl = {https://ui.adsabs.harvard.edu/abs/2003ARA&A..41...57L}
}

@INPROCEEDINGS{Wright2023,
       author = {{Wright}, N.~J. and {Kounkel}, M. and {Zari}, E. and {Goodwin}, S. and {Jeffries}, R.~D.},
        title = "{OB Associations}",
    booktitle = {Astronomical Society of the Pacific Conference Series},
         year = 2023,
       editor = {{Inutsuka}, S. and {Aikawa}, Y. and {Muto}, T. and {Tomida}, K. and {Tamura}, M.},
       series = {Astronomical Society of the Pacific Conference Series},
       volume = {534},
        month = jul,
        pages = {129},
       adsurl = {https://ui.adsabs.harvard.edu/abs/2023ASPC..534..129W}
}

@ARTICLE{EfremovElmegreen1998,
       author = {{Efremov}, Yuri N. and {Elmegreen}, Bruce G.},
        title = "{Hierarchical star formation from the time-space distribution of star clusters in the Large Magellanic Cloud}",
      journal = {\mnras},
         year = 1998,
        month = sep,
       volume = {299},
       number = {2},
        pages = {588-594},
          doi = {10.1046/j.1365-8711.1998.01819.x},
archivePrefix = {arXiv},
       eprint = {astro-ph/9805259},
 primaryClass = {astro-ph},
       adsurl = {https://ui.adsabs.harvard.edu/abs/1998MNRAS.299..588E}
}

@ARTICLE{Murphy2013,
       author = {{Murphy}, Simon J. and {Lawson}, Warrick A. and {Bessell}, Michael S.},
        title = "{Re-examining the membership and origin of the ɛ Cha association}",
      journal = {\mnras},
         year = 2013,
        month = oct,
       volume = {435},
       number = {2},
        pages = {1325-1349},
          doi = {10.1093/mnras/stt1375},
archivePrefix = {arXiv},
       eprint = {1305.4177},
 primaryClass = {astro-ph.SR},
       adsurl = {https://ui.adsabs.harvard.edu/abs/2013MNRAS.435.1325M}
}

@ARTICLE{Sacco2017,
       author = {{Sacco}, G.~G. and {Spina}, L. and {Randich}, S. and {Palla}, F. and {Parker}, R.~J. and {Jeffries}, R.~D. and {Jackson}, R. and {Meyer}, M.~R. and {Mapelli}, M. and {Lanzafame}, A.~C. and {Bonito}, R. and {Damiani}, F. and {Franciosini}, E. and {Frasca}, A. and {Klutsch}, A. and {Prisinzano}, L. and {Tognelli}, E. and {Degl'Innocenti}, S. and {Prada Moroni}, P.~G. and {Alfaro}, E.~J. and {Micela}, G. and {Prusti}, T. and {Barrado}, D. and {Biazzo}, K. and {Bouy}, H. and {Bravi}, L. and {Lopez-Santiago}, J. and {Wright}, N.~J. and {Bayo}, A. and {Gilmore}, G. and {Bragaglia}, A. and {Flaccomio}, E. and {Koposov}, S.~E. and {Pancino}, E. and {Casey}, A.~R. and {Costado}, M.~T. and {Donati}, P. and {Hourihane}, A. and {Jofr{\'e}}, P. and {Lardo}, C. and {Lewis}, J. and {Magrini}, L. and {Monaco}, L. and {Morbidelli}, L. and {Sousa}, S.~G. and {Worley}, C.~C. and {Zaggia}, S.},
        title = "{The Gaia-ESO Survey: Structural and dynamical properties of the young cluster Chamaeleon I}",
      journal = {\aap},
         year = 2017,
        month = may,
       volume = {601},
          eid = {A97},
        pages = {A97},
          doi = {10.1051/0004-6361/201629698},
archivePrefix = {arXiv},
       eprint = {1701.03741},
 primaryClass = {astro-ph.SR},
       adsurl = {https://ui.adsabs.harvard.edu/abs/2017A&A...601A..97S}
}

@ARTICLE{Nguyen2012,
       author = {{Nguyen}, Duy Cuong and {Brandeker}, Alexis and {van Kerkwijk}, Marten H. and {Jayawardhana}, Ray},
        title = "{Close Companions to Young Stars. I. A Large Spectroscopic Survey in Chamaeleon I and Taurus-Auriga}",
      journal = {\apj},
         year = 2012,
        month = feb,
       volume = {745},
       number = {2},
          eid = {119},
        pages = {119},
          doi = {10.1088/0004-637X/745/2/119},
archivePrefix = {arXiv},
       eprint = {1112.0002},
 primaryClass = {astro-ph.SR},
       adsurl = {https://ui.adsabs.harvard.edu/abs/2012ApJ...745..119N}
}

@ARTICLE{JoergensGuenther2001,
       author = {{Joergens}, V. and {Guenther}, E.},
        title = "{UVES spectra of young brown dwarfs in Cha I: Radial and rotational velocities}",
      journal = {\aap},
         year = 2001,
        month = nov,
       volume = {379},
        pages = {L9-L12},
          doi = {10.1051/0004-6361:20011337},
archivePrefix = {arXiv},
       eprint = {astro-ph/0110175},
 primaryClass = {astro-ph},
       adsurl = {https://ui.adsabs.harvard.edu/abs/2001A&A...379L...9J}
}

@ARTICLE{Ratzenboeck2023a,
       author = {{Ratzenb{\"o}ck}, Sebastian and {Gro{\ss}schedl}, Josefa E. and {M{\"o}ller}, Torsten and {Alves}, Jo{\~a}o and {Bomze}, Immanuel and {Meingast}, Stefan},
        title = "{Significance mode analysis (SigMA) for hierarchical structures. An application to the Sco-Cen OB association}",
      journal = {\aap},
         year = {2023a},
        month = sep,
       volume = {677},
          eid = {A59},
        pages = {A59},
          doi = {10.1051/0004-6361/202243690},
       adsurl = {https://ui.adsabs.harvard.edu/abs/2023A&A...677A..59R}
}

@ARTICLE{Ratzenboeck2023b,
       author = {{Ratzenb{\"o}ck}, Sebastian and {Gro{\ss}schedl}, Josefa E. and {Alves}, Jo{\~a}o and {Miret-Roig}, N{\'u}ria and {Bomze}, Immanuel and {Forbes}, John and {Goodman}, Alyssa and {Hacar}, {\'A}lvaro and {Lin}, Doug and {Meingast}, Stefan and {M{\"o}ller}, Torsten and {Piecka}, Martin and {Posch}, Laura and {Rottensteiner}, Alena and {Swiggum}, Cameren and {Zucker}, Catherine},
        title = "{The star formation history of the Sco-Cen association. Coherent star formation patterns in space and time}",
      journal = {\aap},
         year = {2023b},
        month = oct,
       volume = {678},
          eid = {A71},
        pages = {A71},
          doi = {10.1051/0004-6361/202346901},
archivePrefix = {arXiv},
       eprint = {2302.07853},
 primaryClass = {astro-ph.SR},
       adsurl = {https://ui.adsabs.harvard.edu/abs/2023A&A...678A..71R}
}

@ARTICLE{Posch2023,
       author = {{Posch}, Laura and {Miret-Roig}, N{\'u}ria and {Alves}, Jo{\~a}o and {Ratzenb{\"o}ck}, Sebastian and {Gro{\ss}schedl}, Josefa and {Meingast}, Stefan and {Zucker}, Catherine and {Burkert}, Andreas},
        title = "{The Corona Australis star formation complex is accelerating away from the Galactic plane}",
      journal = {\aap},
         year = 2023,
        month = nov,
       volume = {679},
          eid = {L10},
        pages = {L10},
          doi = {10.1051/0004-6361/202347186},
archivePrefix = {arXiv},
       eprint = {2310.14373},
 primaryClass = {astro-ph.GA},
       adsurl = {https://ui.adsabs.harvard.edu/abs/2023A&A...679L..10P}
}

@ARTICLE{Posch2025,
       author = {{Posch}, Laura and {Alves}, Jo{\~a}o and {Miret-Roig}, N{\'u}ria and {Ratzenb{\"o}ck}, Sebastian and {Gro{\ss}schedl}, Josefa and {Meingast}, Stefan and {Swiggum}, Cameren and {Konietzka}, Ralf},
        title = "{The physical properties of cluster chains}",
      journal = {\aap},
         year = 2025,
        month = jan,
       volume = {693},
          eid = {A175},
        pages = {A175},
          doi = {10.1051/0004-6361/202451312},
archivePrefix = {arXiv},
       eprint = {2410.18080},
 primaryClass = {astro-ph.GA},
       adsurl = {https://ui.adsabs.harvard.edu/abs/2025A&A...693A.175P}
}

@ARTICLE{MiretRoig2025,
       author = {{Miret-Roig}, N. and {Alves}, J. and {Ratzenb{\"o}ck}, S. and {Galli}, P.~A.~B. and {Bouy}, H. and {Figueras}, F. and {Gro{\ss}schedl}, J. and {Meingast}, S. and {Posch}, L. and {Rottensteiner}, A. and {Swiggum}, C. and {Wagner}, N.},
        title = "{The TW Hydrae Association is a cluster chain of Sco-Cen}",
      journal = {\aap},
         year = 2025,
        month = feb,
       volume = {694},
          eid = {A60},
        pages = {A60},
          doi = {10.1051/0004-6361/202452558},
archivePrefix = {arXiv},
       eprint = {2501.11716},
 primaryClass = {astro-ph.SR},
       adsurl = {https://ui.adsabs.harvard.edu/abs/2025A&A...694A..60M}
}

@ARTICLE{Alves2025,
       author = {{Alves}, Jo{\~a}o and {Lombardi}, Marco and {Lada}, Charles J.},
        title = "{HP2 Survey: V. Ophiuchus: Filament formation in a dispersing cloud complex}",
      journal = {\aap},
         year = 2025,
        month = may,
       volume = {697},
          eid = {A208},
        pages = {A208},
          doi = {10.1051/0004-6361/202452881},
archivePrefix = {arXiv},
       eprint = {2501.13931},
 primaryClass = {astro-ph.GA},
       adsurl = {https://ui.adsabs.harvard.edu/abs/2025A&A...697A.208A}
}

@ARTICLE{Fang2023,
       author = {{Fang}, Min and {Pascucci}, Ilaria and {Edwards}, Suzan and {Gorti}, Uma and {Hillenbrand}, Lynne A. and {Carpenter}, John M.},
        title = "{A High-resolution Optical Survey of Upper Sco: Evidence for Coevolution of Accretion and Disk Winds}",
      journal = {\apj},
         year = 2023,
        month = mar,
       volume = {945},
       number = {2},
          eid = {112},
        pages = {112},
          doi = {10.3847/1538-4357/acb2c9},
archivePrefix = {arXiv},
       eprint = {2301.09240},
 primaryClass = {astro-ph.SR},
       adsurl = {https://ui.adsabs.harvard.edu/abs/2023ApJ...945..112F}
}

@ARTICLE{Biazzo2012,
       author = {{Biazzo}, K. and {Alcal{\'a}}, J.~M. and {Covino}, E. and {Frasca}, A. and {Getman}, F. and {Spezzi}, L.},
        title = "{The Chamaeleon II low-mass star-forming region: radial velocities, elemental abundances, and accretion properties {\ensuremath{\star}}}",
      journal = {\aap},
         year = 2012,
        month = nov,
       volume = {547},
          eid = {A104},
        pages = {A104},
          doi = {10.1051/0004-6361/201219680},
archivePrefix = {arXiv},
       eprint = {1209.5316},
 primaryClass = {astro-ph.SR},
       adsurl = {https://ui.adsabs.harvard.edu/abs/2012A&A...547A.104B}
}

@ARTICLE{Santana2021,
       author = {{Santana}, Felipe A. and {Beaton}, Rachael L. and {Covey}, Kevin R. and {O'Connell}, Julia E. and {Longa-Pe{\~n}a}, Pen{\'e}lope and {Cohen}, Roger and {Fern{\'a}ndez-Trincado}, Jos{\'e} G. and {Hayes}, Christian R. and {Zasowski}, Gail and {Sobeck}, Jennifer S. and {Majewski}, Steven R. and {Chojnowski}, S.~D. and {De Lee}, Nathan and {Oelkers}, Ryan J. and {Stringfellow}, Guy S. and {Almeida}, Andr{\'e}s and {Anguiano}, Borja and {Donor}, John and {Frinchaboy}, Peter M. and {Hasselquist}, Sten and {Johnson}, Jennifer A. and {Kollmeier}, Juna A. and {Nidever}, David L. and {Price-Whelan}, Adrian M. and {Rojas-Arriagada}, Alvaro and {Schultheis}, Mathias and {Shetrone}, Matthew and {Simon}, Joshua D. and {Aerts}, Conny and {Borissova}, Jura and {Drout}, Maria R. and {Geisler}, Doug and {Law}, C.~Y. and {Medina}, Nicolas and {Minniti}, Dante and {Monachesi}, Antonela and {Mu{\~n}oz}, Ricardo R. and {Poleski}, Rados{\l}aw and {Roman-Lopes}, Alexandre and {Schlaufman}, Kevin C. and {Stutz}, Amelia M. and {Teske}, Johanna and {Tkachenko}, Andrew and {Van Saders}, Jennifer L. and {Weinberger}, Alycia J. and {Zoccali}, Manuela},
        title = "{Final Targeting Strategy for the SDSS-IV APOGEE-2S Survey}",
      journal = {\aj},
         year = 2021,
        month = dec,
       volume = {162},
       number = {6},
          eid = {303},
        pages = {303},
          doi = {10.3847/1538-3881/ac2cbc},
archivePrefix = {arXiv},
       eprint = {2108.11908},
 primaryClass = {astro-ph.GA},
       adsurl = {https://ui.adsabs.harvard.edu/abs/2021AJ....162..303S}
}

@ARTICLE{Abdurrouf2022,
       author = {{Abdurro'uf} and {Accetta}, Katherine and {Aerts}, Conny and {Silva Aguirre}, V{\'\i}ctor and {Ahumada}, Romina and {Ajgaonkar}, Nikhil and {Filiz Ak}, N. and {Alam}, Shadab and {Allende Prieto}, Carlos and {Almeida}, Andr{\'e}s and {Anders}, Friedrich and {Anderson}, Scott F. and {Andrews}, Brett H. and {Anguiano}, Borja and {Aquino-Ort{\'\i}z}, Erik and {Arag{\'o}n-Salamanca}, Alfonso and {Argudo-Fern{\'a}ndez}, Maria and {Ata}, Metin and {Aubert}, Marie and {Avila-Reese}, Vladimir and {Badenes}, Carles and {Barb{\'a}}, Rodolfo H. and {Barger}, Kat and {Barrera-Ballesteros}, Jorge K. and {Beaton}, Rachael L. and {Beers}, Timothy C. and {Belfiore}, Francesco and {Bender}, Chad F. and {Bernardi}, Mariangela and {Bershady}, Matthew A. and {Beutler}, Florian and {Bidin}, Christian Moni and {Bird}, Jonathan C. and {Bizyaev}, Dmitry and {Blanc}, Guillermo A. and {Blanton}, Michael R. and {Boardman}, Nicholas Fraser and {Bolton}, Adam S. and {Boquien}, M{\'e}d{\'e}ric and {Borissova}, Jura and {Bovy}, Jo and {Brandt}, W.~N. and {Brown}, Jordan and {Brownstein}, Joel R. and {Brusa}, Marcella and {Buchner}, Johannes and {Bundy}, Kevin and {Burchett}, Joseph N. and {Bureau}, Martin and {Burgasser}, Adam and {Cabang}, Tuesday K. and {Campbell}, Stephanie and {Cappellari}, Michele and {Carlberg}, Joleen K. and {Wanderley}, F{\'a}bio Carneiro and {Carrera}, Ricardo and {Cash}, Jennifer and {Chen}, Yan-Ping and {Chen}, Wei-Huai and {Cherinka}, Brian and {Chiappini}, Cristina and {Choi}, Peter Doohyun and {Chojnowski}, S. Drew and {Chung}, Haeun and {Clerc}, Nicolas and {Cohen}, Roger E. and {Comerford}, Julia M. and {Comparat}, Johan and {da Costa}, Luiz and {Covey}, Kevin and {Crane}, Jeffrey D. and {Cruz-Gonzalez}, Irene and {Culhane}, Connor and {Cunha}, Katia and {Dai}, Y. Sophia and {Damke}, Guillermo and {Darling}, Jeremy and {Davidson}, James W., Jr. and {Davies}, Roger and {Dawson}, Kyle and {De Lee}, Nathan and {Diamond-Stanic}, Aleksandar M. and {Cano-D{\'\i}az}, Mariana and {S{\'a}nchez}, Helena Dom{\'\i}nguez and {Donor}, John and {Duckworth}, Chris and {Dwelly}, Tom and {Eisenstein}, Daniel J. and {Elsworth}, Yvonne P. and {Emsellem}, Eric and {Eracleous}, Mike and {Escoffier}, Stephanie and {Fan}, Xiaohui and {Farr}, Emily and {Feng}, Shuai and {Fern{\'a}ndez-Trincado}, Jos{\'e} G. and {Feuillet}, Diane and {Filipp}, Andreas and {Fillingham}, Sean P. and {Frinchaboy}, Peter M. and {Fromenteau}, Sebastien and {Galbany}, Llu{\'\i}s and {Garc{\'\i}a}, Rafael A. and {Garc{\'\i}a-Hern{\'a}ndez}, D.~A. and {Ge}, Junqiang and {Geisler}, Doug and {Gelfand}, Joseph and {G{\'e}ron}, Tobias and {Gibson}, Benjamin J. and {Goddy}, Julian and {Godoy-Rivera}, Diego and {Grabowski}, Kathleen and {Green}, Paul J. and {Greener}, Michael and {Grier}, Catherine J. and {Griffith}, Emily and {Guo}, Hong and {Guy}, Julien and {Hadjara}, Massinissa and {Harding}, Paul and {Hasselquist}, Sten and {Hayes}, Christian R. and {Hearty}, Fred and {Hern{\'a}ndez}, Jes{\'u}s and {Hill}, Lewis and {Hogg}, David W. and {Holtzman}, Jon A. and {Horta}, Danny and {Hsieh}, Bau-Ching and {Hsu}, Chin-Hao and {Hsu}, Yun-Hsin and {Huber}, Daniel and {Huertas-Company}, Marc and {Hutchinson}, Brian and {Hwang}, Ho Seong and {Ibarra-Medel}, H{\'e}ctor J. and {Chitham}, Jacob Ider and {Ilha}, Gabriele S. and {Imig}, Julie and {Jaekle}, Will and {Jayasinghe}, Tharindu and {Ji}, Xihan and {Johnson}, Jennifer A. and {Jones}, Amy and {J{\"o}nsson}, Henrik and {Katkov}, Ivan and {Khalatyan}, Arman, Dr. and {Kinemuchi}, Karen and {Kisku}, Shobhit and {Knapen}, Johan H. and {Kneib}, Jean-Paul and {Kollmeier}, Juna A. and {Kong}, Miranda and {Kounkel}, Marina and {Kreckel}, Kathryn and {Krishnarao}, Dhanesh and {Lacerna}, Ivan and {Lane}, Richard R. and {Langgin}, Rachel and {Lavender}, Ramon and {Law}, David R. and {Lazarz}, Daniel and {Leung}, Henry W. and {Leung}, Ho-Hin and {Lewis}, Hannah M. and {Li}, Cheng and {Li}, Ran and {Lian}, Jianhui and {Liang}, Fu-Heng and {Lin}, Lihwai and {Lin}, Yen-Ting and {Lin}, Sicheng and {Lintott}, Chris and {Long}, Dan and {Longa-Pe{\~n}a}, Pen{\'e}lope and {L{\'o}pez-Cob{\'a}}, Carlos and {Lu}, Shengdong and {Lundgren}, Britt F. and {Luo}, Yuanze and {Mackereth}, J. Ted and {de la Macorra}, Axel and {Mahadevan}, Suvrath and {Majewski}, Steven R. and {Manchado}, Arturo and {Mandeville}, Travis and {Maraston}, Claudia and {Margalef-Bentabol}, Berta and {Masseron}, Thomas and {Masters}, Karen L. and {Mathur}, Savita and {McDermid}, Richard M. and {Mckay}, Myles and {Merloni}, Andrea and {Merrifield}, Michael and {Meszaros}, Szabolcs and {Miglio}, Andrea and {Di Mille}, Francesco and {Minniti}, Dante and {Minsley}, Rebecca and {Monachesi}, Antonela and {Moon}, Jeongin and {Mosser}, Benoit and {Mulchaey}, John and {Muna}, Demitri and {Mu{\~n}oz}, Ricardo R. and {Myers}, Adam D. and {Myers}, Natalie and {Nadathur}, Seshadri and {Nair}, Preethi and {Nandra}, Kirpal and {Neumann}, Justus and {Newman}, Jeffrey A. and {Nidever}, David L. and {Nikakhtar}, Farnik and {Nitschelm}, Christian and {O'Connell}, Julia E. and {Garma-Oehmichen}, Luis and {Luan Souza de Oliveira}, Gabriel and {Olney}, Richard and {Oravetz}, Daniel and {Ortigoza-Urdaneta}, Mario and {Osorio}, Yeisson and {Otter}, Justin and {Pace}, Zachary J. and {Padilla}, Nelson and {Pan}, Kaike and {Pan}, Hsi-An and {Parikh}, Taniya and {Parker}, James and {Peirani}, Sebastien and {Pe{\~n}a Ram{\'\i}rez}, Karla and {Penny}, Samantha and {Percival}, Will J. and {Perez-Fournon}, Ismael and {Pinsonneault}, Marc and {Poidevin}, Fr{\'e}d{\'e}rick and {Poovelil}, Vijith Jacob and {Price-Whelan}, Adrian M. and {B{\'a}rbara de Andrade Queiroz}, Anna and {Raddick}, M. Jordan and {Ray}, Amy and {Rembold}, Sandro Barboza and {Riddle}, Nicole and {Riffel}, Rogemar A. and {Riffel}, Rog{\'e}rio and {Rix}, Hans-Walter and {Robin}, Annie C. and {Rodr{\'\i}guez-Puebla}, Aldo and {Roman-Lopes}, Alexandre and {Rom{\'a}n-Z{\'u}{\~n}iga}, Carlos and {Rose}, Benjamin and {Ross}, Ashley J. and {Rossi}, Graziano and {Rubin}, Kate H.~R. and {Salvato}, Mara and {S{\'a}nchez}, Seb{\'a}stian F. and {S{\'a}nchez-Gallego}, Jos{\'e} R. and {Sanderson}, Robyn and {Santana Rojas}, Felipe Antonio and {Sarceno}, Edgar and {Sarmiento}, Regina and {Sayres}, Conor and {Sazonova}, Elizaveta and {Schaefer}, Adam L. and {Schiavon}, Ricardo and {Schlegel}, David J. and {Schneider}, Donald P. and {Schultheis}, Mathias and {Schwope}, Axel and {Serenelli}, Aldo and {Serna}, Javier and {Shao}, Zhengyi and {Shapiro}, Griffin and {Sharma}, Anubhav and {Shen}, Yue and {Shetrone}, Matthew and {Shu}, Yiping and {Simon}, Joshua D. and {Skrutskie}, M.~F. and {Smethurst}, Rebecca and {Smith}, Verne and {Sobeck}, Jennifer and {Spoo}, Taylor and {Sprague}, Dani and {Stark}, David V. and {Stassun}, Keivan G. and {Steinmetz}, Matthias and {Stello}, Dennis and {Stone-Martinez}, Alexander and {Storchi-Bergmann}, Thaisa and {Stringfellow}, Guy S. and {Stutz}, Amelia and {Su}, Yung-Chau and {Taghizadeh-Popp}, Manuchehr and {Talbot}, Michael S. and {Tayar}, Jamie and {Telles}, Eduardo and {Teske}, Johanna and {Thakar}, Ani and {Theissen}, Christopher and {Tkachenko}, Andrew and {Thomas}, Daniel and {Tojeiro}, Rita and {Hernandez Toledo}, Hector and {Troup}, Nicholas W. and {Trump}, Jonathan R. and {Trussler}, James and {Turner}, Jacqueline and {Tuttle}, Sarah and {Unda-Sanzana}, Eduardo and {V{\'a}zquez-Mata}, Jos{\'e} Antonio and {Valentini}, Marica and {Valenzuela}, Octavio and {Vargas-Gonz{\'a}lez}, Jaime and {Vargas-Maga{\~n}a}, Mariana and {Alfaro}, Pablo Vera and {Villanova}, Sandro and {Vincenzo}, Fiorenzo and {Wake}, David and {Warfield}, Jack T. and {Washington}, Jessica Diane and {Weaver}, Benjamin Alan and {Weijmans}, Anne-Marie and {Weinberg}, David H. and {Weiss}, Achim and {Westfall}, Kyle B. and {Wild}, Vivienne and {Wilde}, Matthew C. and {Wilson}, John C. and {Wilson}, Robert F. and {Wilson}, Mikayla and {Wolf}, Julien and {Wood-Vasey}, W.~M. and {Yan}, Renbin and {Zamora}, Olga and {Zasowski}, Gail and {Zhang}, Kai and {Zhao}, Cheng and {Zheng}, Zheng and {Zheng}, Zheng and {Zhu}, Kai},
        title = "{The Seventeenth Data Release of the Sloan Digital Sky Surveys: Complete Release of MaNGA, MaStar, and APOGEE-2 Data}",
      journal = {\apjs},
         year = {2022},
        month = apr,
       volume = {259},
       number = {2},
          eid = {35},
        pages = {35},
          doi = {10.3847/1538-4365/ac4414},
archivePrefix = {arXiv},
       eprint = {2112.02026},
 primaryClass = {astro-ph.GA},
       adsurl = {https://ui.adsabs.harvard.edu/abs/2022ApJS..259...35A}
}

@ARTICLE{Tsantaki2022,
       author = {{Tsantaki}, M. and {Pancino}, E. and {Marrese}, P. and {Marinoni}, S. and {Rainer}, M. and {Sanna}, N. and {Turchi}, A. and {Randich}, S. and {Gallart}, C. and {Battaglia}, G. and {Masseron}, T.},
        title = "{Survey of Surveys. I. The largest compilation of radial velocities for the Galaxy}",
      journal = {\aap},
         year = 2022,
        month = mar,
       volume = {659},
          eid = {A95},
        pages = {A95},
          doi = {10.1051/0004-6361/202141702},
archivePrefix = {arXiv},
       eprint = {2110.09316},
 primaryClass = {astro-ph.GA},
       adsurl = {https://ui.adsabs.harvard.edu/abs/2022A&A...659A..95T}
}

@ARTICLE{Majewski2017,
       author = {{Majewski}, Steven R. and {Schiavon}, Ricardo P. and {Frinchaboy}, Peter M. and {Allende Prieto}, Carlos and {Barkhouser}, Robert and {Bizyaev}, Dmitry and {Blank}, Basil and {Brunner}, Sophia and {Burton}, Adam and {Carrera}, Ricardo and {Chojnowski}, S. Drew and {Cunha}, K{\'a}tia and {Epstein}, Courtney and {Fitzgerald}, Greg and {Garc{\'\i}a P{\'e}rez}, Ana E. and {Hearty}, Fred R. and {Henderson}, Chuck and {Holtzman}, Jon A. and {Johnson}, Jennifer A. and {Lam}, Charles R. and {Lawler}, James E. and {Maseman}, Paul and {M{\'e}sz{\'a}ros}, Szabolcs and {Nelson}, Matthew and {Nguyen}, Duy Coung and {Nidever}, David L. and {Pinsonneault}, Marc and {Shetrone}, Matthew and {Smee}, Stephen and {Smith}, Verne V. and {Stolberg}, Todd and {Skrutskie}, Michael F. and {Walker}, Eric and {Wilson}, John C. and {Zasowski}, Gail and {Anders}, Friedrich and {Basu}, Sarbani and {Beland}, Stephane and {Blanton}, Michael R. and {Bovy}, Jo and {Brownstein}, Joel R. and {Carlberg}, Joleen and {Chaplin}, William and {Chiappini}, Cristina and {Eisenstein}, Daniel J. and {Elsworth}, Yvonne and {Feuillet}, Diane and {Fleming}, Scott W. and {Galbraith-Frew}, Jessica and {Garc{\'\i}a}, Rafael A. and {Garc{\'\i}a-Hern{\'a}ndez}, D. An{\'\i}bal and {Gillespie}, Bruce A. and {Girardi}, L{\'e}o and {Gunn}, James E. and {Hasselquist}, Sten and {Hayden}, Michael R. and {Hekker}, Saskia and {Ivans}, Inese and {Kinemuchi}, Karen and {Klaene}, Mark and {Mahadevan}, Suvrath and {Mathur}, Savita and {Mosser}, Beno{\^\i}t and {Muna}, Demitri and {Munn}, Jeffrey A. and {Nichol}, Robert C. and {O'Connell}, Robert W. and {Parejko}, John K. and {Robin}, A.~C. and {Rocha-Pinto}, Helio and {Schultheis}, Matthias and {Serenelli}, Aldo M. and {Shane}, Neville and {Silva Aguirre}, Victor and {Sobeck}, Jennifer S. and {Thompson}, Benjamin and {Troup}, Nicholas W. and {Weinberg}, David H. and {Zamora}, Olga},
        title = "{The Apache Point Observatory Galactic Evolution Experiment (APOGEE)}",
      journal = {\aj},
         year = 2017,
        month = sep,
       volume = {154},
       number = {3},
          eid = {94},
        pages = {94},
          doi = {10.3847/1538-3881/aa784d},
archivePrefix = {arXiv},
       eprint = {1509.05420},
 primaryClass = {astro-ph.IM},
       adsurl = {https://ui.adsabs.harvard.edu/abs/2017AJ....154...94M}
}

@ARTICLE{Dahm2012,
       author = {{Dahm}, S.~E. and {Slesnick}, Catherine L. and {White}, R.~J.},
        title = "{A Correlation between Circumstellar Disks and Rotation in the Upper Scorpius OB Association}",
      journal = {\apj},
         year = 2012,
        month = jan,
       volume = {745},
       number = {1},
          eid = {56},
        pages = {56},
          doi = {10.1088/0004-637X/745/1/56},
archivePrefix = {arXiv},
       eprint = {1110.0536},
 primaryClass = {astro-ph.SR},
       adsurl = {https://ui.adsabs.harvard.edu/abs/2012ApJ...745...56D}
}

@ARTICLE{James2006,
       author = {{James}, D.~J. and {Melo}, C. and {Santos}, N.~C. and {Bouvier}, J.},
        title = "{Fundamental properties of pre-main sequence stars in young, southern star forming regions: metallicities}",
      journal = {\aap},
         year = 2006,
        month = feb,
       volume = {446},
       number = {3},
        pages = {971-983},
          doi = {10.1051/0004-6361:20053900},
archivePrefix = {arXiv},
       eprint = {astro-ph/0510596},
 primaryClass = {astro-ph},
       adsurl = {https://ui.adsabs.harvard.edu/abs/2006A&A...446..971J}
}

@ARTICLE{Wichmann1999,
       author = {{Wichmann}, R. and {Covino}, E. and {Alcal{\'a}}, J.~M. and {Krautter}, J. and {Allain}, S. and {Hauschildt}, P.~H.},
        title = "{High-resolution spectroscopy of ROSAT-discovered weak-line T Tauri stars near Lupus}",
      journal = {\mnras},
         year = 1999,
        month = aug,
       volume = {307},
       number = {4},
        pages = {909-918},
          doi = {10.1046/j.1365-8711.1999.02666.x},
       adsurl = {https://ui.adsabs.harvard.edu/abs/1999MNRAS.307..909W}
}

@ARTICLE{Chen2011,
       author = {{Chen}, Christine H. and {Mamajek}, Eric E. and {Bitner}, Martin A. and {Pecaut}, Mark and {Su}, Kate Y.~L. and {Weinberger}, Alycia J.},
        title = "{A Magellan MIKE and Spitzer MIPS Study of 1.5-1.0 M $_{sun}$ Stars in Scorpius-Centaurus}",
      journal = {\apj},
         year = 2011,
        month = sep,
       volume = {738},
       number = {2},
          eid = {122},
        pages = {122},
          doi = {10.1088/0004-637X/738/2/122},
       adsurl = {https://ui.adsabs.harvard.edu/abs/2011ApJ...738..122C}
}

@ARTICLE{Krause2018,
       author = {{Krause}, Martin G.~H. and {Burkert}, Andreas and {Diehl}, Roland and {Fierlinger}, Katharina and {Gaczkowski}, Benjamin and {Kroell}, Daniel and {Ngoumou}, Judith and {Roccatagliata}, Veronica and {Siegert}, Thomas and {Preibisch}, Thomas},
        title = "{Surround and Squash: the impact of superbubbles on the interstellar medium in Scorpius-Centaurus OB2}",
      journal = {\aap},
         year = 2018,
        month = nov,
       volume = {619},
          eid = {A120},
        pages = {A120},
          doi = {10.1051/0004-6361/201732416},
archivePrefix = {arXiv},
       eprint = {1808.04788},
 primaryClass = {astro-ph.GA},
       adsurl = {https://ui.adsabs.harvard.edu/abs/2018A&A...619A.120K}
}

@ARTICLE{Meszaros2025,
       author = {{M{\'e}sz{\'a}ros}, Szabolcs and {Jofr{\'e}}, Paula and {Johnson}, Jennifer A. and {Bird}, Jonathan C. and {Bovy}, Jo and {Casey}, Andrew R. and {Chanam{\'e}}, Julio and {Cunha}, Katia and {De Lee}, Nathan and {Frinchaboy}, Peter and {Guiglion}, Guillaume and {Heged{\H{u}}s}, Viola and {Ji}, Alex P. and {Kollmeier}, Juna A. and {Ness}, Melissa K. and {Otto}, Jonah and {Pinsonneault}, Marc H. and {Roman-Lopes}, Alexandre and {Saydjari}, Andrew and {Sinha}, Amaya and {Song}, Ying-Yi and {Stringfellow}, Guy S. and {Stassun}, Keivan G. and {Tayar}, Jamie and {Tkachenko}, Andrew and {Valentini}, Marica and {Way}, Zachary and {Weingrill}, J{\"o}rg},
        title = "{SDSS-V Milky Way Mapper (MWM): ASPCAP Stellar Parameters and Abundances in SDSS-V Data Release 19}",
      journal = {\aj},
         year = 2025,
        month = aug,
       volume = {170},
       number = {2},
          eid = {96},
        pages = {96},
          doi = {10.3847/1538-3881/ade4b9},
archivePrefix = {arXiv},
       eprint = {2506.07845},
 primaryClass = {astro-ph.SR},
       adsurl = {https://ui.adsabs.harvard.edu/abs/2025AJ....170...96M}
}

@ARTICLE{Kollmeier2026,
       author = {{Kollmeier}, Juna A. and {Rix}, Hans-Walter and {Aerts}, Conny and {Aird}, James and {Vera Alfaro}, Pablo and {Almeida}, Andr{\'e}s and {Anderson}, Scott F. and {Arseneau}, Stefan M. and {Assef}, Roberto J. and {Aviram}, Shir and {Aydar}, Catarina and {Badenes}, Carles and {Bandyopadhyay}, Avrajit and {Barger}, Kat and {Barkhouser}, Robert H. and {Bauer}, Franz E. and {Behmard}, Aida and {Bender}, Chad and {Besser}, Felipe and {Bhattarai}, Binod and {Bilgi}, Pavaman and {Bird}, Jonathan and {Bizyaev}, Dmitry and {Blanc}, Guillermo A. and {Blanton}, Michael R. and {Bochanski}, John and {Bovy}, Jo and {Brandon}, Christopher and {Brandt}, William Nielsen and {Brownstein}, Joel R. and {Buchner}, Johannes and {Burchett}, Joseph N. and {Carlberg}, Joleen and {Casey}, Andrew R. and {Castaneda-Carlos}, Lesly and {Chakraborty}, Priyanka and {Chanam{\'e}}, Julio and {Chandra}, Vedant and {Cherinka}, Brian and {Chilingarian}, Igor and {Comparat}, Johan and {Cosens}, Maren and {Covey}, Kevin and {Crane}, Jeffrey D. and {Crumpler}, Nicole R. and {Cruz-Gonzalez}, Irene and {Cunha}, Katia and {Cunningham}, Tim and {Dai}, Xinyu and {Darling}, Jeremy and {Davidson}, Jr., James W. and {Davis}, Megan C. and {De Lee}, Nathan and {Deacon}, Niall and {M{\'e}ndez Delgado}, Jos{\'e} Eduardo and {Demasi}, Sebastian and {Demianenko}, Mariia and {Derwent}, Mark and {D'Onghia}, Elena and {Di Mille}, Francesco and {Dias}, Bruno and {Donor}, John and {Dow}, Peter N. and {Drory}, Niv and {Dwelly}, Tom and {Egorov}, Oleg and {Egorova}, Evgeniya and {El-Badry}, Kareem and {Engelman}, Mike and {Eracleous}, Mike and {Fan}, Xiaohui and {Farr}, Emily and {Fries}, Logan and {Frinchaboy}, Peter and {Froning}, Cynthia S. and {G{\"a}nsicke}, Boris T. and {Garc{\'\i}a}, Pablo and {Gelfand}, Joseph and {Gentile Fusillo}, Nicola Pietro and {Glover}, Simon and {Grabowski}, Katie and {Grebel}, Eva K. and {Green}, Paul J. and {Grier}, Catherine and {Gupta}, Pramod and {Gray}, Aidan C. and {H{\"a}berle}, Maximilian and {Hall}, Patrick B. and {Hammond}, Randolph P. and {Hawkins}, Keith and {Harding}, Albert C. and {Heged{\H{u}}s}, Viola and {Herbst}, Tom and {Hermes}, J.~J. and {Rodr{\'\i}guez Hidalgo}, Paola and {Hilder}, Thomas and {Hogg}, David W. and {Holtzman}, Jon A. and {Horta}, Danny and {Huang}, Yang and {Hwang}, Hsiang-Chih and {Ibarra-Medel}, Hector Javier and {Imig}, Julie and {Inight}, Keith and {Jana}, Arghajit and {Ji}, Alexander P. and {Jim{\'e}nez-Arranz}, {\'O}scar and {Jofre}, Paula and {Johns}, Matt and {Johnson}, Jennifer and {Johnson}, James W. and {Johnston}, Evelyn J. and {Jones}, Amy M. and {Katkov}, Ivan and {Knapp}, Gillian R. and {Koekemoer}, Anton M. and {Kounkel}, Marina and {Kreckel}, Kathryn and {Krishnarao}, Dhanesh and {Krumpe}, Mirko and {Kumari}, Nimisha and {Kupfer}, Thomas and {Lacerna}, Ivan and {Laporte}, Chervin and {Lepine}, Sebastien and {Li}, Jing and {Liu}, Xin and {Loebman}, Sarah and {Long}, Knox and {Roman-Lopes}, Alexandre and {Lu}, Yuxi and {Majewski}, Steven Raymond and {Maoz}, Dan and {McKinnon}, Kevin A. and {Medan}, Ilija and {Merloni}, Andrea and {Minniti}, Dante and {Morrison}, Sean and {Myers}, Natalie and {M{\'e}sz{\'a}ros}, Szabolcs and {Nandra}, Kirpal and {Nayak}, Prasanta K. and {Ness}, Melissa K. and {Nidever}, David L. and {O'Brien}, Thomas and {Oeur}, Micah and {Oravetz}, Audrey and {Oravetz}, Daniel and {Otto}, Jonah and {Pallathadka}, Gautham Adamane and {Palunas}, Povilas and {Pan}, Kaike and {Pappalardo}, Daniel and {Pandey}, Rakesh and {Negrete Pe{\~n}aloza}, Castalia Alenka and {Pinsonneault}, Marc H. and {Pogge}, Richard W. and {Taghizadeh Popp}, Manuchehr and {Price-Whelan}, Adrian M. and {Pulatova}, Nadiia and {Qiu}, Dan and {Ramirez}, Solange and {Rankine}, Amy and {Ricci}, Claudio and {Runnoe}, Jessie C. and {Sanchez}, Sebastian and {Salvato}, Mara and {Sarbadhicary}, Sumit K. and {Sattler}, Natascha and {Saydjari}, Andrew K. and {Sayres}, Conor and {Schinnerer}, Eva and {Schlaufman}, Kevin C. and {Schneider}, Donald P. and {Schreiber}, Matthias R. and {Schwope}, Axel and {Serna}, Javier and {Shen}, Yue and {Sif{\'o}n}, Crist{\'o}bal and {Singh}, Amrita and {Sinha}, Amaya and {Smee}, Stephen and {Song}, Ying-Yi and {Souto}, Diogo and {Stassun}, Keivan G. and {Steinmetz}, Matthias and {Stone-Martinez}, Alexander and {Stringfellow}, Guy and {Stutz}, Amelia and {S{\'a}nchez-Gallego}, Jos{\'e} and {Tan}, Jonathan C. and {Tayar}, Jamie and {Thai}, Riley and {Thakar}, Ani and {Ting}, Yuan-Sen and {Tkachenko}, Andrew and {Tovmassian}, Gagik and {Trakhtenbrot}, Benny and {Fern{\'a}ndez-Trincado}, Jos{\'e} G. and {Troup}, Nicholas},
        title = "{Sloan Digital Sky Survey. V. Pioneering Panoptic Spectroscopy}",
      journal = {\aj},
         year = 2026,
        month = jan,
       volume = {171},
       number = {1},
          eid = {52},
        pages = {52},
          doi = {10.3847/1538-3881/ae0576},
archivePrefix = {arXiv},
       eprint = {2507.06989},
 primaryClass = {astro-ph.IM},
       adsurl = {https://ui.adsabs.harvard.edu/abs/2026AJ....171...52K}
}

@ARTICLE{Kollmeier2017,
       author = {{Kollmeier}, Juna A. and {Zasowski}, Gail and {Rix}, Hans-Walter and {Johns}, Matt and {Anderson}, Scott F. and {Drory}, Niv and {Johnson}, Jennifer A. and {Pogge}, Richard W. and {Bird}, Jonathan C. and {Blanc}, Guillermo A. and {Brownstein}, Joel R. and {Crane}, Jeffrey D. and {De Lee}, Nathan M. and {Klaene}, Mark A. and {Kreckel}, Kathryn and {MacDonald}, Nick and {Merloni}, Andrea and {Ness}, Melissa K. and {O'Brien}, Thomas and {Sanchez-Gallego}, Jose R. and {Sayres}, Conor C. and {Shen}, Yue and {Thakar}, Ani R. and {Tkachenko}, Andrew and {Aerts}, Conny and {Blanton}, Michael R. and {Eisenstein}, Daniel J. and {Holtzman}, Jon A. and {Maoz}, Dan and {Nandra}, Kirpal and {Rockosi}, Constance and {Weinberg}, David H. and {Bovy}, Jo and {Casey}, Andrew R. and {Chaname}, Julio and {Clerc}, Nicolas and {Conroy}, Charlie and {Eracleous}, Michael and {G{\"a}nsicke}, Boris T. and {Hekker}, Saskia and {Horne}, Keith and {Kauffmann}, Jens and {McQuinn}, Kristen B.~W. and {Pellegrini}, Eric W. and {Schinnerer}, Eva and {Schlafly}, Edward F. and {Schwope}, Axel D. and {Seibert}, Mark and {Teske}, Johanna K. and {van Saders}, Jennifer L.},
        title = "{SDSS-V: Pioneering Panoptic Spectroscopy}",
      journal = {arXiv e-prints},
         year = 2017,
        month = nov,
          eid = {arXiv:1711.03234},
        pages = {arXiv:1711.03234},
          doi = {10.48550/arXiv.1711.03234},
archivePrefix = {arXiv},
       eprint = {1711.03234},
 primaryClass = {astro-ph.GA},
       adsurl = {https://ui.adsabs.harvard.edu/abs/2017arXiv171103234K}
}

@ARTICLE{Slesnick2008,
       author = {{Slesnick}, Catherine L. and {Hillenbrand}, Lynne A. and {Carpenter}, John M.},
        title = "{A Large-Area Search for Low-Mass Objects in Upper Scorpius. II. Age and Mass Distributions}",
      journal = {\apj},
         year = 2008,
        month = nov,
       volume = {688},
       number = {1},
        pages = {377-397},
          doi = {10.1086/592265},
archivePrefix = {arXiv},
       eprint = {0809.1436},
 primaryClass = {astro-ph},
       adsurl = {https://ui.adsabs.harvard.edu/abs/2008ApJ...688..377S}
}

@ARTICLE{ElmegreenEfremov1998,
       author = {{Elmegreen}, Bruce G. and {Efremov}, Yuri N.},
        title = "{Hierarchy of Interstellar and Stellar Structures and the Case of the Orion Star-Forming Region}",
      journal = {arXiv e-prints},
         year = 1998,
        month = jan,
          eid = {astro-ph/9801071},
        pages = {astro-ph/9801071},
          doi = {10.48550/arXiv.astro-ph/9801071},
archivePrefix = {arXiv},
       eprint = {astro-ph/9801071},
 primaryClass = {astro-ph},
       adsurl = {https://ui.adsabs.harvard.edu/abs/1998astro.ph..1071E}
}

@ARTICLE{Elmegreen1977,
       author = {{Elmegreen}, B.~G. and {Lada}, C.~J.},
        title = "{Sequential formation of subgroups in OB associations.}",
      journal = {\apj},
         year = 1977,
        month = jun,
       volume = {214},
        pages = {725-741},
          doi = {10.1086/155302},
       adsurl = {https://ui.adsabs.harvard.edu/abs/1977ApJ...214..725E}
}

@ARTICLE{Neuhaeuser2020,
       author = {{Neuh{\"a}user}, R. and {Gie{\ss}ler}, F. and {Hambaryan}, V.~V.},
        title = "{A nearby recent supernova that ejected the runaway star {\ensuremath{\zeta}} Oph, the pulsar PSR B1706-16, and $^{60}$Fe found on Earth}",
      journal = {\mnras},
         year = 2020,
        month = oct,
       volume = {498},
       number = {1},
        pages = {899-917},
          doi = {10.1093/mnras/stz2629},
archivePrefix = {arXiv},
       eprint = {1909.06850},
 primaryClass = {astro-ph.HE},
       adsurl = {https://ui.adsabs.harvard.edu/abs/2020MNRAS.498..899N}
}

@ARTICLE{deAvillezBreitschwerdt2005,
       author = {{de Avillez}, M.~A. and {Breitschwerdt}, D.},
        title = "{Global dynamical evolution of the ISM in star-forming galaxies. I. High resolution 3D simulations: Effect of the magnetic field}",
      journal = {\aap},
         year = 2005,
        month = jun,
       volume = {436},
       number = {2},
        pages = {585-600},
          doi = {10.1051/0004-6361:20042146},
archivePrefix = {arXiv},
       eprint = {astro-ph/0502327},
 primaryClass = {astro-ph},
       adsurl = {https://ui.adsabs.harvard.edu/abs/2005A&A...436..585D}
}

@ARTICLE{Breitschwerdt2016,
       author = {{Breitschwerdt}, D. and {Feige}, J. and {Schulreich}, M.~M. and {Avillez}, M.~A. De. and {Dettbarn}, C. and {Fuchs}, B.},
        title = "{The locations of recent supernovae near the Sun from modelling $^{60}$Fe transport}",
      journal = {\nat},
         year = 2016,
        month = apr,
       volume = {532},
       number = {7597},
        pages = {73-76},
          doi = {10.1038/nature17424},
       adsurl = {https://ui.adsabs.harvard.edu/abs/2016Natur.532...73B}
}

@book{Daniel1990,
  title={Applied Nonparametric Statistics},
  author={Daniel, W.W.},
  isbn={9780534919764},
  lccn={89009463},
  series={Duxbury advanced series in statistics and decision sciences},
  url={https://books.google.de/books?id=0hPvAAAAMAAJ},
  year={1990},
  publisher={PWS-KENT Pub.}
}

@article{Spearman1904,
 ISSN = {00029556},
 URL = {http://www.jstor.org/stable/1412159},
 author = {C. Spearman},
 journal = {The American Journal of Psychology},
 number = {1},
 pages = {72--101},
 publisher = {University of Illinois Press},
 title = {The Proof and Measurement of Association between Two Things},
 urldate = {2024-08-08},
 volume = {15},
 year = {1904}
}

@ARTICLE{Gagne2018a,
       author = {{Gagn{\'e}}, Jonathan and {Roy-Loubier}, Olivier and {Faherty}, Jacqueline K. and {Doyon}, Ren{\'e} and {Malo}, Lison},
        title = "{BANYAN. XII. New Members of Nearby Young Associations from GAIA-Tycho Data}",
      journal = {\apj},
         year = {2018a},
        month = jun,
       volume = {860},
       number = {1},
          eid = {43},
        pages = {43},
          doi = {10.3847/1538-4357/aac2b8},
archivePrefix = {arXiv},
       eprint = {1804.03093},
 primaryClass = {astro-ph.SR},
       adsurl = {https://ui.adsabs.harvard.edu/abs/2018ApJ...860...43G}
}

@ARTICLE{MiretRoig2022b,
       author = {{Miret-Roig}, N. and {Galli}, P.~A.~B. and {Olivares}, J. and {Bouy}, H. and {Alves}, J. and {Barrado}, D.},
        title = "{The star formation history of Upper Scorpius and Ophiuchus. A 7D picture: positions, kinematics, and dynamical traceback ages}",
      journal = {\aap},
         year = {2022b},
        month = nov,
       volume = {667},
          eid = {A163},
        pages = {A163},
          doi = {10.1051/0004-6361/202244709},
       adsurl = {https://ui.adsabs.harvard.edu/abs/2022A&A...667A.163M}
}

@ARTICLE{Hunt2023,
       author = {{Hunt}, Emily L. and {Reffert}, Sabine},
        title = "{Improving the open cluster census. II. An all-sky cluster catalogue with Gaia DR3}",
      journal = {\aap},
         year = 2023,
        month = may,
       volume = {673},
          eid = {A114},
        pages = {A114},
          doi = {10.1051/0004-6361/202346285},
archivePrefix = {arXiv},
       eprint = {2303.13424},
 primaryClass = {astro-ph.GA},
       adsurl = {https://ui.adsabs.harvard.edu/abs/2023A&A...673A.114H}
}

@ARTICLE{Schoenrich2010,
       author = {{Sch{\"o}nrich}, Ralph and {Binney}, James and {Dehnen}, Walter},
        title = "{Local kinematics and the local standard of rest}",
      journal = {\mnras},
         year = 2010,
        month = apr,
       volume = {403},
       number = {4},
        pages = {1829-1833},
          doi = {10.1111/j.1365-2966.2010.16253.x},
archivePrefix = {arXiv},
       eprint = {0912.3693},
 primaryClass = {astro-ph.GA},
       adsurl = {https://ui.adsabs.harvard.edu/abs/2010MNRAS.403.1829S}
}

@ARTICLE{Grossschedl2021,
       author = {{Gro{\ss}schedl}, Josefa E. and {Alves}, Jo{\~a}o and {Meingast}, Stefan and {Herbst-Kiss}, Gabor},
        title = "{3D dynamics of the Orion cloud complex. Discovery of coherent radial gas motions at the 100-pc scale}",
      journal = {\aap},
         year = 2021,
        month = mar,
       volume = {647},
          eid = {A91},
        pages = {A91},
          doi = {10.1051/0004-6361/202038913},
archivePrefix = {arXiv},
       eprint = {2007.07254},
 primaryClass = {astro-ph.SR},
       adsurl = {https://ui.adsabs.harvard.edu/abs/2021A&A...647A..91G}
}

@INPROCEEDINGS{Mamajek2001,
       author = {{Mamajek}, E.~E. and {Feigelson}, E.~D.},
        title = "{The Dispersal of Young Stars and the Greater Sco-Cen Association}",
    booktitle = {Young Stars Near Earth: Progress and Prospects},
         year = 2001,
       editor = {{Jayawardhana}, Ray and {Greene}, Thomas},
       series = {Astronomical Society of the Pacific Conference Series},
       volume = {244},
        month = jan,
        pages = {104-115},
archivePrefix = {arXiv},
       eprint = {astro-ph/0105290},
 primaryClass = {astro-ph},
       adsurl = {https://ui.adsabs.harvard.edu/abs/2001ASPC..244..104M}
}

@ARTICLE{Lada1984,
       author = {{Lada}, C.~J. and {Margulis}, M. and {Dearborn}, D.},
        title = "{The formation and early dynamical evolution of bound stellar systems.}",
      journal = {\apj},
         year = 1984,
        month = oct,
       volume = {285},
        pages = {141-152},
          doi = {10.1086/162485},
       adsurl = {https://ui.adsabs.harvard.edu/abs/1984ApJ...285..141L}
}

@ARTICLE{Swiggum2025,
       author = {{Swiggum}, C. and {Alves}, J. and {D'Onghia}, E.},
        title = "{From moving groups to star formation in the solar neighborhood}",
      journal = {\aap},
         year = 2025,
        month = jul,
       volume = {699},
          eid = {L5},
        pages = {L5},
          doi = {10.1051/0004-6361/202554985},
archivePrefix = {arXiv},
       eprint = {2504.02825},
 primaryClass = {astro-ph.GA},
       adsurl = {https://ui.adsabs.harvard.edu/abs/2025A&A...699L...5S}
}

@ARTICLE{Brown1997,
       author = {{Brown}, A.~G.~A. and {Dekker}, G. and {de Zeeuw}, P.~T.},
        title = "{Kinematic ages of OB associations}",
      journal = {\mnras},
         year = 1997,
        month = mar,
       volume = {285},
       number = {3},
        pages = {479-492},
          doi = {10.1093/mnras/285.3.479},
       adsurl = {https://ui.adsabs.harvard.edu/abs/1997MNRAS.285..479B}
}

@ARTICLE{Wright2018,
       author = {{Wright}, Nicholas J. and {Mamajek}, Eric E.},
        title = "{The kinematics of the Scorpius-Centaurus OB association from Gaia DR1}",
      journal = {\mnras},
         year = 2018,
        month = may,
       volume = {476},
       number = {1},
        pages = {381-398},
          doi = {10.1093/mnras/sty207},
archivePrefix = {arXiv},
       eprint = {1801.08540},
 primaryClass = {astro-ph.SR},
       adsurl = {https://ui.adsabs.harvard.edu/abs/2018MNRAS.476..381W}
}

@ARTICLE{Bovy2015,
       author = {{Bovy}, Jo},
        title = "{galpy: A python Library for Galactic Dynamics}",
      journal = {\apjs},
         year = 2015,
        month = feb,
       volume = {216},
       number = {2},
          eid = {29},
        pages = {29},
          doi = {10.1088/0067-0049/216/2/29},
archivePrefix = {arXiv},
       eprint = {1412.3451},
 primaryClass = {astro-ph.GA},
       adsurl = {https://ui.adsabs.harvard.edu/abs/2015ApJS..216...29B}
}

@ARTICLE{Blaauw1946,
       author = {{Blaauw}, A.},
        title = "{A Study of the Scorpio-Centaurus Cluster}",
      journal = {Publications of the Kapteyn Astronomical Laboratory Groningen},
         year = 1946,
        month = jan,
       volume = {52},
        pages = {1-132},
       adsurl = {https://ui.adsabs.harvard.edu/abs/1946PGro...52....1B}
}

@ARTICLE{Blaauw1964a,
       author = {{Blaauw}, Adriaan},
        title = "{The O Associations in the Solar Neighborhood}",
      journal = {\araa},
         year = {1964a},
        month = jan,
       volume = {2},
        pages = {213},
          doi = {10.1146/annurev.aa.02.090164.001241},
       adsurl = {https://ui.adsabs.harvard.edu/abs/1964ARA&A...2..213B}
}

@INPROCEEDINGS{Blaauw1964b,
       author = {{Blaauw}, A.},
        title = "{The Scorpio-Centaurus association}",
    booktitle = {The Galaxy and the Magellanic Clouds},
         year = {1964b},
       editor = {{Kerr}, Frank J.},
       volume = {20},
        month = jan,
        pages = {50},
       adsurl = {https://ui.adsabs.harvard.edu/abs/1964IAUS...20...50B}
}

@ARTICLE{deZeeuw1999,
       author = {{de Zeeuw}, P.~T. and {Hoogerwerf}, R. and {de Bruijne}, J.~H.~J. and {Brown}, A.~G.~A. and {Blaauw}, A.},
        title = "{A HIPPARCOS Census of the Nearby OB Associations}",
      journal = {\aj},
         year = 1999,
        month = jan,
       volume = {117},
       number = {1},
        pages = {354-399},
          doi = {10.1086/300682},
archivePrefix = {arXiv},
       eprint = {astro-ph/9809227},
 primaryClass = {astro-ph},
       adsurl = {https://ui.adsabs.harvard.edu/abs/1999AJ....117..354D}
}

@ARTICLE{Bressan2012,
       author = {{Bressan}, Alessandro and {Marigo}, Paola and {Girardi}, L{\'e}o. and {Salasnich}, Bernardo and {Dal Cero}, Claudia and {Rubele}, Stefano and {Nanni}, Ambra},
        title = "{PARSEC: stellar tracks and isochrones with the PAdova and TRieste Stellar Evolution Code}",
      journal = {\mnras},
         year = 2012,
        month = nov,
       volume = {427},
       number = {1},
        pages = {127-145},
          doi = {10.1111/j.1365-2966.2012.21948.x},
archivePrefix = {arXiv},
       eprint = {1208.4498},
 primaryClass = {astro-ph.SR},
       adsurl = {https://ui.adsabs.harvard.edu/abs/2012MNRAS.427..127B}
}

@ARTICLE{Castro-Ginard2024,
       author = {{Castro-Ginard}, Alfred and {Penoyre}, Zephyr and {Casey}, Andrew R. and {Brown}, Anthony G.~A. and {Belokurov}, Vasily and {Cantat-Gaudin}, Tristan and {Drimmel}, Ronald and {Fouesneau}, Morgan and {Khanna}, Shourya and {Kurbatov}, Evgeny P. and {Price-Whelan}, Adrian M. and {Rix}, Hans-Walter and {Smart}, Richard L.},
        title = "{Gaia DR3 detectability of unresolved binary systems}",
      journal = {\aap},
         year = 2024,
        month = aug,
       volume = {688},
          eid = {A1},
        pages = {A1},
          doi = {10.1051/0004-6361/202450172},
archivePrefix = {arXiv},
       eprint = {2404.14127},
 primaryClass = {astro-ph.GA},
       adsurl = {https://ui.adsabs.harvard.edu/abs/2024A&A...688A...1C}
}

@ARTICLE{Pearson1895,
       author = {{Pearson}, Karl},
        title = "{Note on Regression and Inheritance in the Case of Two Parents}",
      journal = {Proceedings of the Royal Society of London Series I},
         year = 1895,
        month = jan,
       volume = {58},
        pages = {240-242},
       adsurl = {https://ui.adsabs.harvard.edu/abs/1895RSPS...58..240P}
}

@ARTICLE{Gaczkowski2017,
       author = {{Gaczkowski}, B. and {Roccatagliata}, V. and {Flaischlen}, S. and {Kr{\"o}ll}, D. and {Krause}, M.~G.~H. and {Burkert}, A. and {Diehl}, R. and {Fierlinger}, K. and {Ngoumou}, J. and {Preibisch}, T.},
        title = "{Squeezed between shells? The origin of the Lupus I molecular cloud. II. APEX CO and GASS H I observations}",
      journal = {\aap},
         year = 2017,
        month = dec,
       volume = {608},
          eid = {A102},
        pages = {A102},
          doi = {10.1051/0004-6361/201628508},
archivePrefix = {arXiv},
       eprint = {1710.07446},
 primaryClass = {astro-ph.GA},
       adsurl = {https://ui.adsabs.harvard.edu/abs/2017A&A...608A.102G}
}

@ARTICLE{Galli2013,
       author = {{Galli}, P.~A.~B. and {Bertout}, C. and {Teixeira}, R. and {Ducourant}, C.},
        title = "{A kinematic study and membership analysis of the Lupus star-forming region}",
      journal = {\aap},
         year = 2013,
        month = oct,
       volume = {558},
          eid = {A77},
        pages = {A77},
          doi = {10.1051/0004-6361/201220704},
archivePrefix = {arXiv},
       eprint = {1309.7799},
 primaryClass = {astro-ph.GA},
       adsurl = {https://ui.adsabs.harvard.edu/abs/2013A&A...558A..77G}
}

@ARTICLE{Kerr2021,
       author = {{Kerr}, Ronan M.~P. and {Rizzuto}, Aaron C. and {Kraus}, Adam L. and {Offner}, Stella S.~R.},
        title = "{Stars with Photometrically Young Gaia Luminosities Around the Solar System (SPYGLASS). I. Mapping Young Stellar Structures and Their Star Formation Histories}",
      journal = {\apj},
         year = 2021,
        month = aug,
       volume = {917},
       number = {1},
          eid = {23},
        pages = {23},
          doi = {10.3847/1538-4357/ac0251},
archivePrefix = {arXiv},
       eprint = {2105.09338},
 primaryClass = {astro-ph.GA},
       adsurl = {https://ui.adsabs.harvard.edu/abs/2021ApJ...917...23K}
}

@INPROCEEDINGS{Taylor2005,
   author = {{Taylor}, M.~B.},
    title = "{TOPCAT {\amp} STIL: Starlink Table/VOTable Processing Software}",
booktitle = {Astronomical Data Analysis Software and Systems XIV},
     year = 2005,
   series = {Astronomical Society of the Pacific Conference Series},
   volume = 347,
   editor = {{Shopbell}, P. and {Britton}, M. and {Ebert}, R.},
    month = dec,
    pages = {29},
   adsurl = {http://adsabs.harvard.edu/abs/2005ASPC..347...29T}
}

@ARTICLE{Hunter2007,
       author = {{Hunter}, John D.},
        title = "{Matplotlib: A 2D Graphics Environment}",
      journal = {Computing in Science and Engineering},
         year = 2007,
        month = may,
       volume = {9},
       number = {3},
        pages = {90-95},
          doi = {10.1109/MCSE.2007.55},
       adsurl = {https://ui.adsabs.harvard.edu/abs/2007CSE.....9...90H}
}

@ARTICLE{Ochsenbein2000,
       author = {{Ochsenbein}, F. and {Bauer}, P. and {Marcout}, J.},
        title = "{The VizieR database of astronomical catalogues}",
      journal = {\aaps},
         year = 2000,
        month = apr,
       volume = {143},
        pages = {23-32},
          doi = {10.1051/aas:2000169},
archivePrefix = {arXiv},
       eprint = {astro-ph/0002122},
 primaryClass = {astro-ph},
       adsurl = {https://ui.adsabs.harvard.edu/abs/2000A&AS..143...23O}
}

@ARTICLE{Bonnarel2000,
   author = {{Bonnarel}, F. and {Fernique}, P. and {Bienaym{\'e}}, O. and 
	{Egret}, D. and {Genova}, F. and {Louys}, M. and {Ochsenbein}, F. and 
	{Wenger}, M. and {Bartlett}, J.~G.},
    title = "{The ALADIN interactive sky atlas. A reference tool for identification of astronomical sources}",
  journal = {\aaps},
     year = 2000,
    month = apr,
   volume = 143,
    pages = {33-40},
      doi = {10.1051/aas:2000331},
   adsurl = {http://cdsads.u-strasbg.fr/abs/2000A%26AS..143...33B}
}

@INPROCEEDINGS{Boch2014,
       author = {{Boch}, T. and {Fernique}, P.},
        title = "{Aladin Lite: Embed your Sky in the Browser}",
    booktitle = {Astronomical Data Analysis Software and Systems XXIII},
         year = 2014,
       editor = {{Manset}, N. and {Forshay}, P.},
       series = {Astronomical Society of the Pacific Conference Series},
       volume = {485},
        month = may,
        pages = {277},
       adsurl = {https://ui.adsabs.harvard.edu/abs/2014ASPC..485..277B}
}

@ARTICLE{Walt2011,
       author = {{van der Walt}, St{\'e}fan and {Colbert}, S. Chris and
         {Varoquaux}, Ga{\"e}l},
        title = "{The NumPy Array: A Structure for Efficient Numerical Computation}",
      journal = {Computing in Science and Engineering},
         year = 2011,
        month = mar,
       volume = {13},
       number = {2},
        pages = {22-30},
          doi = {10.1109/MCSE.2011.37},
archivePrefix = {arXiv},
       eprint = {1102.1523},
 primaryClass = {cs.MS},
       adsurl = {https://ui.adsabs.harvard.edu/abs/2011CSE....13b..22V}
}

@ARTICLE{Astropy2013,
   author = {{Astropy Collaboration} and {Robitaille}, T.~P. and {Tollerud}, E.~J. and 
	{Greenfield}, P. and {Droettboom}, M. and {Bray}, E. and {Aldcroft}, T. and 
	{Davis}, M. and {Ginsburg}, A. and {Price-Whelan}, A.~M. and 
	{Kerzendorf}, W.~E. and {Conley}, A. and {Crighton}, N. and 
	{Barbary}, K. and {Muna}, D. and {Ferguson}, H. and {Grollier}, F. and 
	{Parikh}, M.~M. and {Nair}, P.~H. and {Unther}, H.~M. and {Deil}, C. and 
	{Woillez}, J. and {Conseil}, S. and {Kramer}, R. and {Turner}, J.~E.~H. and 
	{Singer}, L. and {Fox}, R. and {Weaver}, B.~A. and {Zabalza}, V. and 
	{Edwards}, Z.~I. and {Azalee Bostroem}, K. and {Burke}, D.~J. and 
	{Casey}, A.~R. and {Crawford}, S.~M. and {Dencheva}, N. and 
	{Ely}, J. and {Jenness}, T. and {Labrie}, K. and {Lim}, P.~L. and 
	{Pierfederici}, F. and {Pontzen}, A. and {Ptak}, A. and {Refsdal}, B. and 
	{Servillat}, M. and {Streicher}, O.},
    title = "{Astropy: A community Python package for astronomy}",
  journal = {\aap},
archivePrefix = "arXiv",
   eprint = {1307.6212},
 primaryClass = "astro-ph.IM",
     year = 2013,
    month = oct,
   volume = 558,
      eid = {A33},
    pages = {A33},
      doi = {10.1051/0004-6361/201322068},
   adsurl = {http://adsabs.harvard.edu/abs/2013A%26A...558A..33A}
}

@ARTICLE{Astropy2022,
       author = {{Astropy Collaboration} and {Price-Whelan}, Adrian M. and {Lim}, Pey Lian and {Earl}, Nicholas and {Starkman}, Nathaniel and {Bradley}, Larry and {Shupe}, David L. and {Patil}, Aarya A. and {Corrales}, Lia and {Brasseur}, C.~E. and {N{\"o}the}, Maximilian and {Donath}, Axel and {Tollerud}, Erik and {Morris}, Brett M. and {Ginsburg}, Adam and {Vaher}, Eero and {Weaver}, Benjamin A. and {Tocknell}, James and {Jamieson}, William and {van Kerkwijk}, Marten H. and {Robitaille}, Thomas P. and {Merry}, Bruce and {Bachetti}, Matteo and {G{\"u}nther}, H. Moritz and {Aldcroft}, Thomas L. and {Alvarado-Montes}, Jaime A. and {Archibald}, Anne M. and {B{\'o}di}, Attila and {Bapat}, Shreyas and {Barentsen}, Geert and {Baz{\'a}n}, Juanjo and {Biswas}, Manish and {Boquien}, M{\'e}d{\'e}ric and {Burke}, D.~J. and {Cara}, Daria and {Cara}, Mihai and {Conroy}, Kyle E. and {Conseil}, Simon and {Craig}, Matthew W. and {Cross}, Robert M. and {Cruz}, Kelle L. and {D'Eugenio}, Francesco and {Dencheva}, Nadia and {Devillepoix}, Hadrien A.~R. and {Dietrich}, J{\"o}rg P. and {Eigenbrot}, Arthur Davis and {Erben}, Thomas and {Ferreira}, Leonardo and {Foreman-Mackey}, Daniel and {Fox}, Ryan and {Freij}, Nabil and {Garg}, Suyog and {Geda}, Robel and {Glattly}, Lauren and {Gondhalekar}, Yash and {Gordon}, Karl D. and {Grant}, David and {Greenfield}, Perry and {Groener}, Austen M. and {Guest}, Steve and {Gurovich}, Sebastian and {Handberg}, Rasmus and {Hart}, Akeem and {Hatfield-Dodds}, Zac and {Homeier}, Derek and {Hosseinzadeh}, Griffin and {Jenness}, Tim and {Jones}, Craig K. and {Joseph}, Prajwel and {Kalmbach}, J. Bryce and {Karamehmetoglu}, Emir and {Ka{\l}uszy{\'n}ski}, Miko{\l}aj and {Kelley}, Michael S.~P. and {Kern}, Nicholas and {Kerzendorf}, Wolfgang E. and {Koch}, Eric W. and {Kulumani}, Shankar and {Lee}, Antony and {Ly}, Chun and {Ma}, Zhiyuan and {MacBride}, Conor and {Maljaars}, Jakob M. and {Muna}, Demitri and {Murphy}, N.~A. and {Norman}, Henrik and {O'Steen}, Richard and {Oman}, Kyle A. and {Pacifici}, Camilla and {Pascual}, Sergio and {Pascual-Granado}, J. and {Patil}, Rohit R. and {Perren}, Gabriel I. and {Pickering}, Timothy E. and {Rastogi}, Tanuj and {Roulston}, Benjamin R. and {Ryan}, Daniel F. and {Rykoff}, Eli S. and {Sabater}, Jose and {Sakurikar}, Parikshit and {Salgado}, Jes{\'u}s and {Sanghi}, Aniket and {Saunders}, Nicholas and {Savchenko}, Volodymyr and {Schwardt}, Ludwig and {Seifert-Eckert}, Michael and {Shih}, Albert Y. and {Jain}, Anany Shrey and {Shukla}, Gyanendra and {Sick}, Jonathan and {Simpson}, Chris and {Singanamalla}, Sudheesh and {Singer}, Leo P. and {Singhal}, Jaladh and {Sinha}, Manodeep and {Sip{\H{o}}cz}, Brigitta M. and {Spitler}, Lee R. and {Stansby}, David and {Streicher}, Ole and {{\v{S}}umak}, Jani and {Swinbank}, John D. and {Taranu}, Dan S. and {Tewary}, Nikita and {Tremblay}, Grant R. and {de Val-Borro}, Miguel and {Van Kooten}, Samuel J. and {Vasovi{\'c}}, Zlatan and {Verma}, Shresth and {de Miranda Cardoso}, Jos{\'e} Vin{\'\i}cius and {Williams}, Peter K.~G. and {Wilson}, Tom J. and {Winkel}, Benjamin and {Wood-Vasey}, W.~M. and {Xue}, Rui and {Yoachim}, Peter and {Zhang}, Chen and {Zonca}, Andrea and {Astropy Project Contributors}},
        title = "{The Astropy Project: Sustaining and Growing a Community-oriented Open-source Project and the Latest Major Release (v5.0) of the Core Package}",
      journal = {\apj},
         year = 2022,
        month = aug,
       volume = {935},
       number = {2},
          eid = {167},
        pages = {167},
          doi = {10.3847/1538-4357/ac7c74},
archivePrefix = {arXiv},
       eprint = {2206.14220},
 primaryClass = {astro-ph.IM},
       adsurl = {https://ui.adsabs.harvard.edu/abs/2022ApJ...935..167A}
}

@ARTICLE{Pecaut2016,
       author = {{Pecaut}, Mark J. and {Mamajek}, Eric E.},
        title = "{The star formation history and accretion-disc fraction among the K-type members of the Scorpius-Centaurus OB association}",
      journal = {\mnras},
         year = 2016,
        month = sep,
       volume = {461},
       number = {1},
        pages = {794-815},
          doi = {10.1093/mnras/stw1300},
archivePrefix = {arXiv},
       eprint = {1605.08789},
 primaryClass = {astro-ph.SR},
       adsurl = {https://ui.adsabs.harvard.edu/abs/2016MNRAS.461..794P}
}

@ARTICLE{Preibisch1999,
       author = {{Preibisch}, Thomas and {Zinnecker}, Hans},
        title = "{The History of Low-Mass Star Formation in the Upper Scorpius OB Association}",
      journal = {\aj},
         year = 1999,
        month = may,
       volume = {117},
       number = {5},
        pages = {2381-2397},
          doi = {10.1086/300842},
       adsurl = {https://ui.adsabs.harvard.edu/abs/1999AJ....117.2381P}
}

@INCOLLECTION{Preibisch2008,
       author = {{Preibisch}, T. and {Mamajek}, E.},
        title = "{The Nearest OB Association: Scorpius-Centaurus (Sco OB2)}",
    booktitle = {Handbook of Star Forming Regions, Volume II},
         year = 2008,
       editor = {{Reipurth}, B.},
    publisher = {Astronomical Society of the Pacific},
       volume = {5},
        pages = {235},
       adsurl = {https://ui.adsabs.harvard.edu/abs/2008hsf2.book..235P}
}

@ARTICLE{Buder2024,
       author = {{Buder}, S. and {Kos}, J. and {Wang}, E.~X. and {McKenzie}, M. and {Howell}, M. and {Martell}, S.~L. and {Hayden}, M.~R. and {Zucker}, D.~B. and {Nordlander}, T. and {Montet}, B.~T. and {Traven}, G. and {Bland-Hawthorn}, J. and {De Silva}, G.~M. and {Freeman}, K.~C. and {Lewis}, G.~F. and {Lind}, K. and {Sharma}, S. and {Simpson}, J.~D. and {Stello}, D. and {Zwitter}, T. and {Amarsi}, A.~M. and {Armstrong}, J.~J. and {Banks}, K. and {Beavis}, M.~A. and {Beeson}, K. and {Chen}, B. and {Ciuc{\u{a}}}, I. and {Da Costa}, G.~S. and {de Grijs}, R. and {Martin}, B. and {Nataf}, D.~M. and {Ness}, M.~K. and {Rains}, A.~D. and {Scarr}, T. and {Vogrin{\v{c}}i{\v{c}}}, R. and {Wang}, Z. and {Wittenmyer}, R.~A. and {Xie}, Y. and {The GALAH Collaboration}},
        title = "{The GALAH Survey: Data Release 4}",
      journal = {arXiv e-prints},
         year = 2024,
        month = sep,
          eid = {arXiv:2409.19858},
        pages = {arXiv:2409.19858},
          doi = {10.48550/arXiv.2409.19858},
archivePrefix = {arXiv},
       eprint = {2409.19858},
 primaryClass = {astro-ph.GA},
       adsurl = {https://ui.adsabs.harvard.edu/abs/2024arXiv240919858B}
}

@ARTICLE{Buder2021,
       author = {{Buder}, Sven and {Sharma}, Sanjib and {Kos}, Janez and {Amarsi}, Anish M. and {Nordlander}, Thomas and {Lind}, Karin and {Martell}, Sarah L. and {Asplund}, Martin and {Bland-Hawthorn}, Joss and {Casey}, Andrew R. and {de Silva}, Gayandhi M. and {D'Orazi}, Valentina and {Freeman}, Ken C. and {Hayden}, Michael R. and {Lewis}, Geraint F. and {Lin}, Jane and {Schlesinger}, Katharine J. and {Simpson}, Jeffrey D. and {Stello}, Dennis and {Zucker}, Daniel B. and {Zwitter}, Toma{\v{z}} and {Beeson}, Kevin L. and {Buck}, Tobias and {Casagrande}, Luca and {Clark}, Jake T. and {{\v{C}}otar}, Klemen and {da Costa}, Gary S. and {de Grijs}, Richard and {Feuillet}, Diane and {Horner}, Jonathan and {Kafle}, Prajwal R. and {Khanna}, Shourya and {Kobayashi}, Chiaki and {Liu}, Fan and {Montet}, Benjamin T. and {Nandakumar}, Govind and {Nataf}, David M. and {Ness}, Melissa K. and {Spina}, Lorenzo and {Tepper-Garc{\'\i}a}, Thor and {Ting}, Yuan-Sen and {Traven}, Gregor and {Vogrin{\v{c}}i{\v{c}}}, Rok and {Wittenmyer}, Robert A. and {Wyse}, Rosemary F.~G. and {{\v{Z}}erjal}, Maru{\v{s}}a and {Galah Collaboration}},
        title = "{The GALAH+ survey: Third data release}",
      journal = {\mnras},
         year = 2021,
        month = sep,
       volume = {506},
       number = {1},
        pages = {150-201},
          doi = {10.1093/mnras/stab1242},
archivePrefix = {arXiv},
       eprint = {2011.02505},
 primaryClass = {astro-ph.GA},
       adsurl = {https://ui.adsabs.harvard.edu/abs/2021MNRAS.506..150B}
}

@ARTICLE{DeSilva2015,
       author = {{De Silva}, G.~M. and {Freeman}, K.~C. and {Bland-Hawthorn}, J. and {Martell}, S. and {de Boer}, E. Wylie and {Asplund}, M. and {Keller}, S. and {Sharma}, S. and {Zucker}, D.~B. and {Zwitter}, T. and {Anguiano}, B. and {Bacigalupo}, C. and {Bayliss}, D. and {Beavis}, M.~A. and {Bergemann}, M. and {Campbell}, S. and {Cannon}, R. and {Carollo}, D. and {Casagrande}, L. and {Casey}, A.~R. and {Da Costa}, G. and {D'Orazi}, V. and {Dotter}, A. and {Duong}, L. and {Heger}, A. and {Ireland}, M.~J. and {Kafle}, P.~R. and {Kos}, J. and {Lattanzio}, J. and {Lewis}, G.~F. and {Lin}, J. and {Lind}, K. and {Munari}, U. and {Nataf}, D.~M. and {O'Toole}, S. and {Parker}, Q. and {Reid}, W. and {Schlesinger}, K.~J. and {Sheinis}, A. and {Simpson}, J.~D. and {Stello}, D. and {Ting}, Y. -S. and {Traven}, G. and {Watson}, F. and {Wittenmyer}, R. and {Yong}, D. and {{\v{Z}}erjal}, M.},
        title = "{The GALAH survey: scientific motivation}",
      journal = {\mnras},
         year = 2015,
        month = may,
       volume = {449},
       number = {3},
        pages = {2604-2617},
          doi = {10.1093/mnras/stv327},
archivePrefix = {arXiv},
       eprint = {1502.04767},
 primaryClass = {astro-ph.GA},
       adsurl = {https://ui.adsabs.harvard.edu/abs/2015MNRAS.449.2604D}
}

@ARTICLE{Jackson2022,
       author = {{Jackson}, R.~J. and {Jeffries}, R.~D. and {Wright}, N.~J. and {Randich}, S. and {Sacco}, G. and {Bragaglia}, A. and {Hourihane}, A. and {Tognelli}, E. and {Degl'Innocenti}, S. and {Prada Moroni}, P.~G. and {Gilmore}, G. and {Bensby}, T. and {Pancino}, E. and {Smiljanic}, R. and {Bergemann}, M. and {Carraro}, G. and {Franciosini}, E. and {Gonneau}, A. and {Jofr{\'e}}, P. and {Lewis}, J. and {Magrini}, L. and {Morbidelli}, L. and {Prisinzano}, L. and {Worley}, C. and {Zaggia}, S. and {Tautvai{\v{s}}iene}, G. and {Guti{\'e}rrez Albarr{\'a}n}, M.~L. and {Montes}, D. and {Jim{\'e}nez-Esteban}, F.},
        title = "{The Gaia-ESO Survey: Membership probabilities for stars in 63 open and 7 globular clusters from 3D kinematics}",
      journal = {\mnras},
         year = 2022,
        month = jan,
       volume = {509},
       number = {2},
        pages = {1664-1680},
          doi = {10.1093/mnras/stab3032},
archivePrefix = {arXiv},
       eprint = {2110.10477},
 primaryClass = {astro-ph.SR},
       adsurl = {https://ui.adsabs.harvard.edu/abs/2022MNRAS.509.1664J}
}

@ARTICLE{Kunder2017,
       author = {{Kunder}, Andrea and {Kordopatis}, Georges and {Steinmetz}, Matthias and {Zwitter}, Toma{\v{z}} and {McMillan}, Paul J. and {Casagrande}, Luca and {Enke}, Harry and {Wojno}, Jennifer and {Valentini}, Marica and {Chiappini}, Cristina and {Matijevi{\v{c}}}, Gal and {Siviero}, Alessandro and {de Laverny}, Patrick and {Recio-Blanco}, Alejandra and {Bijaoui}, Albert and {Wyse}, Rosemary F.~G. and {Binney}, James and {Grebel}, E.~K. and {Helmi}, Amina and {Jofre}, Paula and {Antoja}, Teresa and {Gilmore}, Gerard and {Siebert}, Arnaud and {Famaey}, Benoit and {Bienaym{\'e}}, Olivier and {Gibson}, Brad K. and {Freeman}, Kenneth C. and {Navarro}, Julio F. and {Munari}, Ulisse and {Seabroke}, George and {Anguiano}, Borja and {{\v{Z}}erjal}, Maru{\v{s}}a and {Minchev}, Ivan and {Reid}, Warren and {Bland-Hawthorn}, Joss and {Kos}, Janez and {Sharma}, Sanjib and {Watson}, Fred and {Parker}, Quentin A. and {Scholz}, Ralf-Dieter and {Burton}, Donna and {Cass}, Paul and {Hartley}, Malcolm and {Fiegert}, Kristin and {Stupar}, Milorad and {Ritter}, Andreas and {Hawkins}, Keith and {Gerhard}, Ortwin and {Chaplin}, W.~J. and {Davies}, G.~R. and {Elsworth}, Y.~P. and {Lund}, M.~N. and {Miglio}, A. and {Mosser}, B.},
        title = "{The Radial Velocity Experiment (RAVE): Fifth Data Release}",
      journal = {\aj},
         year = 2017,
        month = feb,
       volume = {153},
       number = {2},
          eid = {75},
        pages = {75},
          doi = {10.3847/1538-3881/153/2/75},
archivePrefix = {arXiv},
       eprint = {1609.03210},
 primaryClass = {astro-ph.SR},
       adsurl = {https://ui.adsabs.harvard.edu/abs/2017AJ....153...75K}
}

@ARTICLE{Steinmetz2020a,
       author = {{Steinmetz}, Matthias and {Matijevi{\v{c}}}, Gal and {Enke}, Harry and {Zwitter}, Toma{\v{z}} and {Guiglion}, Guillaume and {McMillan}, Paul J. and {Kordopatis}, Georges and {Valentini}, Marica and {Chiappini}, Cristina and {Casagrande}, Luca and {Wojno}, Jennifer and {Anguiano}, Borja and {Bienaym{\'e}}, Olivier and {Bijaoui}, Albert and {Binney}, James and {Burton}, Donna and {Cass}, Paul and {de Laverny}, Patrick and {Fiegert}, Kristin and {Freeman}, Kenneth and {Fulbright}, Jon P. and {Gibson}, Brad K. and {Gilmore}, Gerard and {Grebel}, Eva K. and {Helmi}, Amina and {Kunder}, Andrea and {Munari}, Ulisse and {Navarro}, Julio F. and {Parker}, Quentin and {Ruchti}, Gregory R. and {Recio-Blanco}, Alejandra and {Reid}, Warren and {Seabroke}, George M. and {Siviero}, Alessandro and {Siebert}, Arnaud and {Stupar}, Milorad and {Watson}, Fred and {Williams}, Mary E.~K. and {Wyse}, Rosemary F.~G. and {Anders}, Friedrich and {Antoja}, Teresa and {Birko}, Danijela and {Bland-Hawthorn}, Joss and {Bossini}, Diego and {Garc{\'\i}a}, Rafael A. and {Carrillo}, Ismael and {Chaplin}, William J. and {Elsworth}, Yvonne and {Famaey}, Benoit and {Gerhard}, Ortwin and {Jofre}, Paula and {Just}, Andreas and {Mathur}, Savita and {Miglio}, Andrea and {Minchev}, Ivan and {Monari}, Giacomo and {Mosser}, Benoit and {Ritter}, Andreas and {Rodrigues}, Thaise S. and {Scholz}, Ralf-Dieter and {Sharma}, Sanjib and {Sysoliatina}, Kseniia and {RAVE Collaboration}},
        title = "{The Sixth Data Release of the Radial Velocity Experiment (RAVE). I. Survey Description, Spectra, and Radial Velocities}",
      journal = {\aj},
         year = {2020a},
        month = aug,
       volume = {160},
       number = {2},
          eid = {82},
        pages = {82},
          doi = {10.3847/1538-3881/ab9ab9},
archivePrefix = {arXiv},
       eprint = {2002.04377},
 primaryClass = {astro-ph.SR},
       adsurl = {https://ui.adsabs.harvard.edu/abs/2020AJ....160...82S}
}

@ARTICLE{Torres2006,
       author = {{Torres}, C.~A.~O. and {Quast}, G.~R. and {da Silva}, L. and {de La Reza}, R. and {Melo}, C.~H.~F. and {Sterzik}, M.},
        title = "{Search for associations containing young stars (SACY). I. Sample and searching method}",
      journal = {\aap},
         year = 2006,
        month = dec,
       volume = {460},
       number = {3},
        pages = {695-708},
          doi = {10.1051/0004-6361:20065602},
archivePrefix = {arXiv},
       eprint = {astro-ph/0609258},
 primaryClass = {astro-ph},
       adsurl = {https://ui.adsabs.harvard.edu/abs/2006A&A...460..695T}
}

@ARTICLE{Jilinski2006,
       author = {{Jilinski}, E. and {Daflon}, S. and {Cunha}, K. and {de La Reza}, R.},
        title = "{Radial velocity measurements of B stars in the Scorpius-Centaurus association}",
      journal = {\aap},
         year = 2006,
        month = mar,
       volume = {448},
       number = {3},
        pages = {1001-1006},
          doi = {10.1051/0004-6361:20041614},
archivePrefix = {arXiv},
       eprint = {astro-ph/0601643},
 primaryClass = {astro-ph},
       adsurl = {https://ui.adsabs.harvard.edu/abs/2006A&A...448.1001J}
}

@ARTICLE{Gontcharov2006,
       author = {{Gontcharov}, G.~A.},
        title = "{Pulkovo Compilation of Radial Velocities for 35 495 Hipparcos stars in a common system}",
      journal = {Astronomy Letters},
         year = 2006,
        month = nov,
       volume = {32},
       number = {11},
        pages = {759-771},
          doi = {10.1134/S1063773706110065},
archivePrefix = {arXiv},
       eprint = {1606.08053},
 primaryClass = {astro-ph.SR},
       adsurl = {https://ui.adsabs.harvard.edu/abs/2006AstL...32..759G}
}

@ARTICLE{Guenther2007,
       author = {{Guenther}, E.~W. and {Esposito}, M. and {Mundt}, R. and {Covino}, E. and {Alcal{\'a}}, J.~M. and {Cusano}, F. and {Stecklum}, B.},
        title = "{Pre-main sequence spectroscopic binaries suitable for VLTI observations}",
      journal = {\aap},
         year = 2007,
        month = jun,
       volume = {467},
       number = {3},
        pages = {1147-1155},
          doi = {10.1051/0004-6361:20065686},
archivePrefix = {arXiv},
       eprint = {astro-ph/0702268},
 primaryClass = {astro-ph},
       adsurl = {https://ui.adsabs.harvard.edu/abs/2007A&A...467.1147G}
}

@ARTICLE{Gilmore2012,
       author = {{Gilmore}, G. and {Randich}, S. and {Asplund}, M. and {Binney}, J. and {Bonifacio}, P. and {Drew}, J. and {Feltzing}, S. and {Ferguson}, A. and {Jeffries}, R. and {Micela}, G. and {Negueruela}, I. and {Prusti}, T. and {Rix}, H. -W. and {Vallenari}, A. and {Alfaro}, E. and {Allende-Prieto}, C. and {Babusiaux}, C. and {Bensby}, T. and {Blomme}, R. and {Bragaglia}, A. and {Flaccomio}, E. and {Fran{\c{c}}ois}, P. and {Irwin}, M. and {Koposov}, S. and {Korn}, A. and {Lanzafame}, A. and {Pancino}, E. and {Paunzen}, E. and {Recio-Blanco}, A. and {Sacco},  G. and {et al.} and {Gaia-ESO Survey Team}},
        title = "{The Gaia-ESO Public Spectroscopic Survey}",
      journal = {The Messenger},
         year = 2012,
        month = mar,
       volume = {147},
        pages = {25-31},
       adsurl = {https://ui.adsabs.harvard.edu/abs/2012Msngr.147...25G}
}

@ARTICLE{Lindegren2021,
       author = {{Lindegren}, L. and {Klioner}, S.~A. and {Hern{\'a}ndez}, J. and {Bombrun}, A. and {Ramos-Lerate}, M. and {Steidelm{\"u}ller}, H. and {Bastian}, U. and {Biermann}, M. and {de Torres}, A. and {Gerlach}, E. and {Geyer}, R. and {Hilger}, T. and {Hobbs}, D. and {Lammers}, U. and {McMillan}, P.~J. and {Stephenson}, C.~A. and {Casta{\~n}eda}, J. and {Davidson}, M. and {Fabricius}, C. and {Gracia-Abril}, G. and {Portell}, J. and {Rowell}, N. and {Teyssier}, D. and {Torra}, F. and {Bartolom{\'e}}, S. and {Clotet}, M. and {Garralda}, N. and {Gonz{\'a}lez-Vidal}, J.~J. and {Torra}, J. and {Abbas}, U. and {Altmann}, M. and {Anglada Varela}, E. and {Balaguer-N{\'u}{\~n}ez}, L. and {Balog}, Z. and {Barache}, C. and {Becciani}, U. and {Bernet}, M. and {Bertone}, S. and {Bianchi}, L. and {Bouquillon}, S. and {Brown}, A.~G.~A. and {Bucciarelli}, B. and {Busonero}, D. and {Butkevich}, A.~G. and {Buzzi}, R. and {Cancelliere}, R. and {Carlucci}, T. and {Charlot}, P. and {Cioni}, M. -R.~L. and {Crosta}, M. and {Crowley}, C. and {del Peloso}, E.~F. and {del Pozo}, E. and {Drimmel}, R. and {Esquej}, P. and {Fienga}, A. and {Fraile}, E. and {Gai}, M. and {Garcia-Reinaldos}, M. and {Guerra}, R. and {Hambly}, N.~C. and {Hauser}, M. and {Jan{\ss}en}, K. and {Jordan}, S. and {Kostrzewa-Rutkowska}, Z. and {Lattanzi}, M.~G. and {Liao}, S. and {Licata}, E. and {Lister}, T.~A. and {L{\"o}ffler}, W. and {Marchant}, J.~M. and {Masip}, A. and {Mignard}, F. and {Mints}, A. and {Molina}, D. and {Mora}, A. and {Morbidelli}, R. and {Murphy}, C.~P. and {Pagani}, C. and {Panuzzo}, P. and {Pe{\~n}alosa Esteller}, X. and {Poggio}, E. and {Re Fiorentin}, P. and {Riva}, A. and {Sagrist{\`a} Sell{\'e}s}, A. and {Sanchez Gimenez}, V. and {Sarasso}, M. and {Sciacca}, E. and {Siddiqui}, H.~I. and {Smart}, R.~L. and {Souami}, D. and {Spagna}, A. and {Steele}, I.~A. and {Taris}, F. and {Utrilla}, E. and {van Reeven}, W. and {Vecchiato}, A.},
        title = "{Gaia Early Data Release 3. The astrometric solution}",
      journal = {\aap},
         year = 2021,
        month = may,
       volume = {649},
          eid = {A2},
        pages = {A2},
          doi = {10.1051/0004-6361/202039709},
archivePrefix = {arXiv},
       eprint = {2012.03380},
 primaryClass = {astro-ph.IM},
       adsurl = {https://ui.adsabs.harvard.edu/abs/2021A&A...649A...2L}
}

@ARTICLE{Brown2016,
       author = {{Gaia Collaboration} and {Brown}, A.~G.~A. and {Vallenari}, A. and {Prusti}, T. and {de Bruijne}, J.~H.~J. and {Mignard}, F. and {Drimmel}, R. and {Babusiaux}, C. and {Bailer-Jones}, C.~A.~L. and {Bastian}, U. and {Biermann}, M. and {Evans}, D.~W. and {Eyer}, L. and {Jansen}, F. and {Jordi}, C. and {Katz}, D. and {Klioner}, S.~A. and {Lammers}, U. and {Lindegren}, L. and {Luri}, X. and {O'Mullane}, W. and {Panem}, C. and {Pourbaix}, D. and {Randich}, S. and {Sartoretti}, P. and {Siddiqui}, H.~I. and {Soubiran}, C. and {Valette}, V. and {van Leeuwen}, F. and {Walton}, N.~A. and {Aerts}, C. and {Arenou}, F. and {Cropper}, M. and {H{\o}g}, E. and {Lattanzi}, M.~G. and {Grebel}, E.~K. and {Holland}, A.~D. and {Huc}, C. and {Passot}, X. and {Perryman}, M. and {Bramante}, L. and {Cacciari}, C. and {Casta{\~n}eda}, J. and {Chaoul}, L. and {Cheek}, N. and {De Angeli}, F. and {Fabricius}, C. and {Guerra}, R. and {Hern{\'a}ndez}, J. and {Jean-Antoine-Piccolo}, A. and {Masana}, E. and {Messineo}, R. and {Mowlavi}, N. and {Nienartowicz}, K. and {Ord{\'o}{\~n}ez-Blanco}, D. and {Panuzzo}, P. and {Portell}, J. and {Richards}, P.~J. and {Riello}, M. and {Seabroke}, G.~M. and {Tanga}, P. and {Th{\'e}venin}, F. and {Torra}, J. and {Els}, S.~G. and {Gracia-Abril}, G. and {Comoretto}, G. and {Garcia-Reinaldos}, M. and {Lock}, T. and {Mercier}, E. and {Altmann}, M. and {Andrae}, R. and {Astraatmadja}, T.~L. and {Bellas-Velidis}, I. and {Benson}, K. and {Berthier}, J. and {Blomme}, R. and {Busso}, G. and {Carry}, B. and {Cellino}, A. and {Clementini}, G. and {Cowell}, S. and {Creevey}, O. and {Cuypers}, J. and {Davidson}, M. and {De Ridder}, J. and {de Torres}, A. and {Delchambre}, L. and {Dell'Oro}, A. and {Ducourant}, C. and {Fr{\'e}mat}, Y. and {Garc{\'\i}a-Torres}, M. and {Gosset}, E. and {Halbwachs}, J. -L. and {Hambly}, N.~C. and {Harrison}, D.~L. and {Hauser}, M. and {Hestroffer}, D. and {Hodgkin}, S.~T. and {Huckle}, H.~E. and {Hutton}, A. and {Jasniewicz}, G. and {Jordan}, S. and {Kontizas}, M. and {Korn}, A.~J. and {Lanzafame}, A.~C. and {Manteiga}, M. and {Moitinho}, A. and {Muinonen}, K. and {Osinde}, J. and {Pancino}, E. and {Pauwels}, T. and {Petit}, J. -M. and {Recio-Blanco}, A. and {Robin}, A.~C. and {Sarro}, L.~M. and {Siopis}, C. and {Smith}, M. and {Smith}, K.~W. and {Sozzetti}, A. and {Thuillot}, W. and {van Reeven}, W. and {Viala}, Y. and {Abbas}, U. and {Abreu Aramburu}, A. and {Accart}, S. and {Aguado}, J.~J. and {Allan}, P.~M. and {Allasia}, W. and {Altavilla}, G. and {{\'A}lvarez}, M.~A. and {Alves}, J. and {Anderson}, R.~I. and {Andrei}, A.~H. and {Anglada Varela}, E. and {Antiche}, E. and {Antoja}, T. and {Ant{\'o}n}, S. and {Arcay}, B. and {Bach}, N. and {Baker}, S.~G. and {Balaguer-N{\'u}{\~n}ez}, L. and {Barache}, C. and {Barata}, C. and {Barbier}, A. and {Barblan}, F. and {Barrado y Navascu{\'e}s}, D. and {Barros}, M. and {Barstow}, M.~A. and {Becciani}, U. and {Bellazzini}, M. and {Bello Garc{\'\i}a}, A. and {Belokurov}, V. and {Bendjoya}, P. and {Berihuete}, A. and {Bianchi}, L. and {Bienaym{\'e}}, O. and {Billebaud}, F. and {Blagorodnova}, N. and {Blanco-Cuaresma}, S. and {Boch}, T. and {Bombrun}, A. and {Borrachero}, R. and {Bouquillon}, S. and {Bourda}, G. and {Bouy}, H. and {Bragaglia}, A. and {Breddels}, M.~A. and {Brouillet}, N. and {Br{\"u}semeister}, T. and {Bucciarelli}, B. and {Burgess}, P. and {Burgon}, R. and {Burlacu}, A. and {Busonero}, D. and {Buzzi}, R. and {Caffau}, E. and {Cambras}, J. and {Campbell}, H. and {Cancelliere}, R. and {Cantat-Gaudin}, T. and {Carlucci}, T. and {Carrasco}, J.~M. and {Castellani}, M. and {Charlot}, P. and {Charnas}, J. and {Chiavassa}, A. and {Clotet}, M. and {Cocozza}, G. and {Collins}, R.~S. and {Costigan}, G. and {Crifo}, F. and {Cross}, N.~J.~G. and {Crosta}, M. and {Crowley}, C. and {Dafonte}, C. and {Damerdji}, Y. and {Dapergolas}, A. and {David}, P. and {David}, M. and {De Cat}, P. and {de Felice}, F. and {de Laverny}, P. and {De Luise}, F. and {De March}, R. and {de Martino}, D. and {de Souza}, R. and {Debosscher}, J. and {del Pozo}, E. and {Delbo}, M. and {Delgado}, A. and {Delgado}, H.~E. and {Di Matteo}, P. and {Diakite}, S. and {Distefano}, E. and {Dolding}, C. and {Dos Anjos}, S. and {Drazinos}, P. and {Duran}, J. and {Dzigan}, Y. and {Edvardsson}, B. and {Enke}, H. and {Evans}, N.~W. and {Eynard Bontemps}, G. and {Fabre}, C. and {Fabrizio}, M. and {Faigler}, S. and {Falc{\~a}o}, A.~J. and {Farr{\`a}s Casas}, M. and {Federici}, L. and {Fedorets}, G. and {Fern{\'a}ndez-Hern{\'a}ndez}, J. and {Fernique}, P. and {Fienga}, A. and {Figueras}, F. and {Filippi}, F. and {Findeisen}, K. and {Fonti}, A. and {Fouesneau}, M. and {Fraile}, E. and {Fraser}, M. and {Fuchs}, J. and {Gai}, M. and {Galleti}, S. and {Galluccio}, L. and {Garabato}, D. and {Garc{\'\i}a-Sedano}, F. and {Garofalo}, A. and {Garralda}, N. and {Gavras}, P. and {Gerssen}, J. and {Geyer}, R. and {Gilmore}, G. and {Girona}, S. and {Giuffrida}, G. and {Gomes}, M. and {Gonz{\'a}lez-Marcos}, A. and {Gonz{\'a}lez-N{\'u}{\~n}ez}, J. and {Gonz{\'a}lez-Vidal}, J.~J. and {Granvik}, M. and {Guerrier}, A. and {Guillout}, P. and {Guiraud}, J. and {G{\'u}rpide}, A. and {Guti{\'e}rrez-S{\'a}nchez}, R. and {Guy}, L.~P. and {Haigron}, R. and {Hatzidimitriou}, D. and {Haywood}, M. and {Heiter}, U. and {Helmi}, A. and {Hobbs}, D. and {Hofmann}, W. and {Holl}, B. and {Holland}, G. and {Hunt}, J.~A.~S. and {Hypki}, A. and {Icardi}, V. and {Irwin}, M. and {Jevardat de Fombelle}, G. and {Jofr{\'e}}, P. and {Jonker}, P.~G. and {Jorissen}, A. and {Julbe}, F. and {Karampelas}, A. and {Kochoska}, A. and {Kohley}, R. and {Kolenberg}, K. and {Kontizas}, E. and {Koposov}, S.~E. and {Kordopatis}, G. and {Koubsky}, P. and {Krone-Martins}, A. and {Kudryashova}, M. and {Kull}, I. and {Bachchan}, R.~K. and {Lacoste-Seris}, F. and {Lanza}, A.~F. and {Lavigne}, J. -B. and {Le Poncin-Lafitte}, C. and {Lebreton}, Y. and {Lebzelter}, T. and {Leccia}, S. and {Leclerc}, N. and {Lecoeur-Taibi}, I. and {Lemaitre}, V. and {Lenhardt}, H. and {Leroux}, F. and {Liao}, S. and {Licata}, E. and {Lindstr{\o}m}, H.~E.~P. and {Lister}, T.~A. and {Livanou}, E. and {Lobel}, A. and {L{\"o}ffler}, W. and {L{\'o}pez}, M. and {Lorenz}, D. and {MacDonald}, I. and {Magalh{\~a}es Fernandes}, T. and {Managau}, S. and {Mann}, R.~G. and {Mantelet}, G. and {Marchal}, O. and {Marchant}, J.~M. and {Marconi}, M. and {Marinoni}, S. and {Marrese}, P.~M. and {Marschalk{\'o}}, G. and {Marshall}, D.~J. and {Mart{\'\i}n-Fleitas}, J.~M. and {Martino}, M. and {Mary}, N. and {Matijevi{\v{c}}}, G. and {Mazeh}, T. and {McMillan}, P.~J. and {Messina}, S. and {Michalik}, D. and {Millar}, N.~R. and {Miranda}, B.~M.~H. and {Molina}, D. and {Molinaro}, R. and {Molinaro}, M. and {Moln{\'a}r}, L. and {Moniez}, M. and {Montegriffo}, P. and {Mor}, R. and {Mora}, A. and {Morbidelli}, R. and {Morel}, T. and {Morgenthaler}, S. and {Morris}, D. and {Mulone}, A.~F. and {Muraveva}, T. and {Musella}, I. and {Narbonne}, J. and {Nelemans}, G. and {Nicastro}, L. and {Noval}, L. and {Ord{\'e}novic}, C. and {Ordieres-Mer{\'e}}, J. and {Osborne}, P. and {Pagani}, C. and {Pagano}, I. and {Pailler}, F. and {Palacin}, H. and {Palaversa}, L. and {Parsons}, P. and {Pecoraro}, M. and {Pedrosa}, R. and {Pentik{\"a}inen}, H. and {Pichon}, B. and {Piersimoni}, A.~M. and {Pineau}, F. -X. and {Plachy}, E. and {Plum}, G. and {Poujoulet}, E. and {Pr{\v{s}}a}, A. and {Pulone}, L. and {Ragaini}, S. and {Rago}, S. and {Rambaux}, N. and {Ramos-Lerate}, M. and {Ranalli}, P. and {Rauw}, G. and {Read}, A. and {Regibo}, S. and {Reyl{\'e}}, C. and {Ribeiro}, R.~A. and {Rimoldini}, L. and {Ripepi}, V. and {Riva}, A. and {Rixon}, G. and {Roelens}, M. and {Romero-G{\'o}mez}, M. and {Rowell}, N. and {Royer}, F. and {Ruiz-Dern}, L. and {Sadowski}, G. and {Sagrist{\`a} Sell{\'e}s}, T. and {Sahlmann}, J. and {Salgado}, J. and {Salguero}, E. and {Sarasso}, M. and {Savietto}, H. and {Schultheis}, M. and {Sciacca}, E. and {Segol}, M. and {Segovia}, J.~C. and {Segransan}, D. and {Shih}, I. -C. and {Smareglia}, R. and {Smart}, R.~L. and {Solano}, E. and {Solitro}, F. and {Sordo}, R. and {Soria Nieto}, S. and {Souchay}, J. and {Spagna}, A. and {Spoto}, F. and {Stampa}, U. and {Steele}, I.~A. and {Steidelm{\"u}ller}, H. and {Stephenson}, C.~A. and {Stoev}, H. and {Suess}, F.~F. and {S{\"u}veges}, M. and {Surdej}, J. and {Szabados}, L. and {Szegedi-Elek}, E. and {Tapiador}, D. and {Taris}, F. and {Tauran}, G. and {Taylor}, M.~B. and {Teixeira}, R. and {Terrett}, D. and {Tingley}, B. and {Trager}, S.~C. and {Turon}, C. and {Ulla}, A. and {Utrilla}, E. and {Valentini}, G. and {van Elteren}, A. and {Van Hemelryck}, E. and {van Leeuwen}, M. and {Varadi}, M. and {Vecchiato}, A. and {Veljanoski}, J. and {Via}, T. and {Vicente}, D. and {Vogt}, S. and {Voss}, H. and {Votruba}, V. and {Voutsinas}, S. and {Walmsley}, G. and {Weiler}, M. and {Weingrill}, K. and {Wevers}, T. and {Wyrzykowski}, {\L}. and {Yoldas}, A. and {{\v{Z}}erjal}, M. and {Zucker}, S. and {Zurbach}, C. and {Zwitter}, T. and {Alecu}, A. and {Allen}, M. and {Allende Prieto}, C. and {Amorim}, A. and {Anglada-Escud{\'e}}, G. and {Arsenijevic}, V. and {Azaz}, S. and {Balm}, P. and {Beck}, M. and {Bernstein}, H. -H. and {Bigot}, L. and {Bijaoui}, A. and {Blasco}, C. and {Bonfigli}, M. and {Bono}, G. and {Boudreault}, S. and {Bressan}, A. and {Brown}, S. and {Brunet}, P. -M. and {Bunclark}, P. and {Buonanno}, R. and {Butkevich}, A.~G. and {Carret}, C. and {Carrion}, C. and {Chemin}, L. and {Ch{\'e}reau}, F. and {Corcione}, L. and {Darmigny}, E. and {de Boer}, K.~S. and {de Teodoro}, P. and {de Zeeuw}, P.~T. and {Delle Luche}, C. and {Domingues}, C.~D. and {Dubath}, P. and {Fodor}, F. and {Fr{\'e}zouls}, B. and {Fries}, A. and {Fustes}, D. and {Fyfe}, D. and {Gallardo}, E. and {Gallegos}, J. and {Gardiol}, D. and {Gebran}, M. and {Gomboc}, A. and {G{\'o}mez}, A. and {Grux}, E. and {Gueguen}, A. and {Heyrovsky}, A. and {Hoar}, J. and {Iannicola}, G. and {Isasi Parache}, Y. and {Janotto}, A. -M. and {Joliet}, E. and {Jonckheere}, A. and {Keil}, R. and {Kim}, D. -W. and {Klagyivik}, P. and {Klar}, J. and {Knude}, J. and {Kochukhov}, O. and {Kolka}, I. and {Kos}, J. and {Kutka}, A. and {Lainey}, V. and {LeBouquin}, D. and {Liu}, C. and {Loreggia}, D. and {Makarov}, V.~V. and {Marseille}, M.~G. and {Martayan}, C. and {Martinez-Rubi}, O. and {Massart}, B. and {Meynadier}, F. and {Mignot}, S. and {Munari}, U. and {Nguyen}, A. -T. and {Nordlander}, T. and {Ocvirk}, P. and {O'Flaherty}, K.~S. and {Olias Sanz}, A. and {Ortiz}, P. and {Osorio}, J. and {Oszkiewicz}, D. and {Ouzounis}, A. and {Palmer}, M. and {Park}, P. and {Pasquato}, E. and {Peltzer}, C. and {Peralta}, J. and {P{\'e}turaud}, F. and {Pieniluoma}, T. and {Pigozzi}, E. and {Poels}, J. and {Prat}, G. and {Prod'homme}, T. and {Raison}, F. and {Rebordao}, J.~M. and {Risquez}, D. and {Rocca-Volmerange}, B. and {Rosen}, S. and {Ruiz-Fuertes}, M.~I. and {Russo}, F. and {Sembay}, S. and {Serraller Vizcaino}, I. and {Short}, A. and {Siebert}, A. and {Silva}, H. and {Sinachopoulos}, D. and {Slezak}, E. and {Soffel}, M. and {Sosnowska}, D. and {Strai{\v{z}}ys}, V. and {ter Linden}, M. and {Terrell}, D. and {Theil}, S. and {Tiede}, C. and {Troisi}, L. and {Tsalmantza}, P. and {Tur}, D. and {Vaccari}, M. and {Vachier}, F. and {Valles}, P. and {Van Hamme}, W. and {Veltz}, L. and {Virtanen}, J. and {Wallut}, J. -M. and {Wichmann}, R. and {Wilkinson}, M.~I. and {Ziaeepour}, H. and {Zschocke}, S.},
        title = "{Gaia Data Release 1. Summary of the astrometric, photometric, and survey properties}",
      journal = {\aap},
         year = 2016,
        month = nov,
       volume = {595},
          eid = {A2},
        pages = {A2},
          doi = {10.1051/0004-6361/201629512},
archivePrefix = {arXiv},
       eprint = {1609.04172},
 primaryClass = {astro-ph.IM},
       adsurl = {https://ui.adsabs.harvard.edu/abs/2016A&A...595A...2G}
}

@ARTICLE{Squicciarini2021,
       author = {{Squicciarini}, Vito and {Gratton}, Raffaele and {Bonavita}, Mariangela and {Mesa}, Dino},
        title = "{Unveiling the star formation history of the Upper Scorpius association through its kinematics}",
      journal = {\mnras},
         year = 2021,
        month = oct,
       volume = {507},
       number = {1},
        pages = {1381-1400},
          doi = {10.1093/mnras/stab2079},
archivePrefix = {arXiv},
       eprint = {2107.08057},
 primaryClass = {astro-ph.GA},
       adsurl = {https://ui.adsabs.harvard.edu/abs/2021MNRAS.507.1381S}
}

@ARTICLE{Chen2020,
       author = {{Chen}, Boquan and {D'Onghia}, Elena and {Alves}, Jo{\~a}o and {Adamo}, Angela},
        title = "{Discovery of new stellar groups in the Orion complex. Towards a robust unsupervised approach}",
      journal = {\aap},
         year = 2020,
        month = nov,
       volume = {643},
          eid = {A114},
        pages = {A114},
          doi = {10.1051/0004-6361/201935955},
archivePrefix = {arXiv},
       eprint = {1905.11429},
 primaryClass = {astro-ph.GA},
       adsurl = {https://ui.adsabs.harvard.edu/abs/2020A&A...643A.114C}
}

@ARTICLE{Katz2023,
       author = {{Katz}, D. and {Sartoretti}, P. and {Guerrier}, A. and {Panuzzo}, P. and {Seabroke}, G.~M. and {Th{\'e}venin}, F. and {Cropper}, M. and {Benson}, K. and {Blomme}, R. and {Haigron}, R. and {Marchal}, O. and {Smith}, M. and {Baker}, S. and {Chemin}, L. and {Damerdji}, Y. and {David}, M. and {Dolding}, C. and {Fr{\'e}mat}, Y. and {Gosset}, E. and {Jan{\ss}en}, K. and {Jasniewicz}, G. and {Lobel}, A. and {Plum}, G. and {Samaras}, N. and {Snaith}, O. and {Soubiran}, C. and {Vanel}, O. and {Zwitter}, T. and {Antoja}, T. and {Arenou}, F. and {Babusiaux}, C. and {Brouillet}, N. and {Caffau}, E. and {Di Matteo}, P. and {Fabre}, C. and {Fabricius}, C. and {Fragkoudi}, F. and {Haywood}, M. and {Huckle}, H.~E. and {Hottier}, C. and {Lasne}, Y. and {Leclerc}, N. and {Mastrobuono-Battisti}, A. and {Royer}, F. and {Teyssier}, D. and {Zorec}, J. and {Crifo}, F. and {Jean-Antoine Piccolo}, A. and {Turon}, C. and {Viala}, Y.},
        title = "{Gaia Data Release 3. Properties and validation of the radial velocities}",
      journal = {\aap},
         year = 2023,
        month = jun,
       volume = {674},
          eid = {A5},
        pages = {A5},
          doi = {10.1051/0004-6361/202244220},
archivePrefix = {arXiv},
       eprint = {2206.05902},
 primaryClass = {astro-ph.GA},
       adsurl = {https://ui.adsabs.harvard.edu/abs/2023A&A...674A...5K}
}

@ARTICLE{GC-Vallenari2023,
       author = {{Gaia Collaboration} and {Vallenari}, A. and {Brown}, A.~G.~A. and {Prusti}, T. and {de Bruijne}, J.~H.~J. and {Arenou}, F. and {Babusiaux}, C. and {Biermann}, M. and {Creevey}, O.~L. and {Ducourant}, C. and {Evans}, D.~W. and {Eyer}, L. and {Guerra}, R. and {Hutton}, A. and {Jordi}, C. and {Klioner}, S.~A. and {Lammers}, U.~L. and {Lindegren}, L. and {Luri}, X. and {Mignard}, F. and {Panem}, C. and {Pourbaix}, D. and {Randich}, S. and {Sartoretti}, P. and {Soubiran}, C. and {Tanga}, P. and {Walton}, N.~A. and {Bailer-Jones}, C.~A.~L. and {Bastian}, U. and {Drimmel}, R. and {Jansen}, F. and {Katz}, D. and {Lattanzi}, M.~G. and {van Leeuwen}, F. and {Bakker}, J. and {Cacciari}, C. and {Casta{\~n}eda}, J. and {De Angeli}, F. and {Fabricius}, C. and {Fouesneau}, M. and {Fr{\'e}mat}, Y. and {Galluccio}, L. and {Guerrier}, A. and {Heiter}, U. and {Masana}, E. and {Messineo}, R. and {Mowlavi}, N. and {Nicolas}, C. and {Nienartowicz}, K. and {Pailler}, F. and {Panuzzo}, P. and {Riclet}, F. and {Roux}, W. and {Seabroke}, G.~M. and {Sordo}, R. and {Th{\'e}venin}, F. and {Gracia-Abril}, G. and {Portell}, J. and {Teyssier}, D. and {Altmann}, M. and {Andrae}, R. and {Audard}, M. and {Bellas-Velidis}, I. and {Benson}, K. and {Berthier}, J. and {Blomme}, R. and {Burgess}, P.~W. and {Busonero}, D. and {Busso}, G. and {C{\'a}novas}, H. and {Carry}, B. and {Cellino}, A. and {Cheek}, N. and {Clementini}, G. and {Damerdji}, Y. and {Davidson}, M. and {de Teodoro}, P. and {Nu{\~n}ez Campos}, M. and {Delchambre}, L. and {Dell'Oro}, A. and {Esquej}, P. and {Fern{\'a}ndez-Hern{\'a}ndez}, J. and {Fraile}, E. and {Garabato}, D. and {Garc{\'\i}a-Lario}, P. and {Gosset}, E. and {Haigron}, R. and {Halbwachs}, J. -L. and {Hambly}, N.~C. and {Harrison}, D.~L. and {Hern{\'a}ndez}, J. and {Hestroffer}, D. and {Hodgkin}, S.~T. and {Holl}, B. and {Jan{\ss}en}, K. and {Jevardat de Fombelle}, G. and {Jordan}, S. and {Krone-Martins}, A. and {Lanzafame}, A.~C. and {L{\"o}ffler}, W. and {Marchal}, O. and {Marrese}, P.~M. and {Moitinho}, A. and {Muinonen}, K. and {Osborne}, P. and {Pancino}, E. and {Pauwels}, T. and {Recio-Blanco}, A. and {Reyl{\'e}}, C. and {Riello}, M. and {Rimoldini}, L. and {Roegiers}, T. and {Rybizki}, J. and {Sarro}, L.~M. and {Siopis}, C. and {Smith}, M. and {Sozzetti}, A. and {Utrilla}, E. and {van Leeuwen}, M. and {Abbas}, U. and {{\'A}brah{\'a}m}, P. and {Abreu Aramburu}, A. and {Aerts}, C. and {Aguado}, J.~J. and {Ajaj}, M. and {Aldea-Montero}, F. and {Altavilla}, G. and {{\'A}lvarez}, M.~A. and {Alves}, J. and {Anders}, F. and {Anderson}, R.~I. and {Anglada Varela}, E. and {Antoja}, T. and {Baines}, D. and {Baker}, S.~G. and {Balaguer-N{\'u}{\~n}ez}, L. and {Balbinot}, E. and {Balog}, Z. and {Barache}, C. and {Barbato}, D. and {Barros}, M. and {Barstow}, M.~A. and {Bartolom{\'e}}, S. and {Bassilana}, J. -L. and {Bauchet}, N. and {Becciani}, U. and {Bellazzini}, M. and {Berihuete}, A. and {Bernet}, M. and {Bertone}, S. and {Bianchi}, L. and {Binnenfeld}, A. and {Blanco-Cuaresma}, S. and {Blazere}, A. and {Boch}, T. and {Bombrun}, A. and {Bossini}, D. and {Bouquillon}, S. and {Bragaglia}, A. and {Bramante}, L. and {Breedt}, E. and {Bressan}, A. and {Brouillet}, N. and {Brugaletta}, E. and {Bucciarelli}, B. and {Burlacu}, A. and {Butkevich}, A.~G. and {Buzzi}, R. and {Caffau}, E. and {Cancelliere}, R. and {Cantat-Gaudin}, T. and {Carballo}, R. and {Carlucci}, T. and {Carnerero}, M.~I. and {Carrasco}, J.~M. and {Casamiquela}, L. and {Castellani}, M. and {Castro-Ginard}, A. and {Chaoul}, L. and {Charlot}, P. and {Chemin}, L. and {Chiaramida}, V. and {Chiavassa}, A. and {Chornay}, N. and {Comoretto}, G. and {Contursi}, G. and {Cooper}, W.~J. and {Cornez}, T. and {Cowell}, S. and {Crifo}, F. and {Cropper}, M. and {Crosta}, M. and {Crowley}, C. and {Dafonte}, C. and {Dapergolas}, A. and {David}, M. and {David}, P. and {de Laverny}, P. and {De Luise}, F. and {De March}, R. and {De Ridder}, J. and {de Souza}, R. and {de Torres}, A. and {del Peloso}, E.~F. and {del Pozo}, E. and {Delbo}, M. and {Delgado}, A. and {Delisle}, J. -B. and {Demouchy}, C. and {Dharmawardena}, T.~E. and {Di Matteo}, P. and {Diakite}, S. and {Diener}, C. and {Distefano}, E. and {Dolding}, C. and {Edvardsson}, B. and {Enke}, H. and {Fabre}, C. and {Fabrizio}, M. and {Faigler}, S. and {Fedorets}, G. and {Fernique}, P. and {Fienga}, A. and {Figueras}, F. and {Fournier}, Y. and {Fouron}, C. and {Fragkoudi}, F. and {Gai}, M. and {Garcia-Gutierrez}, A. and {Garcia-Reinaldos}, M. and {Garc{\'\i}a-Torres}, M. and {Garofalo}, A. and {Gavel}, A. and {Gavras}, P. and {Gerlach}, E. and {Geyer}, R. and {Giacobbe}, P. and {Gilmore}, G. and {Girona}, S. and {Giuffrida}, G. and {Gomel}, R. and {Gomez}, A. and {Gonz{\'a}lez-N{\'u}{\~n}ez}, J. and {Gonz{\'a}lez-Santamar{\'\i}a}, I. and {Gonz{\'a}lez-Vidal}, J.~J. and {Granvik}, M. and {Guillout}, P. and {Guiraud}, J. and {Guti{\'e}rrez-S{\'a}nchez}, R. and {Guy}, L.~P. and {Hatzidimitriou}, D. and {Hauser}, M. and {Haywood}, M. and {Helmer}, A. and {Helmi}, A. and {Sarmiento}, M.~H. and {Hidalgo}, S.~L. and {Hilger}, T. and {H{\l}adczuk}, N. and {Hobbs}, D. and {Holland}, G. and {Huckle}, H.~E. and {Jardine}, K. and {Jasniewicz}, G. and {Jean-Antoine Piccolo}, A. and {Jim{\'e}nez-Arranz}, {\'O}. and {Jorissen}, A. and {Juaristi Campillo}, J. and {Julbe}, F. and {Karbevska}, L. and {Kervella}, P. and {Khanna}, S. and {Kontizas}, M. and {Kordopatis}, G. and {Korn}, A.~J. and {K{\'o}sp{\'a}l}, {\'A}. and {Kostrzewa-Rutkowska}, Z. and {Kruszy{\'n}ska}, K. and {Kun}, M. and {Laizeau}, P. and {Lambert}, S. and {Lanza}, A.~F. and {Lasne}, Y. and {Le Campion}, J. -F. and {Lebreton}, Y. and {Lebzelter}, T. and {Leccia}, S. and {Leclerc}, N. and {Lecoeur-Taibi}, I. and {Liao}, S. and {Licata}, E.~L. and {Lindstr{\o}m}, H.~E.~P. and {Lister}, T.~A. and {Livanou}, E. and {Lobel}, A. and {Lorca}, A. and {Loup}, C. and {Madrero Pardo}, P. and {Magdaleno Romeo}, A. and {Managau}, S. and {Mann}, R.~G. and {Manteiga}, M. and {Marchant}, J.~M. and {Marconi}, M. and {Marcos}, J. and {Marcos Santos}, M.~M.~S. and {Mar{\'\i}n Pina}, D. and {Marinoni}, S. and {Marocco}, F. and {Marshall}, D.~J. and {Martin Polo}, L. and {Mart{\'\i}n-Fleitas}, J.~M. and {Marton}, G. and {Mary}, N. and {Masip}, A. and {Massari}, D. and {Mastrobuono-Battisti}, A. and {Mazeh}, T. and {McMillan}, P.~J. and {Messina}, S. and {Michalik}, D. and {Millar}, N.~R. and {Mints}, A. and {Molina}, D. and {Molinaro}, R. and {Moln{\'a}r}, L. and {Monari}, G. and {Mongui{\'o}}, M. and {Montegriffo}, P. and {Montero}, A. and {Mor}, R. and {Mora}, A. and {Morbidelli}, R. and {Morel}, T. and {Morris}, D. and {Muraveva}, T. and {Murphy}, C.~P. and {Musella}, I. and {Nagy}, Z. and {Noval}, L. and {Oca{\~n}a}, F. and {Ogden}, A. and {Ordenovic}, C. and {Osinde}, J.~O. and {Pagani}, C. and {Pagano}, I. and {Palaversa}, L. and {Palicio}, P.~A. and {Pallas-Quintela}, L. and {Panahi}, A. and {Payne-Wardenaar}, S. and {Pe{\~n}alosa Esteller}, X. and {Penttil{\"a}}, A. and {Pichon}, B. and {Piersimoni}, A.~M. and {Pineau}, F. -X. and {Plachy}, E. and {Plum}, G. and {Poggio}, E. and {Pr{\v{s}}a}, A. and {Pulone}, L. and {Racero}, E. and {Ragaini}, S. and {Rainer}, M. and {Raiteri}, C.~M. and {Rambaux}, N. and {Ramos}, P. and {Ramos-Lerate}, M. and {Re Fiorentin}, P. and {Regibo}, S. and {Richards}, P.~J. and {Rios Diaz}, C. and {Ripepi}, V. and {Riva}, A. and {Rix}, H. -W. and {Rixon}, G. and {Robichon}, N. and {Robin}, A.~C. and {Robin}, C. and {Roelens}, M. and {Rogues}, H.~R.~O. and {Rohrbasser}, L. and {Romero-G{\'o}mez}, M. and {Rowell}, N. and {Royer}, F. and {Ruz Mieres}, D. and {Rybicki}, K.~A. and {Sadowski}, G. and {S{\'a}ez N{\'u}{\~n}ez}, A. and {Sagrist{\`a} Sell{\'e}s}, A. and {Sahlmann}, J. and {Salguero}, E. and {Samaras}, N. and {Sanchez Gimenez}, V. and {Sanna}, N. and {Santove{\~n}a}, R. and {Sarasso}, M. and {Schultheis}, M. and {Sciacca}, E. and {Segol}, M. and {Segovia}, J.~C. and {S{\'e}gransan}, D. and {Semeux}, D. and {Shahaf}, S. and {Siddiqui}, H.~I. and {Siebert}, A. and {Siltala}, L. and {Silvelo}, A. and {Slezak}, E. and {Slezak}, I. and {Smart}, R.~L. and {Snaith}, O.~N. and {Solano}, E. and {Solitro}, F. and {Souami}, D. and {Souchay}, J. and {Spagna}, A. and {Spina}, L. and {Spoto}, F. and {Steele}, I.~A. and {Steidelm{\"u}ller}, H. and {Stephenson}, C.~A. and {S{\"u}veges}, M. and {Surdej}, J. and {Szabados}, L. and {Szegedi-Elek}, E. and {Taris}, F. and {Taylor}, M.~B. and {Teixeira}, R. and {Tolomei}, L. and {Tonello}, N. and {Torra}, F. and {Torra}, J. and {Torralba Elipe}, G. and {Trabucchi}, M. and {Tsounis}, A.~T. and {Turon}, C. and {Ulla}, A. and {Unger}, N. and {Vaillant}, M.~V. and {van Dillen}, E. and {van Reeven}, W. and {Vanel}, O. and {Vecchiato}, A. and {Viala}, Y. and {Vicente}, D. and {Voutsinas}, S. and {Weiler}, M. and {Wevers}, T. and {Wyrzykowski}, {\L}. and {Yoldas}, A. and {Yvard}, P. and {Zhao}, H. and {Zorec}, J. and {Zucker}, S. and {Zwitter}, T.},
        title = "{Gaia Data Release 3. Summary of the content and survey properties}",
      journal = {\aap},
         year = 2023,
        month = jun,
       volume = {674},
          eid = {A1},
        pages = {A1},
          doi = {10.1051/0004-6361/202243940},
archivePrefix = {arXiv},
       eprint = {2208.00211},
 primaryClass = {astro-ph.GA},
       adsurl = {https://ui.adsabs.harvard.edu/abs/2023A&A...674A...1G}
}

@ARTICLE{Kos2019,
       author = {{Kos}, Janez and {Bland-Hawthorn}, Joss and {Asplund}, Martin and {Buder}, Sven and {Lewis}, Geraint F. and {Lin}, Jane and {Martell}, Sarah L. and {Ness}, Melissa K. and {Sharma}, Sanjib and {De Silva}, Gayandhi M. and {Simpson}, Jeffrey D. and {Zucker}, Daniel B. and {Zwitter}, Toma{\v{z}} and {{\v{C}}otar}, Klemen and {Spina}, Lorenzo},
        title = "{Discovery of a 21 Myr old stellar population in the Orion complex{\ensuremath{\star}}}",
      journal = {\aap},
         year = 2019,
        month = nov,
       volume = {631},
          eid = {A166},
        pages = {A166},
          doi = {10.1051/0004-6361/201834710},
archivePrefix = {arXiv},
       eprint = {1811.11762},
 primaryClass = {astro-ph.SR},
       adsurl = {https://ui.adsabs.harvard.edu/abs/2019A&A...631A.166K}
}

@ARTICLE{Kuhn2019,
       author = {{Kuhn}, Michael A. and {Hillenbrand}, Lynne A. and {Sills}, Alison and {Feigelson}, Eric D. and {Getman}, Konstantin V.},
        title = "{Kinematics in Young Star Clusters and Associations with Gaia DR2}",
      journal = {\apj},
         year = 2019,
        month = jan,
       volume = {870},
       number = {1},
          eid = {32},
        pages = {32},
          doi = {10.3847/1538-4357/aaef8c},
archivePrefix = {arXiv},
       eprint = {1807.02115},
 primaryClass = {astro-ph.GA},
       adsurl = {https://ui.adsabs.harvard.edu/abs/2019ApJ...870...32K}
}

\begin{appendix}

\section{Used data and parameters} \label{apx:data}

\subsection{The Sco-Cen cluster sample} \label{apx:cluster-sample}

As in \citetalias{Ratzenboeck2023b}, we consider only 34 out of the total 37 \texttt{SigMA} selected clusters from \citetalias{Ratzenboeck2023a} as probable members of the Sco-Cen association\footnote{Three clusters are likely unrelated: Norma-North, Oph-SE, and Oph-NF \citepalias[see][]{Ratzenboeck2023b}.}. 
We initially included the three removed clusters (Oph-Southeast, Oph-NorthFar, and Norma-North) in our analysis, and we can confirm that these are unrelated, since they show significantly different kinematics when compared to the rest of the 34 Sco-Cen clusters.
The 34 clusters contain a total of 12,972 stellar members. An overview is given in Table~\ref{tab:overview}, where we list the cluster names, ages, and number statistics. 
Moreover, each cluster is assigned to a traditional region (TR), as defined in 2D projection. The TRs include the classical \citet{Blaauw1946} regions Upper-Scorpius (US), Upper Centaurus Lupus (UCL), and Lower Centaurus Crux (LCC), which have often been treated as the three stellar sub-groups of Sco-Cen \citep[e.g.,][]{deZeeuw1999, Mamajek2001, Preibisch2008}. \citetalias{Ratzenboeck2023a} further identified and included clusters toward Pipe, Corona Australis (CrA), and Chamaeleon (Cham), and to the Galactic North-East (NE) of Sco-Cen. The two clusters in the TW Hydrae association (TWA-a, TWA-b) were added to Sco-Cen by \citetalias{MiretRoig2025}, while stellar members were previously reported in \citet{Gagne2018a} or \citet{Luhman2023}. With this, the 36 clusters contain a total of 13,011 stellar members. 

In our analysis, we do not include the Chamaeleon clusters (Cham-I, Cham-II, Centaurus-Far), since these are connected to a background structure, as discussed in \citet{Edenhofer2024b}. The authors identified a ``C'' structure in a 3D dust map \citep{Edenhofer2024a}. The shell-like structure was likely created by a supernova, creating a rim that contains the Chameleon, Musca, and Coalsack clouds, located at the back of the main Sco-Cen complex. The history of this structure is likely connected to Sco-Cen, where remaining gas was possibly influenced by massive stellar feedback originating from the OB association. Nevertheless, we remove the clusters that are associated with the ``C'' structure from our current analysis, since the relative motions show peculiarities compared to the rest of the Sco-Cen clusters, and they are also located slightly detached from the main body of Sco-Cen. 
Moreover, we do not include the cluster $\mu$\,Sco in our analysis, since it has poor RV statistics; only two stellar members pass our RV quality criteria (Appendix~\ref{apx:quality}), which additionally show a large dispersion in RV space. 

We combine the \texttt{SigMA} Sco-Cen sample with the TWA sample (66 stellar members). We find that 27 sources are in common, which are candidate members of $\sigma$\,Cen, a cluster which is connected to TWA in a cluster chain. The overlapping sources have low statistical stability for $\sigma$\,Cen, as given by the \texttt{SigMA} algorithm, and they are scattered suspiciously in front of the cluster along the line-of-sight, which makes it more likely that these sources belong to TWA (see also \citetalias{MiretRoig2025}). Hence, these 27 sources are counted to TWA in this paper. TWA was tentatively split into two clusters in \citetalias{MiretRoig2025}, named \mbox{TWA-a} and \mbox{TWA-b}, containing 44 and 22 stellar members, respectively. 

Eventually, we analyse 32 Sco-Cen clusters in this paper that contain a total of 12,612 stellar members, with 12,546 retrieved from the 30 \texttt{SigMA} Sco-Cen clusters \citepalias{Ratzenboeck2023a} and 66 from the two TWA clusters \citepalias{MiretRoig2025}. For completeness, we also report statistics for the four removed \texttt{SigMA} clusters in the final rows of Table~\ref{tab:overview}.

\subsection{Parameters and parameter transformations} \label{apx:parameters}

We use the astrometric parameters from \textit{Gaia} DR3 and supplementary radial velocity data (see Appendix~\ref{apx:rv_comparison}): 

\begin{itemize}
  \item $\alpha$, $\delta$ (deg): Right ascension and Declination
  \item $l$, $b$ (deg): Galactic longitude and latitude
  \item $\varpi$\,(mas): parallax
  \item $d$\,(pc): distance ($1000/\varpi_\mathrm{corr}$)
  \item $\mu_{\alpha},\,\mu_\delta \,\si{(mas~yr^{-1})}$: proper motions along $\alpha$ and $\delta$ (for simplicity we denote here $\mu_{\alpha} \cos(\delta)$ as $\mu_{\alpha}$)
  \item $\mu_{l},\,\mu_b \,\si{(mas~yr^{-1})}$: proper motions along $l$ and $b$ (for simplicity we denote here $\mu_{l} \cos(b)$ as $\mu_{l}$)
  \item $v_\text{RV}\,\si{(km~s^{-1})}$: Heliocentric radial velocity (line-of-sight motion relative to the Sun, determined from stellar spectra)
\end{itemize}
\noindent We correct the parallaxes ($\varpi_\mathrm{corr}$) for the parallax bias identified in \citet{Lindegren2021}, using the recommended zero-point correction (python package \texttt{zero\_point})\footnote{\url{https://gitlab.com/icc-ub/public/gaiadr3_zeropoint}}. 
In this work, we use the inverse of the corrected parallax as a distance estimate, which is reasonable for targets within about 200\,pc from the Sun and for sources with low uncertainties (see \citetalias{Ratzenboeck2023a}).

The velocities are corrected for the standard solar motion \citep{Schoenrich2010}, hence transformed to velocities relative to the local standard of rest (LSR) with the help of \texttt{astropy.coordinates} \citep{Astropy2022}. 
 They are given as $\mu_{\alpha,\mathit{lsr}}$, $\mu_{\delta,\mathit{lsr}}$, $\mu_{l,\mathit{lsr}}$, $\mu_{b,\mathit{lsr}}$ (mas/yr) and $v_{\text{RV},\mathit{lsr}}\,\si{(km/s)}$. 
The transformation from $\mu_x$ to $\mu_{x,lsr}$ can alternatively be achieved by the transformations outlined in \citet{Poleski2013}.
The tangential velocities ($v_{\alpha}$, $v_{\delta}$, $v_{l}$, $v_{b}$, or $v_{\alpha,\mathit{lsr}}$, $v_{\delta,\mathit{lsr}}$, $v_{l,\mathit{lsr}}$, $v_{b,\mathit{lsr}}$) are derived from proper motions and parallaxes as follows:
\begin{equation}
    \label{eq:vtan}
    \begin{aligned}
        v_x \, (\mathrm{km}\,\mathrm{s}^{-1}) = 4.74047 \cdot \mu_x / \varpi_\mathrm{corr} \\
    \end{aligned}
\end{equation}
The proper motion can be entered heliocentric or LSR-corrected in the above formula, while the latter results in tangential velocities in the LSR frame. 
$XYZ$ are the Heliocentric Galactic Cartesian coordinates in parsec (pc):
\begin{small}
\begin{equation}
    \label{eq:xyz}
    X = d  \cos(l) \cos(b),\, Y = d  \sin(l) \cos(b),\, Z = d \sin(b) 
\end{equation}
\end{small}
$X$ is positive toward the Galactic centre, $Y$ is positive in the direction of Galactic rotation, and $Z$ is positive toward the Galactic North Pole. 
The corresponding Heliocentric Galactic Cartesian velocities in the directions of $XYZ$ are given as $UVW$ (in km\,s$^{-1}$):
\begin{small}
\begin{equation}
    \label{eq:uvw}
    \begin{aligned}
        &U = v_\text{RV}  \cos(l) \cos(b) - v_l \sin(l) - v_b \cos(l) \sin(b) \\
        &V = v_\text{RV}  \sin(l) \cos(b) + v_l \cos(l) - v_b \sin(l) \sin(b) \\
        &W = v_\text{RV}  \sin(b) + v_b \cos(b)
    \end{aligned}
\end{equation}
\end{small}
When substituting the given velocities with $v_{\text{RV},\mathit{lsr}}$, $v_{l,\mathit{lsr}}$,  and $v_{b,\mathit{lsr}}$, we get the Cartesian velocities relative to the local standard of rest (LSR), denoted as $U_\mathrm{LSR}$, $V_\mathrm{LSR}$, and $W_\mathrm{LSR}$ (km\,s$^{-1}$). Alternatively, $UVW_\mathrm{LSR}$ is calculated from $UVW$ by adding the values of the standard Solar motion from \citet{Schoenrich2010}.  
In our analysis, one can use both the Heliocentric or LSR corrected Galactic Cartesian velocities, since it does not make a difference when investigating relative motions within a region.

To get uncertainties for the derived parameters, we use an error sampling approach. We sample from the various parameters by randomly drawing from Gaussians, using the measurements and their uncertainties as the mean and standard deviation. For instance, to get the uncertainties of the tangential velocities, we use Eq.~(\ref{eq:vtan}), and we sample over the parallaxes and proper motions using their uncertainties as the standard deviation of the Gaussian. We then use the median and standard deviation as the derived parameters and their uncertainties.

\subsection{The radial velocity surveys} \label{apx:rv_comparison}

Table~\ref{tab:ref} lists 22 radial velocity surveys or literature studies that are combined in this work to get more complete 6D phase space information per cluster. 
The RV surveys are combined with the \textit{Gaia} DR3 Sco-Cen catalogue (13\,011 sources) by using the source IDs or sky cross-matches. A \textit{Gaia} \texttt{source\_id} match is possible for catalogues that contain information on \textit{Gaia}, 2MASS, or HIP IDs, which can be matched within the \textit{Gaia} archive (see also Appendix~A in \citetalias{Ratzenboeck2023a}). For other catalogues, we use a \SI{1}{\arcsecond} cross-match radius, using the closest match. 

The list in Table~\ref{tab:ref} includes several well-known large-scale surveys and also smaller-scale spectroscopic studies. This creates a heterogeneous RV sample, observed with different instruments at different resolutions, and processed with different pipelines. 
Hence, we are dealing with variable data coverage and possible systematic differences between the RV surveys. Moreover, the measurement uncertainties could have additional systematics and might have been estimated differently, further complicating a comparison at face value. 
One way forward could be to use an already homogenised catalogue, like the Survey of Surveys (SoS) by \citet{Tsantaki2022}, which is a compilation of six spectroscopic surveys (\textit{Gaia}\,DR2, APOGEE\,DR16, GALAH\,DR2, \textit{Gaia}-ESO\,DR3, LAMOST\,DR5, RAVE\,DR6). However, most of the data releases (DR) of the included surveys have already been superseded by more recent releases. Hence, we decided to use the most recent DRs and other original literature data, which were not used by SoS.

Small systematic variations between surveys are not highly critical when averaging the motions per cluster to get each cluster's median or mean motion, while for the calculation of the velocity dispersion, we aim at higher quality. Hence, we apply quality criteria, outlined in Appendix~\ref{apx:quality}, and we compare the individually measured RVs to each other to identify possible systematic shifts between the surveys (for this, we matched the full external catalogues with {\it Gaia} DR3). 
Binaries could also cause discrepancies between measurements from different surveys; however, we would not expect a systematic shift in one direction between two surveys to be caused by binaries.
We use \textit{Gaia} DR3 RVs as zero-point, since this is the largest survey in our list, covering the whole sky; hence, we can compare it to the full available catalogues of the other surveys. 
We only identify significant global shifts in four cases: SDSS, GALAH, RAVE, and PCRV.  
For the rest of the surveys, we either do not find a significant shift or there are not enough sources in common to identify a clear trend. 
The average shifts are given in Table~\ref{tab:ref} as $\Delta \mathrm{RV}$.
We do not correct for these shifts, since we did not test for dependencies (relative to T$_\mathit{eff}$, $\log g$, or metallicity; see, e.g., \citealt{Tsantaki2022}), which would go beyond the scope of this paper. To account for this additional uncertainty, we add the shift quadratically to the measurement uncertainties of these four surveys. 
Concerning the measurement uncertainties in general, we do not consider any systematics between the surveys, and we use the surveys' reported RV uncertainties ($e_\mathrm{RV}$) for simplicity. We correct the RV errors in the four mentioned cases as follows, $e_\mathrm{RV,corr} = \sqrt{e_\mathrm{RV}^2 + \Delta \mathrm{RV}^2}$, and we use these values in the quality criteria below.

In the following, we give some details concerning RVs from the recently released SDSS DR19, which are part of the Milky Way Mapper (MWM) program, while we use observations from the APOGEE instruments. 
We use the MWM/Astra/ASPCAP catalogue\footnote{\texttt{astraAllStarASPCAP-0.6.0.fits} from \url{https://dr19.sdss.org/sas/dr19/spectro/astra/0.6.0/summary/}.
}, which contains APOGEE RVs from \mbox{SDSS-IV} DR17 and \mbox{SDSS-V} DR19 \citep{Kollmeier2026}. The column ``\texttt{release}''  specifies if a source originates from the earlier (\texttt{dr17}) or the more recent (\texttt{sdss5}) data release.   
SDSS also provides a scatter parameter (\texttt{std\_v\_rad}) for sources observed more than once within the MWM survey, which can be used to remove potential multiple stellar systems. 
We recompute the SDSS RV uncertainties as follows: $e_\mathrm{RV,corr} = \sqrt{\mathtt{e\_v\_rad}^2 + \mathtt{std\_v\_rad}^2 + \Delta \mathrm{RV}^2}$. 
Using the corrected RV errors, sources with larger scatter (potential binary candidates) get automatically removed when applying the error cuts as outlined in the next section. Sources without the scatter information, which were observed only once within SDSS, lack this information, and their errors might be underestimated. At this stage, we can not know about any additional scatter for such sources, if only observed by one survey. Some sources are observed by multiple RV surveys that are listed in Table~\ref{tab:ref}, and we outline the scatter estimate for such sources in the next section.

\begin{table*}[!t] 
\begin{small}
\caption{Data references for radial velocities.}
\renewcommand{\arraystretch}{1.4}
\resizebox{1\textwidth}{!}{
\centering
\begin{tabular}{l
                >{\raggedright}p{48mm}
                >{\raggedright}p{25mm}
                >{\raggedright}p{78mm}
                r
                r
                r
                c
                } 

\hline \hline

\multicolumn{1}{c}{Ref.} &
\multicolumn{1}{c}{References} &
\multicolumn{1}{c}{Region} &  
\multicolumn{1}{c}{Survey Name / Notes / Comments} &
\multicolumn{3}{c}{Number of sources} &
\multicolumn{1}{c}{$\Delta\mathrm{RV}$} \\

\cmidrule(lr){5-7}

\multicolumn{1}{c}{} &
\multicolumn{1}{c}{} &
\multicolumn{1}{l}{} &  
\multicolumn{1}{c}{} &
\multicolumn{1}{c}{All} &
\multicolumn{1}{c}{Matched} &
\multicolumn{1}{c}{Used} &
\multicolumn{1}{c}{(km/s)} \\

\cmidrule(lr){1-8}

1 & \citet{GC-Vallenari2023}, \citet{Katz2023} & All-Sky & Gaia DR3 & 33,812,183 & 4978 (3243) & 2091 (750) & 0 \\

2 & \citet{Majewski2017}, \citet{Santana2021}, \citet{Abdurrouf2022}, \citet{Kollmeier2017, Kollmeier2019, Kollmeier2026}, \citet{Meszaros2025} & USco, sub-parts of Sco-Cen & 
SDSS Milky Way Mapper (MWM), SDSSDR19/Astra from the APOGEE Stellar Parameter and Chemical Abundances Pipeline (ASPCAP), containing RVs from SDSS-IV DR17 or SDSS-V DR19 & 1,095,480 & 1215 (310) & 1008 (252) & $-0.2$ \\

3 & \citet{DeSilva2015,Buder2021,Buder2024} &  Parts of Southern Hemisphere & GALAH DR4 (Galactic Archaeology with HERMES) & 906,661 & 2035 (557) & 1528 (362) & $-0.22$ \\

4 & \citet{Gilmore2012}, \citet{Sacco2017}, \citet{Jackson2022} & Targeted clusters& Gaia-ESO iDR6 (GES) & 107,779 & 125 (28) &  111 (26) & 0 \\

5 & \citet{Kunder2017}, \citet{Steinmetz2020a}  & Southern Hemisphere& RAVE DR6 (The Radial Velocity Experiment)  & 451,636 & 152 (11) &  106 (2) & $0.34$ \\ 

6 & \citet{Gontcharov2006} & All-Sky & PCRV (Pulkovo Compilation of Radial Velocities) for 35,495 Hipparcos stars, standard stars compiled from other literature, homogenised RVs & 35,495 & 111 (48) & 55 (19) &  $0.1$ \\

7 & \citet{Torres2006} & Southern Hemisphere & SACY (Search for associations containing young stars), 1511 observed by them, 115 from other literature & 1626 & 239 (26) & 202 (19) & 0 \\

\cmidrule(lr){1-8}

8 & \citet{Jilinski2006} & Sco-Cen & B-stars in Sco-Cen & 119 & 25 (10) & 13 (6) & 0 \\

9 & \citet{Chen2011} & Sco-Cen &Magellan MIKE and Spitzer MIPS Study &  192 & 91 (11) & 67 (7) & 0 \\

10 & \citet{Dahm2012} & USco & HIRES for 50 members, MIKE for 44 members  & 131 & 79 (8) & 64 (8) &  0 \\

11 & \citet{MiretRoig2022b} & USco & Compilation of observed and archival spectra, R$\sim$20,000--115,000 &  157 & 154 (9) &  120 (4) & 0 \\

12 & \citet{Fang2023} & USco & Collected from Keck/HIRES Optical Survey, R$\sim$34,000, typical RV-uncertainty $\sim$\,1.4\,km/s &  115 & 97 (5) & 90 (3) &  0 \\

13 & \citet{James2006} & Cham, Lup, CrA & ESO/FEROS, R$\sim$32,000 & 53 & 22 (0) & 19 (0) & 0 \\

14 & \citet{Guenther2007} & Cham, Lup, CrA, Oph & ESO/FEROS, R$\sim$48,000 &  96 & 39 (5) & 32 (5) &  0 \\

15 & \citet{Wichmann1999} & Lupus & ROSAT-discovered WTTS near Lupus & 71 & 44 (4) &  30 (4) &  0 \\

16 & \citet{Galli2013} & Lupus, Oph & Kinematic study of the Lupus star-forming region &  52 & 42 (4) & 42 (4) &  0 \\

17 & \citet{JoergensGuenther2001} & Cham I &UVES spectra at 6000--10,400\,\si{\angstrom} (R$\sim$40,000)  & 12 &  11 (5) & 11 (5) &  0 \\

18 & \citet{Nguyen2012} & Cham I &MIKE spectra at 4800--9400\,\si{\angstrom} (R$\sim$60,000) &  67 & 46 (4) & 43 (4) & 0 \\

19 & \citet{Biazzo2012} & Cham II &UVES/GIRAFFE &  83 & 29 (21) & 19 (16) &   0 \\

20 & \citet{Murphy2013} & eps Cham &WiFeS multi-epoch spectroscopy, R$\sim$7000 &  45 &  20 (9) & 20 (9) &  0 \\

21 & \citet{Luhman2023} & TWA & Literature compilation from 16 references (we exclude {\it Gaia} DR3 and GALAH DR2 from their compilation) &  38 & 26 (15) & 21 (13) &  0 \\

22 & \citet{MiretRoig2025} & TWA & Archival compilation of spectra from ESO, CFHT, and NOIRLab, named RV$_\mathrm{ispec}$  &  24 & 17 (2) & 13 (2) &  0 \\

\hline
\end{tabular}
} 
\renewcommand{\arraystretch}{1}
\label{tab:ref}
\tablefoot{The first seven references are large spectroscopic survey programs or RV compilations, partially covering the whole sky or a full hemisphere. The rest are targeted observations collected from the mentioned literature, loosely grouped by observed region.  
Col.\,5 lists the total number of sources available in each survey. 
Col.\,6 \& 7 list the number of sources matched with the full Sco-Cen sample (13,011 contained in 36 clusters) and the number of used RVs within the 36 clusters from the given survey; in brackets, we provide the number of sources that only have an RV measurement from the given survey but not in any of the other surveys.
Col.\,8 gives the global RV shift between the given survey and \textit{Gaia} DR3 RVs, if identified. 
}
\end{small}
\end{table*}

\begin{table*}[!ht]
\begin{small}
\caption{Overview of 36 stellar clusters toward Sco-Cen from \citetalias{Ratzenboeck2023a} (\texttt{SigMA} selected clusters) and \citetalias{MiretRoig2025} (TWA clusters), including radial velocity data statistics, sorted by cluster age.}
\centering
\renewcommand{\arraystretch}{1.3}
\resizebox{0.78\textwidth}{!}{

\begin{tabular}{llllcrrrrr}

\hline \hline
\multicolumn{1}{l}{Idx} &
\multicolumn{1}{l}{\texttt{SigMA}} &
\multicolumn{1}{l}{TR} &
\multicolumn{1}{l}{Name} &
\multicolumn{1}{c}{\texttt{stab\_lim}}&
\multicolumn{1}{c}{Age} & 
\multicolumn{4}{c}{Number of sources} \\
[-0.9mm] 

\cmidrule(lr){7-10}

\multicolumn{1}{l}{} &
\multicolumn{1}{l}{} &
\multicolumn{1}{l}{} &
\multicolumn{1}{l}{} &
\multicolumn{1}{l}{} &
\multicolumn{1}{c}{(Myr)} &
\multicolumn{1}{c}{\textit{XYZ}$_\mathrm{all}$} &
\multicolumn{1}{c}{\textit{XYZ}$_\mathrm{used}$ (\%)} &
\multicolumn{1}{r}{\textit{UVW}$_\mathrm{all}$ (\%)} &
\multicolumn{1}{r}{\textit{UVW}$_\mathrm{used}$ (\%)} \\

\cmidrule(lr){1-10}

1 & 17 & UCL & $e$ Lup & 8.2 & 20.9$^{+0.7}_{-0.8}$ & 516 & 468\,(\phantom{0}90.7) & 199\,(38.6) & 72\,(14.0) \\
2 & 12 & UCL & Libra-South & 3.9 & 20.0$^{+2.5}_{-2.2}$ & 71 & 65\,(\phantom{0}91.5) & 26\,(36.6) & 11\,(15.5) \\
3 & 9 & US & US-foreground & 4.6 & 19.1$^{+2.4}_{-1.3}$ & 276 & 254\,(\phantom{0}92.0) & 173\,(62.7) & 113\,(40.9) \\
4 & 15 & UCL & $\phi$ Lup & 6.8 & 16.9$^{+0.9}_{-0.6}$ & 1114 & 1047\,(\phantom{0}94.0) & 514\,(46.1) & 242\,(21.7) \\
5 & 27 & Pipe & Pipe-North & 1.9 & 15.9$^{+1.6}_{-2.1}$ & 42 & 40\,(\phantom{0}95.2) & 21\,(50.0) & 9\,(21.4) \\
6 & 20 & UCL & $\nu$ Cen & 7.8 & 15.7$^{+0.3}_{-0.9}$ & 1737 & 1522\,(\phantom{0}87.6) & 709\,(40.8) & 290\,(16.7) \\
7 & 21 & LCC & $\sigma$ Cen & 9.3 & 15.5$^{+0.6}_{-0.5}$ & 1778 & 1618\,(\phantom{0}91.0) & 881\,(49.6) & 445\,(25.0) \\
8 & 28 & Pipe & $\theta$ Oph & 8.8 & 15.4$^{+0.8}_{-1.9}$ & 98 & 90\,(\phantom{0}91.8) & 40\,(40.8) & 11\,(11.2) \\
9 & 14 & UCL & $\eta$ Lup & 6.4 & 15.3$^{+0.6}_{-0.3}$ & 769 & 722\,(\phantom{0}93.9) & 386\,(50.2) & 147\,(19.1) \\
10 & 10 & UCL & V1062-Sco & 9.9 & 15.0$^{+0.9}_{-1.4}$ & 1029 & 963\,(\phantom{0}93.6) & 339\,(32.9) & 126\,(12.2) \\
11 & 8 & US & Scorpio-Body & 5.7 & 14.7$^{+0.8}_{-0.7}$ & 373 & 344\,(\phantom{0}92.2) & 147\,(39.4) & 51\,(13.7) \\
12 & 31 & CrA & Scorpio-Sting & 1.6 & 14.5$^{+0.6}_{-0.6}$ & 132 & 129\,(\phantom{0}97.7) & 42\,(31.8) & 14\,(10.6) \\
13 & 19 & UCL & $\rho$ Lup & 8.6 & 14.4$^{+0.4}_{-0.9}$ & 246 & 235\,(\phantom{0}95.5) & 101\,(41.1) & 49\,(19.9) \\
14 & 7 & US & $\rho$ Sco & 7.3 & 13.7$^{+1.3}_{-0.6}$ & 240 & 227\,(\phantom{0}94.6) & 132\,(55.0) & 84\,(35.0) \\
15 & 18 & UCL & UPK606 & 8.7 & 13.4$^{+1.4}_{-0.7}$ & 131 & 115\,(\phantom{0}87.8) & 60\,(45.8) & 27\,(20.6) \\
16 & 6 & US & Antares & 9.7 & 12.7$^{+0.4}_{-1.7}$ & 502 & 468\,(\phantom{0}93.2) & 316\,(62.9) & 212\,(42.2) \\
17 & 30 & CrA & CrA-North & 9.9 & 11.6$^{+0.5}_{-0.8}$ & 351 & 333\,(\phantom{0}94.9) & 143\,(40.7) & 48\,(13.7) \\
18 & 22 & LCC & Acrux & 9.9 & 11.2$^{+1.0}_{-1.0}$ & 394 & 385\,(\phantom{0}97.7) & 222\,(56.3) & 115\,(29.2) \\
19 & 23 & LCC & Musca-foreground & 9.9 & 10.2$^{+1.0}_{-0.7}$ & 95 & 88\,(\phantom{0}92.6) & 46\,(48.4) & 22\,(23.2) \\
20 & 5 & US & $\sigma$ Sco* & 9.7 & 10.0$^{+1.0}_{-0.5}$ & 544 & 508\,(\phantom{0}93.4) & 317\,(58.3) & 225\,(41.4) \\
21 & 3 & US & $\delta$ Sco* & 9.9 & 9.8$^{+1.2}_{-1.4}$ & 691 & 674\,(\phantom{0}97.5) & 512\,(74.1) & 382\,(55.3) \\
22 & 35 & NE & L134/L183 & 1.0 & 9.6$^{+1.7}_{-2.2}$ & 24 & 24\,(100.0) & 8\,(33.3) & 3\,(12.5) \\
23 & 25 & LCC & $\eta$ Cham & 7.8 & 9.4$^{+1.4}_{-0.9}$ & 30 & 23\,(\phantom{0}76.7) & 15\,(50.0) & 9\,(30.0) \\
24 & -- & TWA & TWA-a & 9.9 & 9.0$^{+2.0}_{-1.0}$ & 44 & 44\,(100.0) & 36\,(81.8) & 25\,(56.8) \\
25 & 24 & LCC & $\epsilon$ Cham & 9.9 & 8.8$^{+0.6}_{-0.4}$ & 39 & 36\,(\phantom{0}92.3) & 32\,(82.1) & 25\,(64.1) \\
26 & 29 & CrA & CrA-Main & 9.9 & 8.5$^{+2.0}_{-2.4}$ & 96 & 85\,(\phantom{0}88.5) & 72\,(75.0) & 47\,(49.0) \\
27 & 4 & US & $\beta$ Sco* & 9.8 & 7.6$^{+0.8}_{-0.7}$ & 285 & 263\,(\phantom{0}92.3) & 206\,(72.3) & 166\,(58.2) \\
28 & -- & TWA & TWA-b & 9.9 & 6.0$^{+2.0}_{-1.0}$ & 22 & 22\,(100.0) & 22\,(100.0) & 19\,(86.4) \\
29 & 13 & UCL & Lupus 1-4* & 7.7 & 6.0$^{+0.6}_{-0.9}$ & 226 & 199\,(\phantom{0}88.1) & 113\,(50.0) & 61\,(27.0) \\
30 & 2 & US & $\nu$ Sco* & 10.0 & 5.8$^{+1.8}_{-0.5}$ & 150 & 147\,(\phantom{0}98.0) & 118\,(78.7) & 98\,(65.3) \\
31 & 1 & US & $\rho$ Oph/L1688* & 9.9 & 3.8$^{+0.4}_{-0.4}$ & 535 & 507\,(\phantom{0}94.8) & 346\,(64.7) & 250\,(46.7) \\
32 & 26 & Pipe & B59* & 4.7 & 3.4$^{+3.1}_{-0.9}$ & 32 & 28\,(\phantom{0}87.5) & 9\,(28.1) & 4\,(12.5) \\
\cmidrule(lr){1-10} 
33 & 11 & UCL & $\mu$ Sco & 9.9 & 17.2$^{+0.9}_{-2.4}$ & 54 & 52\,(\phantom{0}96.3) & 14\,(25.9) & 2\,(\phantom{0}3.7) \\
34 & 32 & Cham & Centaurus-Far & 0.1 & 8.5$^{+1.1}_{-1.3}$ & 99 & 99\,(100.0) & 25\,(25.3) & 9\,(\phantom{0}9.1) \\
35 & 33 & Cham & Chamaeleon-I & 9.9 & 3.8$^{+1.9}_{-0.9}$ & 192 & 184\,(\phantom{0}95.8) & 106\,(55.2) & 89\,(46.4) \\
36 & 34 & Cham & Chamaeleon-II & 7.7 & 2.8$^{+0.7}_{-1.1}$ & 54 & 49\,(\phantom{0}90.7) & 32\,(59.3) & 19\,(35.2) \\

\hline
\end{tabular}
} 
\renewcommand{\arraystretch}{1}
\label{tab:overview}
\tablefoot{
Col.\,1: Running index for the listed clusters, ordered by decreasing age for the top 32 rows. The four clusters at the bottom are not included in our analysis but are listed here for completeness. 
Col.\,2: \texttt{SigMA} label from \citetalias{Ratzenboeck2023a}. 
Col.\,3: Position of clusters within traditional regions (TR), see Appendix~\ref{apx:cluster-sample}.  
Col.\,4: Cluster names from \citetalias{Ratzenboeck2023a} or \citetalias{MiretRoig2025}.
Col.\,5: Stability threshold for each cluster used for quality criteria. 
Col.\,6: Cluster ages and their uncertainties from \citetalias{Ratzenboeck2023b} and \citetalias{MiretRoig2025} (PARSEC, BPRP).
Cols.\,7-10 list the number of sources for: 
all stellar members per cluster from the literature; 
used sources after applying stability cut for the position sample; 
all stellar members with available RV data; 
used sources after applying quality criteria for the velocity sample; 
the percentage (\%) of remaining sources relative to \textit{XYZ}$_\mathrm{all}$ is given in brackets.
The seven clusters marked with an asterisk (*) in Col.~4 are identified as ``peculiar'' concerning their relative motions within Sco-Cen (see Sect.~\ref{sec:tangential}).
}
\end{small}
\end{table*}

\subsection{Quality criteria} \label{apx:quality}

First, we use the stability criterion per stellar cluster member that is provided in the \texttt{SigMA} catalogue \citepalias{Ratzenboeck2023a}. This criterion indicates how often individual sources appear throughout the ensemble of clustering solutions per cluster, denoted as stability in the range from 0--100\%; hence, sources with low stability are less reliable cluster members. The maximum stability per cluster is variable, while generally higher for richer clusters, since they have been more often retained by \texttt{SigMA}. Hence, we apply a variable stability cut per cluster, which is set as follows. \citetalias{Ratzenboeck2023a} find that 
stars in more ``stable'' clusters (generally the richer and more massive ones) have generally higher stability values ($\gtrsim$80\%). \citetalias{Ratzenboeck2023a} suggest that a cut at about \texttt{stability}\,$>$\,11\% would be a reasonable discriminator for such clusters to select more reliable stellar cluster members.
We decide to create a variable stability cut to account for the fact that the stability values of stars within sparser clusters are generally lower. First, we calculate the median stability per cluster, and then we use 10\% of the median as the threshold, which gives less conservative values than the suggested 11\%. The resulting value is used as the individual stability threshold per cluster: 
$\mathtt{stab\_lim} = \mathtt{median\_stability} \,\text{(per cluster)} \cdot 0.1$. 
This value is listed in Table~\ref{tab:overview} (Col.~5). We then only keep stars with: $\mathtt{stability} > \mathtt{stab\_lim}$.
This procedure gives lower stability thresholds than 11\% for all clusters, which ensures that we only remove the most obvious outliers, which is in particular critical for the less prominent clusters. 
This variable stability cut is applied both to the positional and the velocity data, while the latter requires additional quality checks for the RV data, as follows. 

The sample size of the 3D velocity data ($UVW$) strongly depends on the available RV data (Table~\ref{tab:ref}). We find that about 50\% of the sources (6540/13\,011) have been targeted by at least one RV survey. Considering these sources, approximately 34\% have been targeted by more than one survey (2205/6540). This sample can be used to identify binary candidates. Additionally, we investigate the \textit{Gaia} DR3 RUWE parameter \citep{Castro-Ginard2024} that indicates if a source is a probable binary or multiple stellar system, further discussed in Appendix~\ref{apx:methods:calc-veldisp}.

To obtain a clean sample of RVs per cluster while retaining a significant number of sources, we apply the following quality criteria. In a first step, we retain only sources with low RV uncertainties, using a cut at $e_\mathrm{RV}<3.1$\,km\,s$^{-1}$ (using $e_\mathrm{RV,corr}<3.1$\,km\,s$^{-1}$ when a correction was applied). 
We decided on a relatively non-conservative cut to ensure that clusters with few RV measurements contain significant numbers of sources with RV information. 

For sources with multiple observations from different surveys, we first calculate the weighted mean and weighted standard deviation as follows:
\begin{small}
\begin{equation}
    \label{eq:weighted-mean}
    \begin{aligned}
        &\bar{v}_{\mathrm{RV}\mathit{w}} = \frac{\sum{w_i v_{\mathrm{RV}i}}}{\sum{w_i}}, \\
        &e_{\mathrm{RV}\mathit{w}} = \sqrt{\frac{\sum{w_i}(v_{\mathrm{RV}i} - \bar{v}_{\mathrm{RV}w})^2}{\sum{w_i}} \cdot  \frac{N}{N-1}},  
    \end{aligned}
\end{equation}
\end{small}
where $w_i = 1/e_{\mathrm{RV}i}^2$ are the weights.
For our analysis, if a source was observed multiple times, we use the RV measurement from the survey with the lowest error (named $v_{\mathrm{RV,best}}$). 
The weighted mean and standard deviation are used to test for significant deviations between the surveys, to remove potential binary or multiple candidates (similar to the scatter in SDSS). 
We keep sources with $|v_{\mathrm{RV,best}} - \bar{v}_{\mathrm{RV}\mathit{w}}| < 3.0$\,km\,s$^{-1}$ and $e_{\mathrm{RV}\mathit{w}}<3.1$\,km\,s$^{-1}$.
We decided for this cut to be comparable to the used RV error cut introduced above.

After these applied cuts, there are still some RVs per cluster that are clearly scattered beyond the clusters' median RVs. Significant deviations in the motions of single stars from the bulk motion of the parent cluster could have several reasons, such as unresolved binaries, mismatches, or contamination from unrelated stellar populations or field stars. 
To clean the RV sample per cluster, we exclude potential contaminants or outliers first via a global outlier cut and second via sigma clipping. 
We use the LSR corrected RVs ($v_{\mathrm{RV,}\mathit{lsr}}$); they show less scatter, since artificial trends are removed that appear when using the heliocentric RVs.
First, when considering the RV distribution of the whole Sco-Cen association, we find that sources scattered beyond $-2.5 \leq v_{\mathrm{RV,}\mathit{lsr}}\,(\mathrm{km}\,\mathrm{s}^{-1}) \leq 18.0$ are general outliers, and we globally remove sources beyond these boundaries. Next, we remove outliers with significant deviations from the bulk motion of each cluster by applying sigma clipping at $3\sigma$ around each cluster's median($v_{\mathrm{RV,}\mathit{lsr}}$). In most cases, this additional step removes only a few outliers or none.  

Finally, the combined quality criteria, including the stability cut, the error cuts, removal of binary candidates, and sigma-clipping, leave 3240 (25\%) sources out of 13,011. The individual numbers and fractions per cluster are given in Table~\ref{tab:overview}.
We compare the weighted mean with the finally selected {\it best} RVs per source (when observed multiple times) after applying the mentioned quality criteria. 
We do not find a significant deviation for the majority of these sources: 
$\mathrm{median}( v_{\mathrm{RV,best}} - \bar{v}_{\mathrm{RV}\mathit{w}} )  = (0.002^{+0.112}_{-0.134})$\,km/s, underlining the robustness of these selected RV values.

\section{Detailed methods descriptions} \label{apx:methods}

\subsection{Calculating the cumulative velocity dispersion} \label{apx:methods:calc-veldisp}

Here, we give more details on the construction of Fig.~\ref{fig:cumul-vel-disp}, which shows the change of the cumulative 3D velocity dispersion with cluster age or formation time. The cumulative \sigdrei is calculated chronologically by incorporating the stellar members of the clusters that lie within a certain age threshold. In more detail, we start to calculate \sigdrei by using the stellar members of the oldest cluster, \elup, and then add chronologically at each step the stars of the next youngest cluster, until we reach the final step that contains all stars in Sco-Cen, from old to young. At each step, we calculate the total velocity dispersion of the stellar members that formed before the time threshold as given on the x-axis of Fig.~\ref{fig:cumul-vel-disp}, which results in the cumulative velocity dispersion over a time span of about 20\,Myr. 

To account for the different sample sizes of each cluster---several clusters contain only a handful of RVs (see Table~\ref{tab:overview})---we use at each time-step a random subsample of 20 sources per cluster. This sub-sampling is done to avoid the dispersion estimations being dominated by the sample size, which would naturally give more weight to the more massive clusters or the clusters that contain a higher number of observed RVs. The relatively low number of 20 is chosen to account for the fact that several clusters have very low number statistics, especially when requiring good RV data quality. If a cluster has fewer than 20 members, then we randomly sample 20 data points with replacement. This up-sampling concerns nine out of the 32 clusters.  
We repeat this procedure 1000 times, and each time randomly draw another 20 stars from each cluster, which delivers a median and uncertainty for \sigdrei at each time step, as plotted in Fig.~\ref{fig:cumul-vel-disp}.
By iteratively adding random sub-samples of stellar members of each cluster one after another, we obtain an increase in velocity dispersion, until the total velocity dispersion of the whole association is reached with 4.64$\pm$0.04\,km\,s$^{-1}$ (see also Table~\ref{tab:sig3D}).

\begin{figure}[!t]
    \centering
        \includegraphics[width=1\linewidth]{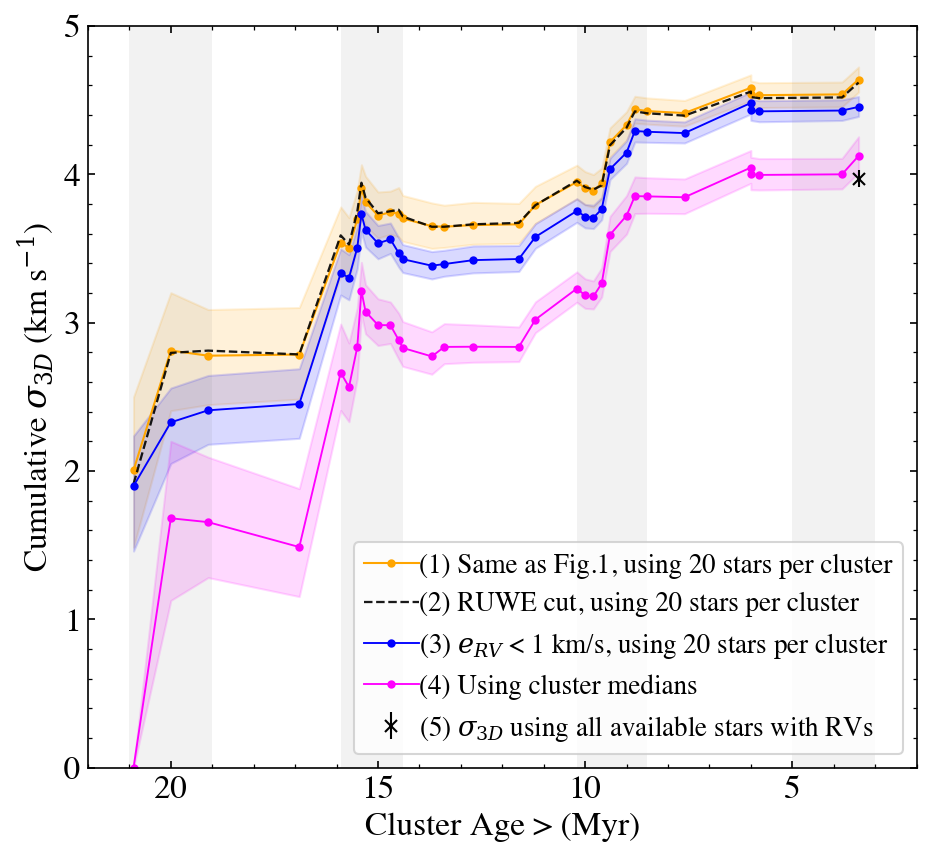}
    \caption{Influence of measurement errors or binaries on the cumulative\,\sigdrei. (1) The upper orange curve is the same as in Fig.~\ref{fig:cumul-vel-disp}. (2) The black, dashed curve shows the cumulative \sigdrei when applying a cut on the \textit{Gaia} RUWE parameter, to test the influence of binaries. (3) The middle blue curve shows the cumulative \sigdrei when applying an additional error cut at $e_{\mathrm{RV}} \leq 1$\,km\,s$^{-1}$ on the stellar members of each cluster. (4) The lower magenta curve shows the cumulative \sigdrei when using the cluster medians. (5) The black cross marks the total \sigdrei as calculated from all stars in Sco-Cen with valid RV measurements. The error bars (shaded regions) show the 95\% interquartile ranges (2$\sigma$).
    }
    \label{fig:cumul-vel-clusters}
\end{figure}

As outlined in Sect.~\ref{sec:results}, we compare the total \sigdrei from the sub-sampling approach to the total \sigdrei when all available stellar members of the 32 clusters in Sco-Cen are used for its calculation. Hence, we use all stars in the 32 clusters that pass our quality criteria (i.e., 3123 out of 12,612), instead of 20 sub-samples from each cluster (the latter results in a sample of only 640 stars when the final time-step is reached). We bootstrap over the 3123 available sources with replacement 10,000 times. The mean of the bootstrap samples results in a 3D velocity dispersion of 3.97$\pm$0.03\,km\,s$^{-1}$,
which is about 0.7\,km/s lower than the final value of the cumulative \sigdrei. This difference might be caused by the undersampling of the velocity space when only using sub-samples, which might also increase the importance of individual measurement errors, which could inflate the final \sigdrei. At the same time, the total velocity dispersion using all available stars might be dominated by the most massive clusters, which contain overall more stars with valid RV measurements.

\begin{table}[!t]
\begin{small}
\caption{
Results of the total \sigdrei of the whole Sco-Cen association when using the different, mentioned data samples. 
See also the final step of each trend in Fig.~\ref{fig:cumul-vel-clusters}.
}
\renewcommand{\arraystretch}{1.2}
\centering
\resizebox{1\columnwidth}{!}{
\begin{tabular}{lcc}

\hline \hline
\multicolumn{1}{c}{Used sample} &
\multicolumn{1}{c}{\sigdrei} &
\multicolumn{1}{c}{$e_{\sigma_\mathrm{3D}}$} \\

\multicolumn{1}{c}{} &
\multicolumn{1}{c}{km/s} &
\multicolumn{1}{c}{km/s} \\

\cmidrule(lr){1-3}
(1) Using fiducial sample as in Fig.\,1 ($e_\text{RV} < 3.1$\,km/s) & 4.638 & 0.044 \\ 
(2) Using the RUWE-cut (exclude binary candidates) & 4.620 & 0.045 \\
(3) Using stricter error-cut ($e_\text{RV} < 1.0$\,km/s) & 4.457 & 0.034 \\ 
(4) Using the cluster medians & 4.126 & 0.060 \\
(5) Using all stars in Sco-Cen with valid RVs & 3.970 & 0.029 \\

\hline
\end{tabular}
} 
\renewcommand{\arraystretch}{1}
\label{tab:sig3D}
\tablefoot{
The first three values are calculated with the method of using iteratively 20 subsamples per cluster, the fourth value is calculated by using the cluster medians, and the final value is calculated by using all stars with valid RVs in the 32 Sco-Cen clusters (3123 sources), after the applied quality criteria. The given values are the mean and its uncertainty (standard deviation) as obtained from bootstrap or error sampling.
}
\end{small}
\end{table}

To test the influence of RV measurement errors, we recalculate the cumulative \sigdrei using two alternative data samples.
First, we apply an additional RV error cut at $e_{\mathrm{RV}} < 1$\,km\,s$^{-1}$ (and $e_{\mathrm{RV}\mathit{w}} < 1$\,km\,s$^{-1}$ if a source was observed by more than one survey). In some cases, this reduces the number of stellar members per cluster significantly; for 13 clusters, there are now fewer than 20 members available, so their numbers have to be up-sampled. 
Nevertheless, when comparing the resulting trend to the original one form Fig.~\ref{fig:cumul-vel-disp}, we find that the cumulative velocity dispersion increases similarly over time, while there is only a shift to overall lower values, on average about $-0.2$\,km\,s$^{-1}$ (see Fig.~\ref{fig:cumul-vel-clusters}, blue curve), and the total \sigdrei results in 4.46$\pm$0.03\,km\,s$^{-1}$.

Second, we calculate the cumulative velocity dispersion by using only the median $UVW$ values of the clusters. With this approach, we remove any internal velocity scatter of each cluster, and we only focus on the more robust cluster velocities; so we get a measure of the lower limit of \sigdrei. We sample from cluster median $UVW$ velocities and their uncertainties (assuming Gaussian errors) to get a measure of uncertainty for the clusters' cumulative \sigdrei.
Naturally, this approach results in low-number statistics, especially for the first few time-steps (at the start, only one velocity is used, from \elup, hence, it is zero).
Nevertheless, we find that the shape of the curve is similar, as highlighted in Fig.~\ref{fig:cumul-vel-clusters} (magenta curve), with an average shift of about $-0.5$~to~$-0.6$\,km\,s$^{-1}$. The total 3D velocity dispersion of all 32 cluster medians results in 4.13$\pm$0.06\,km\,s$^{-1}$. 

Finally, we test the potential influence of unresolved binaries or multiples. We use \textit{Gaia}'s Renormalised Unit Weight Error (RUWE), and a variable limit (RUWE$_\text{var-lim}$) as given by \citet[][]{Castro-Ginard2024} (see their Appendix~A). The suggested RUWE$_\text{var-lim}$ mainly depends on crowding at the location of a star in the field-of-view of \textit{Gaia}. We repeat the calculation of the cumulative \sigdrei by additionally requiring RUWE\,<\,RUWE$_\text{var-lim}$ and compare it to the cumulative \sigdrei from Fig.~\ref{fig:cumul-vel-disp}. We find that the RUWE cut does not make a significant difference (Fig.~\ref{fig:cumul-vel-clusters}, black, dashed line). 
Overall, we only find small deviations of about 0.01--0.03\,km\,s$^{-1}$ around the original trend. 
The total \sigdrei when using the RUWE cut is then 4.62$\pm$0.05\,km\,s$^{-1}$.
We conclude that the influence of unresolved binaries is negligible concerning our selected \textit{UVW} sample and that the influence of RV measurement errors is dominating any potential shifts of \sigdrei.

Considering these different tests, we conclude that the present-day velocity dispersion of Sco-Cen is between 4.0--4.7\,km\,s$^{-1}$ (see the overview in Table~\ref{tab:sig3D}). 
The measurement errors appear to influence the magnitude of the cumulative \sigdrei, while the increasing trend, with the visible jumps and plateaus in between, appears to be unaffected and stays essentially the same.

\subsection{Calculating the cumulative size of the region} \label{apx:methods:calc-size}

To estimate the cumulative size ($S$, pc) of Sco-Cen, we estimate the change of the extent of the region when ordering the stars by the ages of their parent clusters, starting with the oldest cluster, similar to the cumulative \sigdrei. 
We like to note that orbital tracebacks of the star and cluster positions have not been considered in this work, which will be part of future work.
To estimate the size of a collection of subgroups at any age step, we determine the largest distance between any two sources at this step. Since Euclidean distances ignore the shape of Sco-Cen, we use a graph constructed from individual members (as nodes) to approximate the manifold of the association. We construct the graph via a k-neighbours graph in positional ($XYZ$) space and use Dijkstra's algorithm to compute shortest paths between any two sources in the graph \citep{Dijkstra1959}. At each age step, we compute the maximum distance for all possible shortest paths between members of respective subgroups. We use the Python packages:  \texttt{sklearn.neighbors.kneighbors\_graph},  \texttt{networkx}, \texttt{nx.single\_source\_dijkstra\_path\_length}.

This is done cumulatively, by starting with the stellar members from the oldest cluster, and adding at each step stellar members of the next youngest cluster, while we use only stars that fulfil the stability criterion from Appendix~\ref{apx:quality}. 
We iterate over randomised subsamples per cluster of equal size to get an estimate of the uncertainty and to account for the different cluster sizes. We always add 30 randomly drawn stellar members of each cluster at each cumulative age step. For clusters with fewer members, we add the full set of cluster members of these sparse clusters (no up-sampling, no source appears more than once).  
Using at least 30 sources per cluster allows us to get a similar weight for sparser clusters compared to the richer clusters. Moreover, using fewer stars significantly reduces the computation time of Dijkstra.
We iterate over this procedure 70 times, which delivers a median extent and its uncertainty per age step. With this method, the total present-day size of Sco-Cen yields a value of 205$^{+4}_{-20}$\,pc. See the cumulative $S$ in Fig.~\ref{fig:cumul-size}.

\subsection{Calculating cluster 3D bulk positions and motions} \label{apx:methods-average-3D}

We use the Galactic Cartesian positions ($XYZ$) and motions ($UVW$ or $UVW_\mathrm{LSR}$) of the stellar members of each cluster to calculate the clusters' average properties in the 6D phase space. 
For $XYZ$, we only apply the stability cut as described above. For $UVW$, we use the stellar members that pass our additional quality criteria, with a particular focus on RVs (see Appendix~\ref{apx:quality}). 

The median and mean of the clusters' 3D positions and motions are determined via bootstrapping with replacement (1000 times) by sampling from the stellar members of each cluster. The resulting averaged median, averaged mean, and averaged standard deviation (STD) from the bootstrapped distributions are used to determine the 6D properties of each cluster, reported in an online table (see Appendix~\ref{apx:tables}).
The given uncertainties of median, mean, and STD are the standard deviations of the bootstrapped medians, means, and STDs. For our analysis, we use the medians and their uncertainties.

\begin{table}[!t]
\begin{small}
\caption{Central position and rest velocity of \scocen.}
\renewcommand{\arraystretch}{1.2}
\centering
\begin{tabular}{lrrc}

\hline \hline
\multicolumn{1}{c}{Parameter} &
\multicolumn{1}{c}{Median} &
\multicolumn{1}{c}{Mean} &
\multicolumn{1}{c}{STD} \\

\cmidrule(lr){1-4}
$X$ (pc) & 106.0$\pm$0.6 & 105.0$\pm$0.4 & 35.4 \\ 
$Y$ (pc) & -71.0$\pm$0.6 & -70.3$\pm$0.3 & 25.1 \\ 
$Z$ (pc) &  28.4$\pm$0.3 &  28.6$\pm$0.2 & 14.6 \\

\cmidrule(lr){1-4}
$U$ (km/s) &  -5.98$\pm$0.11 & -6.05$\pm$0.07 & 2.38 \\
$V$ (km/s) & -19.99$\pm$0.06 &-20.04$\pm$0.05 & 1.67 \\
$W$ (km/s) &  -5.28$\pm$0.04 & -5.27$\pm$0.03 & 1.15 \\

\hline
\end{tabular}
\renewcommand{\arraystretch}{1}
\label{tab:bulk}
\tablefoot{The given Median and Mean are the averaged medians and means, which were calculated by bootstrapping 5000 times over the selected stellar members. The given errors are the standard deviations of the 5000 medians or means. The given STD is the average scatter of the $XYZ$ or $UVW$ distribution. The $UVW_\mathrm{LSR}$ values can be derived from $UVW$ by adding the standard solar motion form \citet{Schoenrich2010}.
}
\end{small}
\end{table}

\begin{table}[!t]
\begin{small}
\caption{The difference in positions and motions when comparing the three reference points that are investigated in Figs.~\ref{fig:age-speed-distance}, \ref{fig:hubble-flow}, and \ref{fig:speed+hubble-other}.}
\renewcommand{\arraystretch}{1.2}
\centering
\begin{tabular}{lccc}

\hline \hline
\multicolumn{1}{c}{} &
\multicolumn{1}{c}{\scocen--\elup} &
\multicolumn{1}{c}{\scocen--\philup} &
\multicolumn{1}{c}{\elup--\philup}  \\

\cmidrule(lr){1-4}
$\Delta X$\,(pc) & $-$13.2 & \phantom{0}$-$5.9 & \phantom{$-$0}7.3 \\
$\Delta Y$\,(pc) & \phantom{$-$0}4.8 & $-$17.2 & $-$22.0 \\
$\Delta Z$\,(pc) & \phantom{0}$-$0.7 & $-$11.6& $-$10.9 \\
$\Delta \mathit{Center}$\,(pc) & \phantom{$-$}14.1 & \phantom{$-$}21.6 & \phantom{$-$}25.6 \\

\cmidrule(lr){1-4}
$\Delta U$\,(km/s) & $-$1.11           & $-$1.20 & $-$0.09 \\
$\Delta V$\,(km/s) & \phantom{$-$}0.72 & $-$0.22 & $-$0.94 \\
$\Delta W$\,(km/s) & $-$1.17           & $-$0.67 & \phantom{$-$}0.49 \\
$\Delta \mathit{Speed}$\,(km/s) & \phantom{$-$}1.76 & \phantom{$-$}1.39 & \phantom{$-$}1.07 \\

\hline
\end{tabular}
\renewcommand{\arraystretch}{1}
\label{tab:delta-speed}
\end{small}
\end{table}

\subsection{Testing different reference points} \label{apx:methods-bulk}

We investigate three points of reference in more detail, which are the two clusters \elup and \philup, and \scocen. The latter represents the bulk position and motion of older clusters with \mbox{ages\,>\,15\,Myr}. From these, we remove two sparse clusters, Pipe-North and $\theta$\,Oph, since they show peculiar motions relative to the rest of the older clusters. Hence, we use eight clusters for the calculation of \scocen, which are   
$e$\,Lup, 
Libra-South, 
US-foreground, 
$\phi$\,lup, 
$\eta$\,Lup, 
V1062-Sco, 
$\nu$\,Cen, 
and $\sigma$\,Cen, 
containing a total of 7290 stellar members.  
To determine the centre and rest velocity, we use the Galactic Cartesian positions and motions of the stellar members of the eight clusters with applied quality criteria as outlined in Appendix~\ref{apx:quality}. This leaves 6659 stellar members in $XYZ$ and 1288 in $UVW$ space. 

The median and mean position and velocity of \scocen are obtained via bootstrapping (5000 times with replacement) over the selected stellar $XYZ$ positions or $UVW$ velocities. 
Table~\ref{tab:bulk} reports the averages of the bootstrapped medians, means, and standard deviations in $XYZ$ and $UVW$. The given uncertainties are the standard deviations of the bootstrap medians or means. For similar estimates, see \citetalias{Posch2025} and \citetalias{MiretRoig2025}, while they used slightly different data, in particular concerning the selected RVs.
The difference in position and motion between \elup, \philup, and \scocen are given in Table~\ref{tab:delta-speed}.

Figure~\ref{fig:speed+hubble-other} shows the resulting speed--time and radial-motion--distance relations, when using the two alternative reference points, \philup and \scocen, compared to when using the oldest cluster, \elup, as used in the main part of the paper (Figs.~\ref{fig:age-speed-distance} \& \ref{fig:hubble-flow}). We see similar, while slightly shallower, trends of speed with time and radial motion with distance.

\subsection{Testing the robustness of the linear relations} \label{apx:test-random-speed}

In this section, we test the robustness of the speed--time and radial-motion--distance relations. In the main part of this paper, we use as a first choice the oldest cluster, \elup, as a reference point. Alternatively, we also set \philup or the bulk motion of \scocen as reference. We find similar correlations (see Figs.~\ref{fig:age-speed-distance}, \ref{fig:hubble-flow}, \ref{fig:speed+hubble-other}, \& Table~\ref{tab:acceleration}), while these three reference points are generally at central locations with similar 3D space motions (see Table~\ref{tab:delta-speed}).

We further test these relations by using each of the 32 clusters once as a reference point to calculate the relative speed or distance. In all cases, the reference cluster is excluded from the linear fit. We fit linear slopes to the resulting distributions in the speed--time and radial-motion--distance spaces, and we test the correlations with the coefficient of determination (r$^2$-value), the Spearman rank correlation coefficient \citep[SRCC;][]{Spearman1904, Daniel1990}, and the Pearson correlation coefficient \citep[PCC;][]{Pearson1895}. SRCC tests for monotonic relationships between two variables and ignores the scale, and PCC evaluates linear relationships using the physical scales.  

For the speed--time relation, we find that the time-ordered case with \elup as reference point delivers one of the cleanest linear trends, with the steepest slope and the highest r$^2$, SRCC, and PCC values. We find that if using several of the older clusters as reference, we get similar linear trends compared to \elup; these are \philup, $\nu$\,Cen, $\eta$\,Lup, V1062-Sco, Sco-Sting, $\rho$\,Lup, $\rho$\,Sco, and UPK606 (all older than 13\,Myr). 

\begin{figure}
\centering
\begin{minipage}[c]{\linewidth}
\centering
    \includegraphics[width=0.97\textwidth]{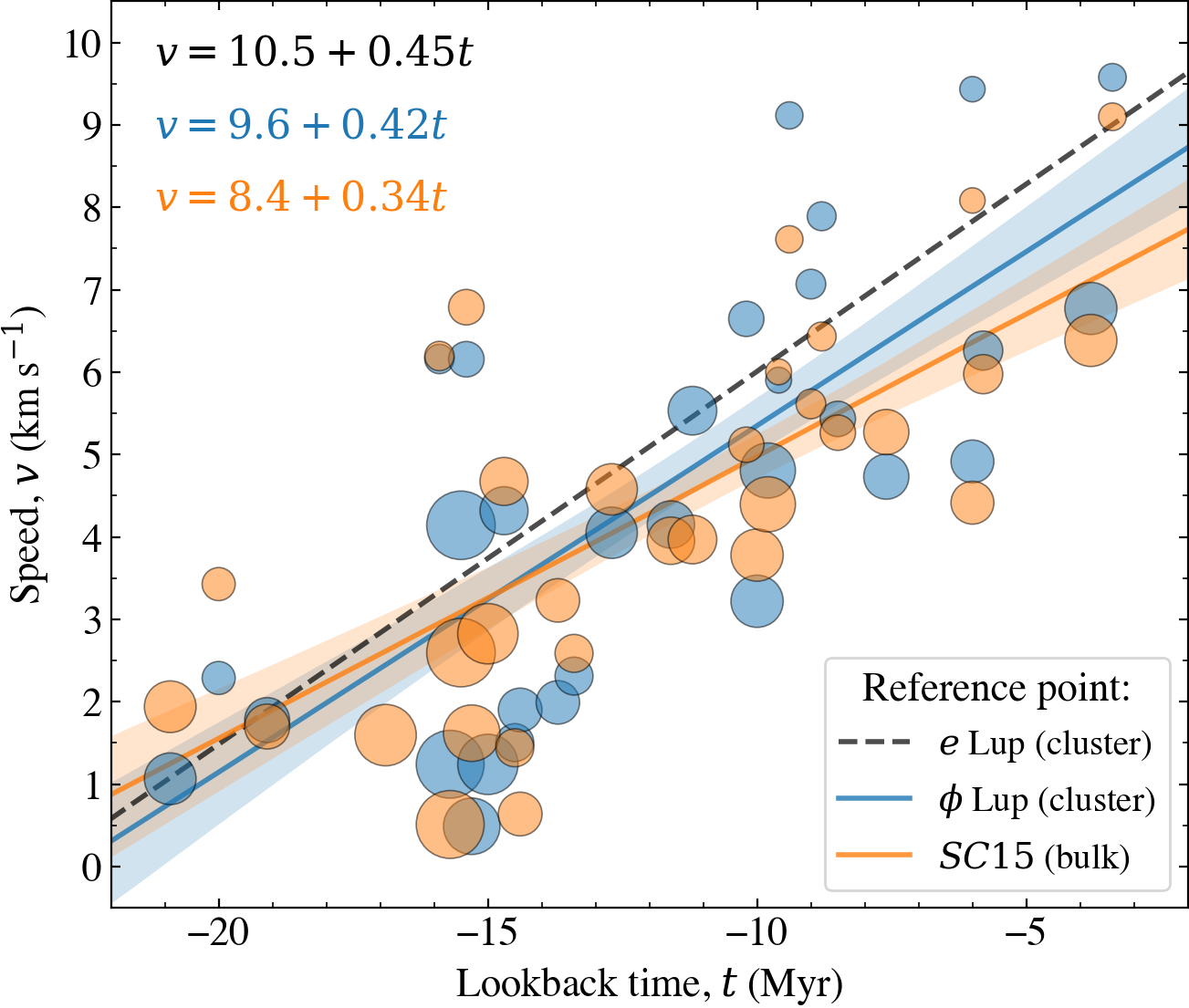}
\end{minipage}
\begin{minipage}[c]{\linewidth}
\centering
    \includegraphics[width=0.97\textwidth]{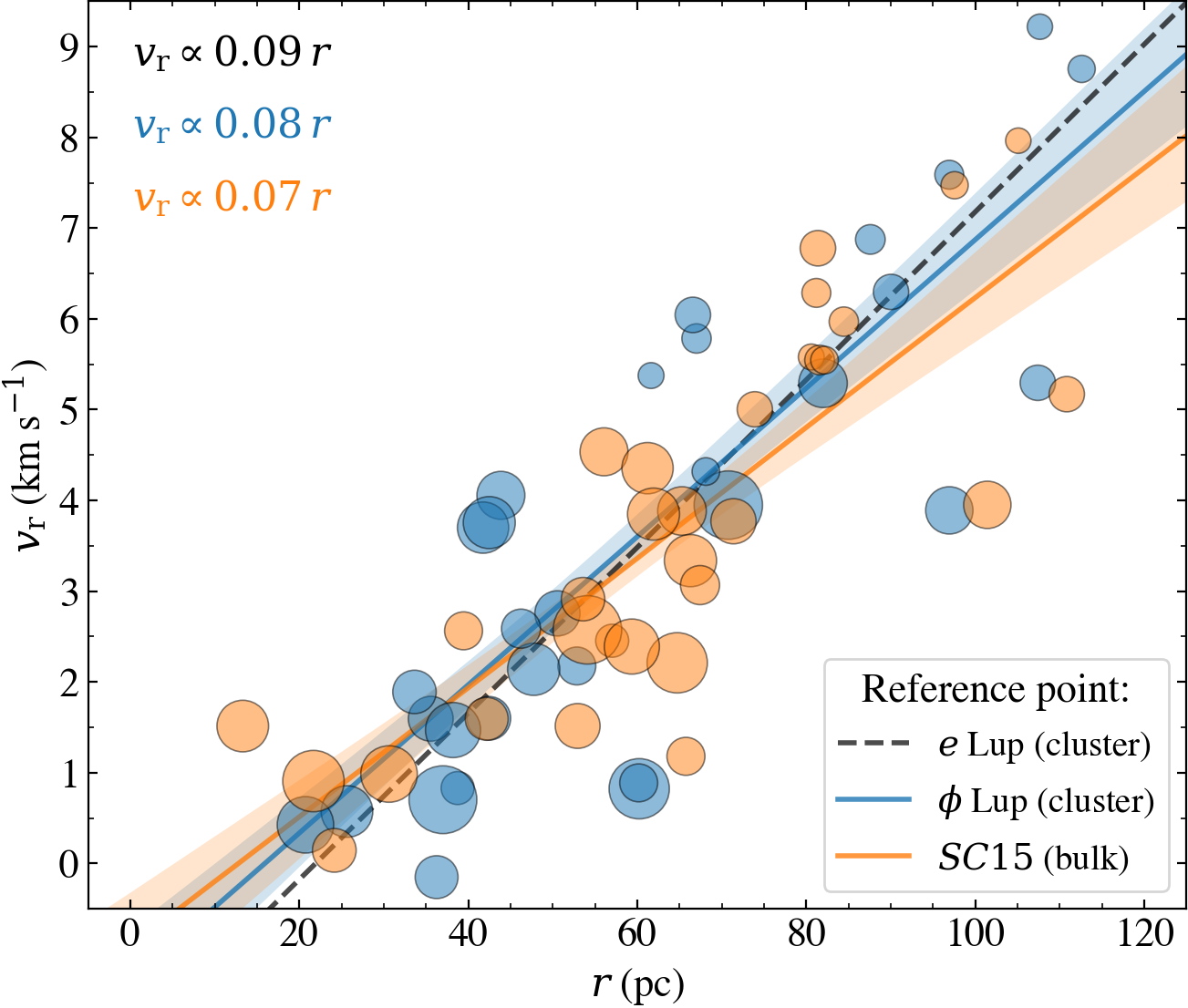}
\end{minipage}
\caption{
    Speed--time relation ({\it top panel}, similar as Fig.~\ref{fig:age-speed-distance}) and Radial-component--distance relation ({\it bottom panel}, similar as Fig.~\ref{fig:hubble-flow}), with \philup (blue) or \scocen (orange) as reference points. 
    For comparison, we also show the slopes when \elup is set as the reference point, from Fig.~\ref{fig:age-speed-distance} or \ref{fig:hubble-flow} (black, dashed lines). The resulting median slopes are given in the upper left corners in the respective colours (see also Table~\ref{tab:acceleration}).
    The shaded areas show the 1-$\sigma$ uncertainties around the median slopes.
    }
\label{fig:speed+hubble-other}
\end{figure}

For the radial-motion--distance relation, we see a similar behaviour, while here \elup stands out more clearly as the best fit with the steepest linear slope. Under ideal conditions, the best reference point (kinematic centre) does correspond to the steepest (and cleanest) positive slope for an approximately isotropic radial expansion pattern, because any offset from the true centre introduces non-radial components, which would flatten the correlation and increase the scatter. 

That several of the older clusters appear to be also relatively good reference points, compared to \elup, is consistent with the fact that the older clusters make up the bulk of Sco-Cen, they have generally similar space motions, and are located at central locations. The youngest of these older clusters is UPK606, with an age of about 13--14\,Myr. This still fits a formation scenario, where feedback originating from the older clusters was mainly affecting younger clusters with ages younger than about 12\,Myr, which approximately marks the onset of the formation of cluster chains. 

Finally, we test if the speed--time relation is indeed strongly linked to the formation time, since we find linear trends when using \elup or several of the older clusters as reference points. To this end, we randomise the order of the time-axis, while we keep the speed-order as calculated for a given reference point. We set each of the 32 clusters once as a reference point, while the reference cluster is then removed from the correlation.   
We test if the time-ordered relations are significant by randomly shuffling the measured formation times (or cluster ages) on the x-axis. Additionally, we create a uniform distribution of times between the formation time limits of the Sco-Cen clusters, $(-20.9,\,-3.4)$\,Myr, to create cases that are within the physical boundaries of the observed case. This allows us to determine the linear regression statistics. For a more robust test, we determine the SRCC, which ranks the order of the values on the x and y-axis, and is now independent of the true physical scale.   

We find that the results from the time-ordered cases---with \elup, \philup, or SC15 as reference point---are only reproduced with a probability that is beyond 4$\sigma$. The positive linear correlation of time versus speed is not reproduced by the vast majority of the randomly ordered cases. About half of the random cases show negative trends (reversed slopes), which we also find when setting some of the mostly younger clusters as reference point (concerning Pipe-N, Musca-fg, $\eta$\,Cham, $\epsilon$\,Cham, TWA-a,b, $\delta$\,Sco, $\nu$\,Sco, $\rho$\,Oph, B59).
This test supports the hypothesis that the time-ordered behaviour of the motions in Sco-Cen, where younger clusters move faster compared to older ones, is unlikely created by a random process in which clusters would form independently from each other, but it lies at the bottom of the formation history of the region.

\section{Online Material} \label{apx:tables}

We provide an online source catalogue (available on Vizier), listing all stellar members of the investigated Sco-Cen clusters, including information on cluster labels and names, ages, derived parameters, and the used RV data. The final used RV value is either the measurement with the lowest RV uncertainty or the only available RV measurement. 
In column \texttt{Ref}, the respective RV References are listed, which indicate from which survey the measurement was taken (see Table~\ref{tab:ref}). Additionally, we give the weighted mean and weighted standard deviation if a source has more than one RV measurement from different surveys. This gives information on possible multiples. We also list the scatter parameter from SDSS if available (\texttt{std\_v\_rad}). Finally, the derived parameters, including \textit{XYZ} and \textit{UVW}, are listed as well (see also Sect.~\ref{apx:parameters}). We flag the sources that pass our quality criteria in column \texttt{used\_stab} and \texttt{used\_rvs}. The first is used for the 5D sample ($\texttt{used\_stab} = 1$) and the second for the 6D sample ($\texttt{used\_rvs}=1$). 

The cluster properties are reported in a second online catalogue (available on Vizier), containing median, mean, standard deviation, and their errors for the investigated parameters (as obtained via bootstrapping), which are the positions and motions (heliocentric and LSR corrected), including l, b, X, Y, Z, RV\_HEL, RV\_LSR, v\_l\_lsr, v\_b\_lsr, U, V, W, U\_LSR, V\_LSR, W\_LSR. Moreover, we list in a third online catalogue (available on Vizier) the medians and uncertainties of the cumulative \sigdrei and size for each time step, which were used to construct Figs.~\ref{fig:cumul-vel-disp} and \ref{fig:cumul-vel-clusters}.

\end{appendix}
\end{document}